\documentclass[12pt]{article}

\pdfoutput=1
\usepackage{color}
\usepackage{epsfig, palatino}
\usepackage{pstricks,pst-node,pst-tree}
\usepackage{epic}
\usepackage{mathrsfs}
\usepackage{ae} 
\usepackage[T1]{fontenc}
\usepackage[ansinew]{inputenc}
\usepackage{amsmath}
\usepackage{amssymb}
\usepackage{graphicx}
\usepackage{ulem}
\usepackage{color}
\definecolor{darkblue}{cmyk}{0.9,0.9,0,0}
\usepackage[colorlinks=true,linkcolor=darkblue,citecolor=darkblue,urlcolor=darkblue]{hyperref}
\usepackage{cite}
\usepackage{hyperref}
\usepackage{wasysym}
\usepackage{varioref}
\usepackage{makeidx}
\usepackage[english]{babel}
\usepackage{simplewick}
\usepackage{array}
\usepackage{multirow}

\usepackage[font={small}]{caption}

\usepackage{pdflscape}

\usepackage{epsfig, palatino}
\usepackage{pstricks,pst-node,pst-tree}
\usepackage{rotating}

\usepackage{subfigure}

\usepackage{tikz}
\usetikzlibrary{automata}
\usetikzlibrary{arrows}
\usetikzlibrary{calc}
\usetikzlibrary{decorations.markings}
\usetikzlibrary{decorations.pathreplacing}
\usetikzlibrary{intersections}
\usetikzlibrary{positioning}
\usetikzlibrary{topaths}
\usetikzlibrary{shapes.geometric}
\usetikzlibrary{shapes.misc}
\usetikzlibrary{arrows,decorations.markings}
\tikzset{->-/.style = {
    decoration = {markings, mark = at position #1 with {\arrow{>}}},
    postaction = {decorate}}}

\usepackage{float}

\newcommand{\comment}[1]{}

\newcommand{\beq}{\begin{equation}}
\newcommand{\eeq}{\end{equation}}
\newcommand{\beqq}{\begin{equation*}}
\newcommand{\eeqq}{\end{equation*}}
\newcommand\beqa{\begin{eqnarray}}
\newcommand\eeqa{\end{eqnarray}}
\newcommand\beqaa{\begin{eqnarray*}}
\newcommand\eeqaa{\end{eqnarray*}}
\newcommand\bea{\begin{array}}
\newcommand\eea{\end{array}}

\newcommand{\nn}{\nonumber}

\newcommand{\neqa}{\nonumber\end{eqnarray}} 
\newcommand{\la}[1]{\label{#1}}

\renewcommand{\d}{\partial}

\newcommand{\<}{{\langle}}
\renewcommand{\>}{{\rangle}}

\newcommand{\re}{\relax{\rm I\kern-.18em R}}

\renewcommand{\sp}{p\hspace{-.40em}/}

\definecolor{darkgreen}{rgb}{0.0, 0.45, 0.0}

\def\XXint#1#2#3{{\setbox0=\hbox{$#1{#2#3}{\int}$}
\vcenter{\hbox{$#2#3$}}\kern-.5\wd0}}

\def\tr{{\rm tr~}}

\def\su2{{SU(2)}}

\def\a{{\alpha}}

\def\[{\left[}
\def\]{\right]}

\def\a{\alpha}

\def\({\left(}
\def\){\right)}
\def\[{\left[}
\def\]{\right]}

\def\<{\langle}
\def\>{\rangle}

\def\i2{\frac{i}{2}}

\def\spi{\relax{\rm \pi\kern-0.5em /}}
\def\sA{\relax{\rm A\kern-0.5em /}}
\def\sp{\relax{\rm p\kern-0.5em /}}
\def\sd{\relax{\rm \d\kern-0.5em /}}
\def\sk{\relax{\rm k\kern-0.5em /}}
\def\sn{\relax{\rm n\kern-0.5em /}}
\def\sl{\relax{\rm l\kern-0.5em /}}
\def\sP{\relax{\rm P\kern-0.7em /}}
\def\sBethe{\relax{\rm \Bethe\kern-0.5em /}}

\def\2F1{\,_2{\rm F}_1}

\makeindex

\begin{document}

\thispagestyle{empty}

\renewcommand{\thefootnote}{\fnsymbol{footnote}}
\setcounter{page}{1}
\setcounter{footnote}{0}
\setcounter{figure}{0}

\begin{center}
$$$$
{\Large\textbf{\mathversion{bold}
Asymptotic Four Point Functions
}\par}

\vspace{1.0cm}

\textrm{Benjamin Basso$^\text{\tiny 1}$, Frank Coronado$^\text{\tiny 2,\tiny 3}$, Shota Komatsu$^\text{\tiny 2}$, \\
Ho Tat Lam$^\text{\tiny 2,\tiny 4}$, Pedro Vieira$^\text{\tiny 2,\tiny 3}$, De-liang Zhong$^\text{\tiny 1}$}
\\ \vspace{1.2cm}
\footnotesize{\textit{
$^\text{\tiny 1}$Laboratoire de Physique Th\'eorique de l'\'Ecole Normale Sup\'erieure, CNRS, Universit\'e de Recherche PSL, \\ Sorbonne Universit\'es, Universit\'e Pierre et Marie Curie, 24 rue Lhomond, 75005 Paris, France\\
$^\text{\tiny 2}$Perimeter Institute for Theoretical Physics, Waterloo, Ontario N2L 2Y5, Canada \\
$^\text{\tiny 3}$ICTP South American Institute for Fundamental Research, IFT-UNESP, S\~ao Paulo, SP Brazil 01440-070 \\
$^{\text{\tiny 4}}$Department of Physics, Princeton University, Princeton, NJ 08544
\vspace{4mm}
}}

\par\vspace{1.5cm}

\textbf{Abstract}\vspace{2mm}
\end{center}
We initiate the study of four-point functions of large BPS operators at any value of the coupling. We do it by casting it as a sum over exchange of superconformal primaries and computing the structure constants using integrability. Along the way, we incorporate the nested Bethe ansatz structure to the hexagon formalism for the three-point functions and obtain a compact formula for the asymptotic structure constant of a non-BPS operator in a higher rank sector.

\noindent

\setcounter{page}{1}
\renewcommand{\thefootnote}{\arabic{footnote}}
\setcounter{footnote}{0}

\setcounter{tocdepth}{2}

 \def\nref#1{{(\ref{#1})}}

\newpage

\tableofcontents

\parskip 5pt plus 1pt   \jot = 1.5ex

\newpage
\section{Introduction} 

Four-point correlation functions are probably the most interesting entities in a conformal field theory. While two- and three- point functions are kinematically constrained by conformal symmetry, four point functions can depend on conformal cross-ratios and will be strikingly different for different conformal theories with different physics. 

\textit{In principle}, the spectrum and operator product expansion (OPE) coefficients of a conformal field theory entail a full non-perturbative solution of a conformal field theory since they can be put together to construct any higher point function. \textit{In practice}, it is usually unpractical to compute \textit{all} needed spectra and three point functions and then preform the sum over \textit{all} possible exchanged operators appearing in the OPE to finally obtain the four point correlator. 

In planar $\mathcal{N}=4$ Super Yang-Mills theory integrability comes to the rescue and renders this task feasible. In this paper, we will construct planar four point functions of \textit{large} BPS operators at any value of the 't Hooft coupling from the knowledge of two- and three- point functions which in turn can be computed by means of integrability. We shall be dealing with large enough external operators so that so-called wrapping corrections can be discarded; we denote such four point functions as \textit{Asymptotic Four Point Functions}.

To compute these four point functions we need to compute the three point functions between two BPS operators and \textit{any} non-BPS operator appearing in its OPE. These non-BPS operators are described using integrability by a set of (at most) seven different type of Nested Bethe roots \cite{BS}. Here we will show that this intimidating Nested Bethe ansatz can actually be described very simply within the hexagon formalism \cite{C123Paper} leading to very compact expressions for the relevant three-point functions and hence for the asymptotic four point functions alluded to above. 

It would be fascinating to take our final expressions for the four-point correlation functions and initiate a systematic exploration of their various interesting mathematical limits thus extracting various relevant physical regimes, many of which with a relevant holographic interpretation. We look forward to performing these analysis in the near future. 

In section \ref{sec2} we discuss four point functions, their operator product decomposition and the precise limits which allow one to discard finite size corrections. In section \ref{sec3} we use the Hexagon approach to conjecture all loop expressions for those asymptotic correlators. In section \ref{sec:comparison} we check the integrability predictions against perturbative data and we conclude in section \ref{sec5}. Various appendices complement the main text. 

\noindent \textbf{Note:} While this paper was being prepared, we learned of the forthcoming paper \cite{Till}, which discusses similar subjects (the OPE of four-point functions and its relation to integrability) from a different perspective. We decided to coordinate the submissions to the arXiv.

\section{Super OPE and Finite Bethe Roots} \la{sec2}
Defined as
\beq
G^{(p)}(z,\bar z,\alpha,\bar \alpha) \equiv \frac{\< \mathcal{O}_1 \mathcal{O}_2 \mathcal{O}_3 \mathcal{O}_4 \>}{ \< \mathcal{O}_1 \mathcal{O}_2  \>  \< \mathcal{O}_3 \mathcal{O}_4 \>} 
\qquad \text{where}\qquad \mathcal{O}_i\equiv \tr((y_i \cdot \phi(x_i))^p)\,, \la{4pt}
\eeq
the reduced correlator is a nice conformal invariant quantify. It is a function of the $SO(2,4)$ and $SO(6)$ cross-ratios
\beq
z\bar z \equiv \frac{x_{12}^2 x_{34}^2}{x_{13}^2 x_{24}^2} \,,\qquad (1-z)(1-\bar z) \equiv \frac{x_{14}^2 x_{23}^2}{x_{13}^2 x_{24}^2}\,,\qquad \alpha\bar \alpha \equiv \frac{y_{12} y_{34}}{y_{13} y_{24}} \,,\qquad (1-\alpha)(1-\bar \alpha) \equiv \frac{y_{14} y_{23}}{y_{13} y_{24}} \,,
\eeq
where $x_{ij}^2=(x_i-x_j)^2$ and $y_{ij}=y_i \cdot y_j$ with $y_j$ being the standard six-dimensional null vectors parametrizing the orientation of the external BPS operators which are inserted at four-dimensional positions $x_j$. 

\subsection{Reservoir Picture and Asymptotic Four-Point Functions}

Let us recall some well-understood facts about the correlator (\ref{4pt}). We will describe it through its infinite OPE series governing what flows from operators $1,2$ to operators $3,4$. In principle, all the multi-trace operators can show up in this OPE representation.
However, at large $N$  
there is an important simplification: only single- and double-trace operators contribute. Then the four-point function can be expanded as
\beqa\la{largeNG}
G^{(p)}&=&1+\overbrace{N^{-2}\!\!\!\!\!\!\!\!\!\!\!\!\!\!\! \!\!\!\!\!\!\!\!\!\!\!\!\!\!\! 
\sum_{\begin{array}{c}\footnotesize\text{single-trace}\\\footnotesize \text{BPS super-conformal} \\ \footnotesize\text{primaries of twist $L=2,4,\dots,2p-2$}\end{array}} 
\!\!\!\!\!\!\!\!\!\!\!\!\!\!\! \!\!\!\!\!\!\!\!\!\!\!\!\!\!\!
L \times \mathcal{F}^\text{BPS}_{L}(z,\bar z,\alpha,\bar \alpha)  }^\text{SUSY protected, coupling independent part}\nn
+
\overbrace{N^{-2}\!\!\!\!\!\!\!\!\!\!\!\!\!\!\!  \sum_{\begin{array}{c}\footnotesize\text{single-trace}\\\footnotesize\text{non-BPS  super-conformal}\\\footnotesize \text{primaries}\end{array}}
\!\!\!\!\!\!\!\!\!\!\!\!\!\!\!
 \(C_p^{\circ\circ\bullet}\)^2  \mathcal{F}_{\Delta,s,n,m}(z,\bar z,\alpha,\bar \alpha)}^\text{more interesting coupling dependent part} \\
&+& \texttt{extremal and double trace contribution} \,.
\eeqa
where the conformal blocks $\mathcal{F}$ in the first line are fixed by super-conformal symmetry and are summarized in appendix \ref{CBAppendix}. Our main focus here is on the last term in the first line corresponding to the contribution of single-trace non-protected operators, whose three-point functions can be computed by the hexagon approach \cite{C123Paper}. A priori, it is non-trivial to disentangle the double-trace contribution from the single-trace contributions since, at finite coupling, they can have the same twist and mix with each other.\footnote{At large $N$ the corresponding anomalous dimensions can cross, in this integrable theory. At finite $N$ this crossing is resolved as discussed in \cite{levelcrossing}.} 
However, in perturbation theory the twist of each exchanged operator is close to its classical value and this allows us to neatly separate the single- and double-trace contributions -- especially if we consider large external operators with $p\gg 1$ -- since the exchanged double traces will have classical twist $\tau \ge 2p$.  Hence, in the OPE limit where $z,\bar{z}$ are small and the ratio $z/\bar{z}$ is fixed, for all twists~$\tau \le 2p-2$ we can safely restrict our attention to single-trace operators as schematically depicted in figure~\ref{OPEcontributions}.

\begin{figure}[t]
\begin{center}
\includegraphics[scale=0.25]{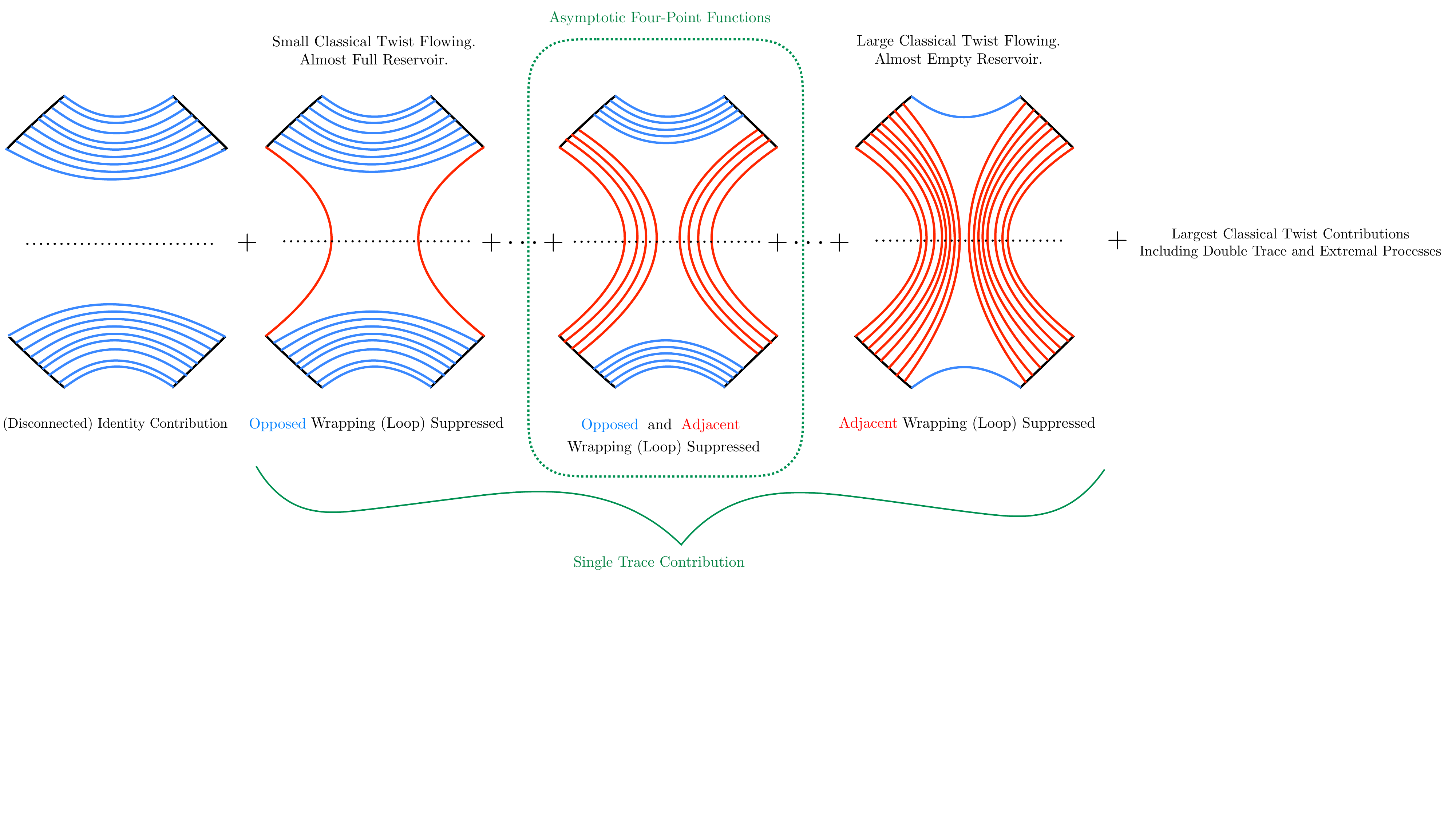} 
\end{center}
\vspace{-3 cm}
\caption{Various contributions to the 4pt function.} \la{OPEcontributions}
\end{figure}

As illustrated in that figure, we can think of operators in the OPE, organized by twist, as originating from a big ``reservoir'' of propagators at the bottom (and top). Operators with a small twist $\tau$ flowing in the OPE arise from opening up a few links at the bottom. As such, they will have small side bridges but very large bottom and top bridges. For these operators wrapping in the so-called opposed channel is greatly suppressed in perturbation theory \cite{C123Paper,3loop}. (The adjacent wrapping does matter eventually, at $\tau/2+2$ loops to be precise.) In the other extreme case we have the contribution of operators with twist close to the double trace threshold, $\tau=2p-O(1)$. Those have huge side bridges which soak up the reservoirs almost completely. For these large twist operators it is thus the adjacent wrapping which is greatly suppressed. (On the other hand, the bottom and top bridges can now be small so that  opposed wrapping eventually kicks in at $p-\tau/2+1$ loops.)

Finally we have the intermediate regime which is the most relevant one for the present paper. For operators whose twist is very large and yet far from emptying the reservoir, $1 \ll \tau \ll 2p$ -- as depicted in the middle of figure \ref{OPEcontributions} -- wrapping is suppressed in both the adjacent and the opposed channels. For such contributions we can thus ignore wrapping contributions altogether and use only the so-called asymptotic prediction for the three point coefficients in the OPE expansion. 

By playing with the polarization vectors we can easily make sure the adjacent bridges are very large, see e.g. \cite{C123SL2}. The basic idea is that if operators $\mathcal{O}_1$ and $\mathcal{O}_2$ have a large non-zero combined R-charge then by R-charge conservation the operators in their OPE must have a large twist, at least as large as the R-charge. For example, we could choose $\mathcal{O}_1$ to be
\beq
\mathcal{O}_1^{ZX} = {\rm tr}(Z^{p-q}X^q)+\text{permutations}= \left.\(\frac{\partial}{\partial \beta_1}\)^q {\rm tr}(y_1\cdot \phi)^p \right|_{\beta_1=0} \,\,\, \text{where} \,\,\, y_1=(1,i,\beta_1,i\beta_1,0,0)\,,
\eeq
and $\mathcal{O}_2$ to be made out of the same $Z$'s and the complex conjugate $\bar X$'s. For the top we proceed similarly using the remaining complex scalars $Y$'s and $\bar Y$.\footnote{All in all, 
\beqa
O_2^{Z\bar X} &=&  \left.\(\frac{\partial}{\partial \beta_2}\)^q {\rm tr}(y_2\cdot \phi)^p \right|_{\beta_2=0} \,\,\, \text{where} \,\,\, y_2=(1,i,\beta_2,-i\beta_2,0,0) \,, \nn \\
O_3^{\bar Z Y} &=&  \left.\(\frac{\partial}{\partial \beta_3}\)^q {\rm tr}(y_3\cdot \phi)^p \right|_{\beta_3=0} \,\,\, \text{where} \,\,\, y_3=(1,-i,0,0,\beta_3,i\beta_3) \,, \\
O_4^{\bar Z\bar Y} &=&  \left.\(\frac{\partial}{\partial \beta_4}\)^q {\rm tr}(y_4\cdot \phi)^p \right|_{\beta_4=0} \,\,\, \text{where} \,\,\, y_4=(1,-i,0,0,\beta_4,-i\beta_4) \,. \nn
\eeqa}
Combined, the total $X$ $U(1)$ charge of operators $1$ and $2$ cancels out but the $Z$ $U(1)$ charge does not. Instead there are $2p-2q$ units of such $R$-charge. As such, in the OPE of such operators we have operators whose twist is at least $\tau=2(p-q)$. Those leading twist operators would have side bridges of length $l=p-q$. Operators with subleading twists will have even larger side bridges. In sum, for the correlator
\beq
\<O_1^{ZX} O_2^{Z \bar X}O_3^{\bar Z Y} O_4^{\bar Z \bar Y} \> = \left.\( \frac{\partial}{\partial \beta_1}\frac{\partial}{\partial \beta_2}\frac{\partial}{\partial \beta_3}\frac{\partial}{\partial \beta_4}\)^q \<O_1 O_2 O_3 O_4 \> \right|_{\beta_i=0} \la{4ptA}
\eeq
with $p$ and $p-q$ both very large, the side wrapping effects in the OPE channel $12$ can be delayed tremendously as they will only kick in at $p-q+2$ loops. Furthermore, if $q$ is also very large then the bottom wrapping is also very suppressed since there will be a huge bottom bridge connecting the $X$'s and $\bar X$'s which requires a lot of twist to eat up. More precisely, for a flowing twist $\tau =2p-2q+2n<2p-2$ bottom wrapping corrections will only show up at $q+1-n$ loops. Only for very subleading twist with $n$ very large will these effect become relevant. To summarize: At weak coupling, for most practical purposes we can ignore \textit{all} wrapping corrections when computing (\ref{4ptA}). Such four point functions are thus dubbed \textit{asymptotic four-point functions}. 

\subsection{Super Operator Product Expansion}
In the OPE (\ref{largeNG}) we sum over super-conformal primaries only. The descendants are automatically taken into account by the super-conformal blocks $\mathcal{F}$ which we summarized out in appendix \ref{CBAppendix}. 

In the integrability context each single trace operator is described by a set of seven kind of Bethe roots satisfying so called Beisert-Staudacher Bethe equations \cite{BS,BES}. 
This description breaks down at some point due to so called wrapping or finite size corrections at which point one must switch to more sophisticated machinery such as the Y-system \cite{grave2}, the Thermodynamic Bethe ansatz \cite{grave1}, finite integral equations \cite{grave3} or,  the current \textit{spectrum problem Ferrari},  the quantum spectral curve \cite{QSC}. In this paper we can disregard finite size corrections as explained in the previous section so that the Beisert-Staudacher equations will suffice in what follows. 
\begin{figure}[t]
\begin{center}
\includegraphics[scale=0.25]{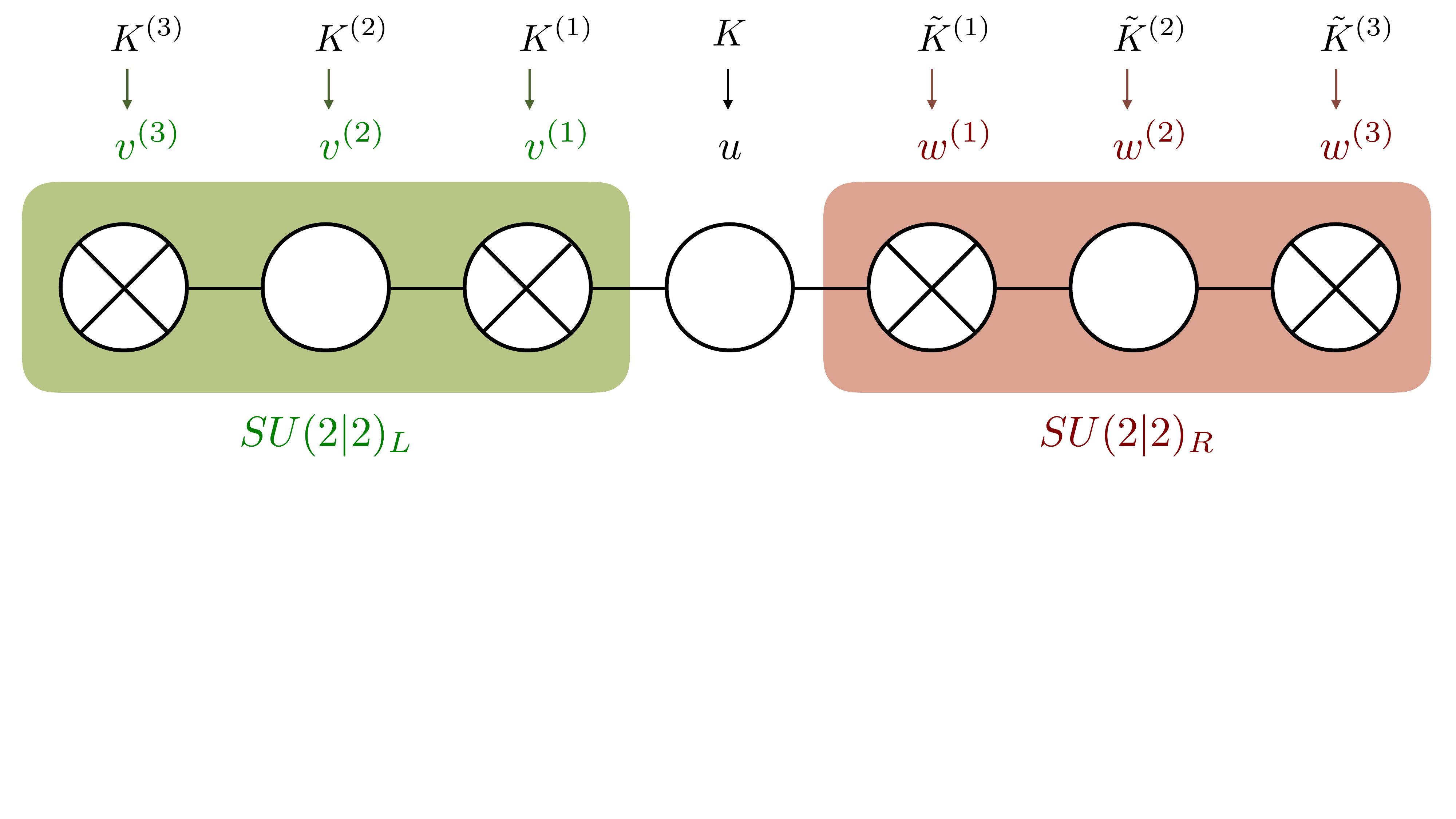} 
\end{center}
\vspace{-4 cm}
\caption{$PSU(2,2|4)$ Dynkin diagram and Bethe roots.}\la{Dynkin} 
\end{figure}

The notation for the Bethe roots is depicted in figure \ref{Dynkin}.
We use $u_{j}$  to denote the \textit{middle node} Bethe roots which obey the middle node equations with a spin-chain length $L$. Then we have a set of three type of Bethe roots~$v_j^{(1)}, v_j^{(2)}, v_j^{(3)}$ describing one of the $\mathfrak{su}(2|2)$ wings and another set of roots~$w_j^{(1)},w_j^{(2)},w_j^{(3)}$ describing the other $\mathfrak{su}(2|2)$ wing. We use~$K$ to denote the number of middle node roots and~$K^{(a)}$ and~$\tilde K^{(a)}$ with $a=1,2,3$ to indicate the number of Bethe roots in each of the wings.\footnote{The notation here differs from the one in \cite{BS} as
\beqa
&& \{K_1,K_2,K_3,K_4,K_5,K_6,K_7\}_\text{there}=\{K^{(3)},K^{(2)},K^{(1)},K,\tilde K^{(1)}, \tilde K^{(2)},\tilde K^{(3)}\}_\text{here} \,, \nn \\
&&\{u_{1,j},u_{2,j},u_{3,j},u_{4,j},u_{5,j},u_{6,j},u_{7,j}\}_\text{there}=\{v^{(3)}_j,v^{(2)}_j,v^{(1)}_j,u_j,w_j^{(1)}, w_j^{(2)},w_j^{(3)}\}_\text{here} \,.
\eeqa
Throughout the paper we will use the $\mathfrak{sl}(2)$ grading which corresponds to $\eta_1=\eta_2=-1$ in \cite{BS}.}  

Now, not all solutions to Bethe equations suit our purpose. Super-conformal primaries are solutions to Bethe equations where all Bethe roots are finite. Furthermore, we should exclude solutions where $x(u_3)=x(u_5)=0$ which also correspond to super descendants \textit{unless} these solutions are part of exact strings in which case the corresponding solutions are denoted as \textit{singular solutions} and should a priori be considered.\footnote{Coincidentally or not we found out that -- on all examples we checked -- these singular solutions yield a vanishing three-point function.} The sum in (\ref{largeNG}) stand therefore for a sum over such finite Bethe roots configurations. 

The number of Bethe roots of each kind can be read of from the quantum numbers of the exchanged operator. 
Since our external operators are all BPS, the three-point functions preserve a diagonal $\mathfrak{su}(2|2)$ subgroup \cite{C123Paper,plefka} which immediately implies that the occupation numbers of the wings must be identified to yield a non-zero result, $\tilde K_a=K_a$. The relation between the Bethe ansatz occupation numbers and length and the labels
\beq
\begin{array}{c|c|c}
\text{Scaling dimension}&\qquad \text{Lorentz}\;\mathfrak{su}(2)\times \mathfrak{su}(2)& \mathfrak{so}(6)\;R\text{-charge}\\ \hline
 \Delta & [s,s] &  [n-m,2m,n-m] \end{array} \nn
\eeq
which show up in (\ref{largeNG}) is then
\beqa
\Delta - \delta \Delta &=& L-K^{(1)}+K^{(3)}+K-2 \la{R1}\\
s & = & K-K^{(1)}-K^{(3)}-2\\
n &= & L/2+K^{(3)}-K^{(2)}-1\\
m & = & L/2 + K^{(2)} - K^{(1)} - 1 \la{R4}
\eeqa
where the anomalous dimension $\delta \Delta= \sum_j \frac{2ig}{x^+(u_j)}- \frac{2ig}{x^-(u_j)}$ up to higher loop finite size corrections which, as mentioned above, we are discarding throughout this work. The quantum numbers in (\ref{R1}--\ref{R4}) are all non-negative integers. This puts restrictions over the length $L$ and the occupation numbers $K$ and $K^{(a)}$. Also, for the Bethe equations to admit finite solutions the occupation numbers usually need to decrease as we go from the middle node occupation $K$ into the wing extremities $K^{(3)}$, see e.g.,\cite{BS,BeisertThesis,Didina}. 

To summarize, we should a priori find \textit{all} finite solutions to Bethe equations with~$K^{(a)}=\tilde K^{(a)}$, read of their quantum numbers from (\ref{R1}--\ref{R4}), compute their three-point functions using the hexagon approach \cite{C123Paper} and add them up as in (\ref{largeNG}) using the super-conformal blocks summarized in appendix \ref{CBAppendix}.

In the next section we will analyze the integrability computation in more detail. We will then observe a remarkable implication of integrability which dramatically simplifies even further the computation of the four-point correlator. It turns out that a version of Yangian symmetry actually implies much more than the global symmetry constraint $K^{(a)}=\tilde K^{(a)}$. To get a non-vanishing OPE contribution we must in fact have absolutely symmetrical wings \textit{root by root}, that is $v_j^{(a)}=w_j^{(a)}$. This is a very sharp and novel space-time implication of the world-sheet integrability. 

\section{Hexagon Wings and Yangian} \la{sec3}
\subsection{Nested Bethe Wave Function} 
The crux of the hexagon formalism lies in cutting a pair of pants, which represents the structure constant, into two hexagonal patches.  Upon cutting, magnons in each operator are divided between two hexagons, and we sum over all such possibilities with appropriate weights, namely a propagation phase $e^{ip\ell}$ and S-matrices. When magnons belong to a rank $1$ sector, things are rather simple since the S-matrix is just a scalar phase. In general however, one has to deal with a complicated index structure: For instance, if the operator has two excitations with indices $A$ and $B$, it produces a complicated set of states as shown in figure~\ref{Splitting} after being cut. In addition, the hexagon form factor itself is defined through the S-matrix and is a complicated object to compute if magnons carry nontrivial indices.
\begin{figure}[t]
\begin{center}
\includegraphics[scale=0.25]{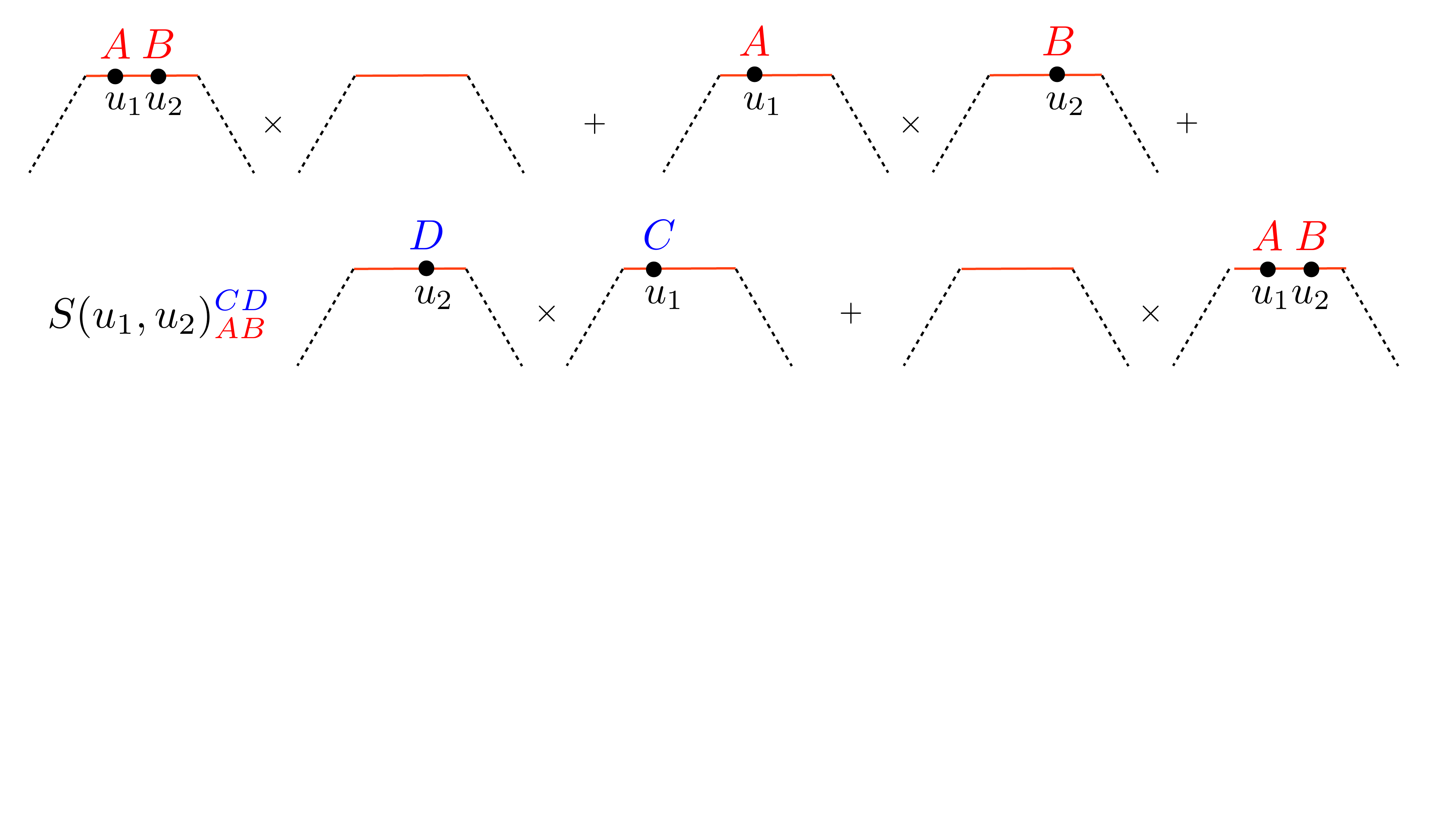} 
\end{center}
\vspace{-5 cm}
\caption{Splitting a two-particle state with indices. When the first particle passes through the second particle, it gets multiplied by a S-matrix $S(u_1,u_2)$. The resulting state is a complicated object which includes a summation over indices ($C$ and $D$ in the figure).} \la{Splitting}
\end{figure}

The way to circumvent such complication of indices is, as is well-known, the Nested Bethe Ansatz. In the Nested Bethe Ansatz, we first make an ansatz for the wave function of the flavour indices, which depends on the order of momentum-carrying roots $u_i$, and a set of roots at higher levels ${\bf w}$. It has an important property that it ``diagonalizes'' the action of the S-matrix,
\beq
\mathbb{S}_{\red i,i+1}|\Psi_{u_1,\ldots, u_{K}}^{\bf w}\rangle=S(u_i, u_{i+1})|\Psi^{\bf w}_{u_1,\cdots, {\red u_{i+1},u_{i}},\ldots,u_{K}}\rangle\,,
\eeq
with $S(u,v)$ being an abelian phase. When ${\bf w}$ satisfy the Bethe equation for higher levels, the wave function has an additional ``nested periodicity'' property,
\beq\label{nestedperiodicity}
|\Psi^{\bf w}_{u_{k+1},\ldots,u_K,{\red u_1,\ldots,u_k}}\rangle =\left(\prod_{i=1}^{k}f(u_i,{\bf w})\right)|\Psi^{\bf w}_{u_1,\ldots,u_K}\rangle \,,
\eeq
where $f$ is a theory-dependent phase factor. With these two properties, one can rewrite the right hand side of the periodicity condition of the full wave function,
\beq
\begin{aligned}
|\Psi^{\bf w}_{u_1,\ldots,u_K}\rangle&= e^{i p_1 L}\left( \prod_{i=2}^{K}\mathbb{S}_{1,i}\right) |\Psi^{\bf w}_{u_1,\ldots,u_K}\rangle\,,
\end{aligned}
\eeq
in the following way:
\beq
\begin{aligned}
e^{i p_1 L}\left( \prod_{i=2}^{K}\mathbb{S}_{1,i}\right) |\Psi^{\bf w}_{u_1,\ldots,u_K}\rangle&=e^{i p_1 L}\left( \prod_{i=2}^{K}S(u_1,u_i)\right) |\Psi^{\bf w}_{u_2,\ldots,u_K,u_1}\rangle\\
&=e^{i p_1 L}f(u_1,{\bf w})\left( \prod_{i=2}^{K}S(u_1,u_i)\right) |\Psi^{\bf w}_{u_1,\ldots,u_K}\rangle\,.
\end{aligned}
\eeq
This leads to the Bethe equation for momentum-carrying roots
\beq
e^{i p_1 L}f(u_1,{\bf w})\left( \prod_{i=2}^{K}S(u_1,u_i)\right)=1\,.
\eeq
Alternatively, we can use this relation to read off the phase factor $f$ from the Bethe equation.
\subsection{Nested Hexagon}\label{NH}
We now use the aforementioned properties to compute general asymptotic three-point functions for two BPS and one non-BPS operators. Bethe states in $\mathcal{N}=4$ SYM spin chain are characterized by seven sets of roots, which are split into two ``wings'' by the momentum-carrying roots $u$ (see figure \ref{Dynkin}). Owing to this structure, the wave function at the nested level is given by a product of two wave functions $\Psi$ and $\dot{\Psi}$. 
\begin{figure}[t]
\vspace{-1. cm}
\begin{center}
\includegraphics[scale=0.55]{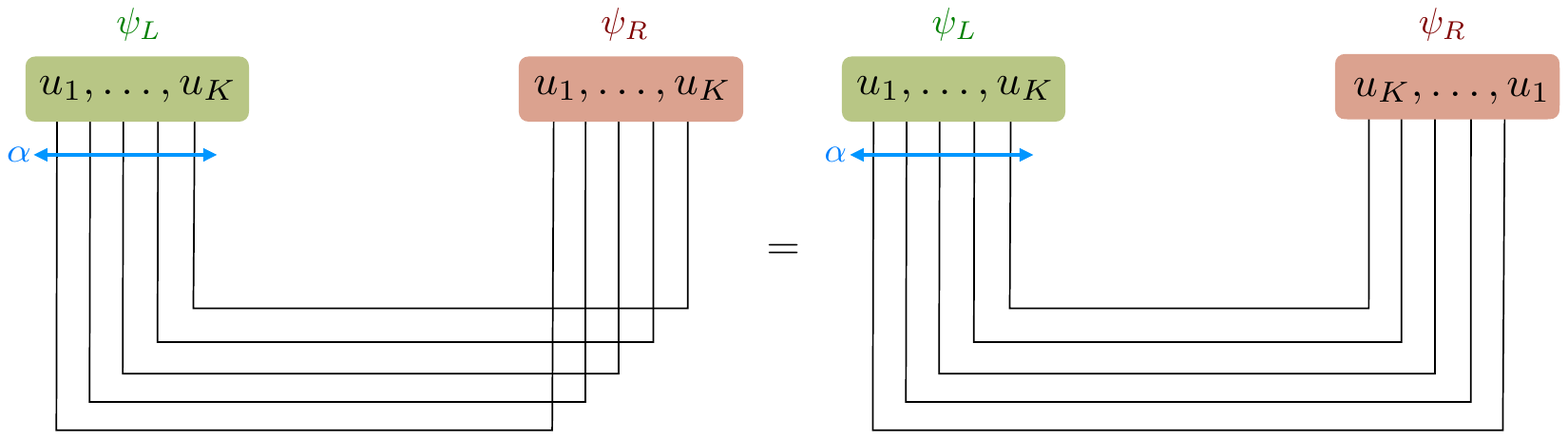} 
\end{center}
\vspace{-6. cm}
\caption{Matrix part for $\bar{\alpha}=\varnothing$: One can simply act the S-matrix to the right wave function $\psi_{R}$ and simplify the structure. Note we are using the normalization of the matrix part, in which the abelian part ($\mathfrak{sl}(2)$ $S$-matrix) is unity.} \la{Undoing}
\end{figure}

To apply the hexagon formalism, we first reorder the magnons (or equivalently the momentum-carrying roots) and split them into two subsets $\alpha$ and $\bar{\alpha}$. Thanks to the property of the nested Bethe wave function, the state we get after reordering is as simple as the original one:
\beq
|\Psi_{{\bf u}}\rangle = \left(\prod_{\substack{i<j\\{u_i\in\bar{\alpha}\,,u_j\in\alpha}}}S(u_i,u_j)\right)|\Psi_{\alpha \bar{\alpha}}\rangle\,.
\eeq
Here $S(u,v)$ is the S-matrix in the $\mathfrak{sl}(2)$ sector. The next step is to contract the nested wave function with the hexagon form factor. When $\bar{\alpha}$ is empty, this can be done easily since the hexagon is essentially given by a product of S-matrices which are already diagonalized by the wave function. As a result, we obtain
\beq
\mathcal{H}_{{\bf u},\varnothing} = \left(\prod_{i<j}h(u_i,u_{j})\right)\langle \overleftarrow{\dot{\Psi}}_{{\bf u}}|\Psi_{{\bf u}}\rangle
\eeq
where $\langle \overleftarrow{\dot{\Psi}}|\Psi\rangle$ is a contraction of the wave functions in two wings, which is defined pictorially in figure \ref{Undoing}.
\begin{figure}[t]
\begin{center}
\includegraphics[scale=0.25]{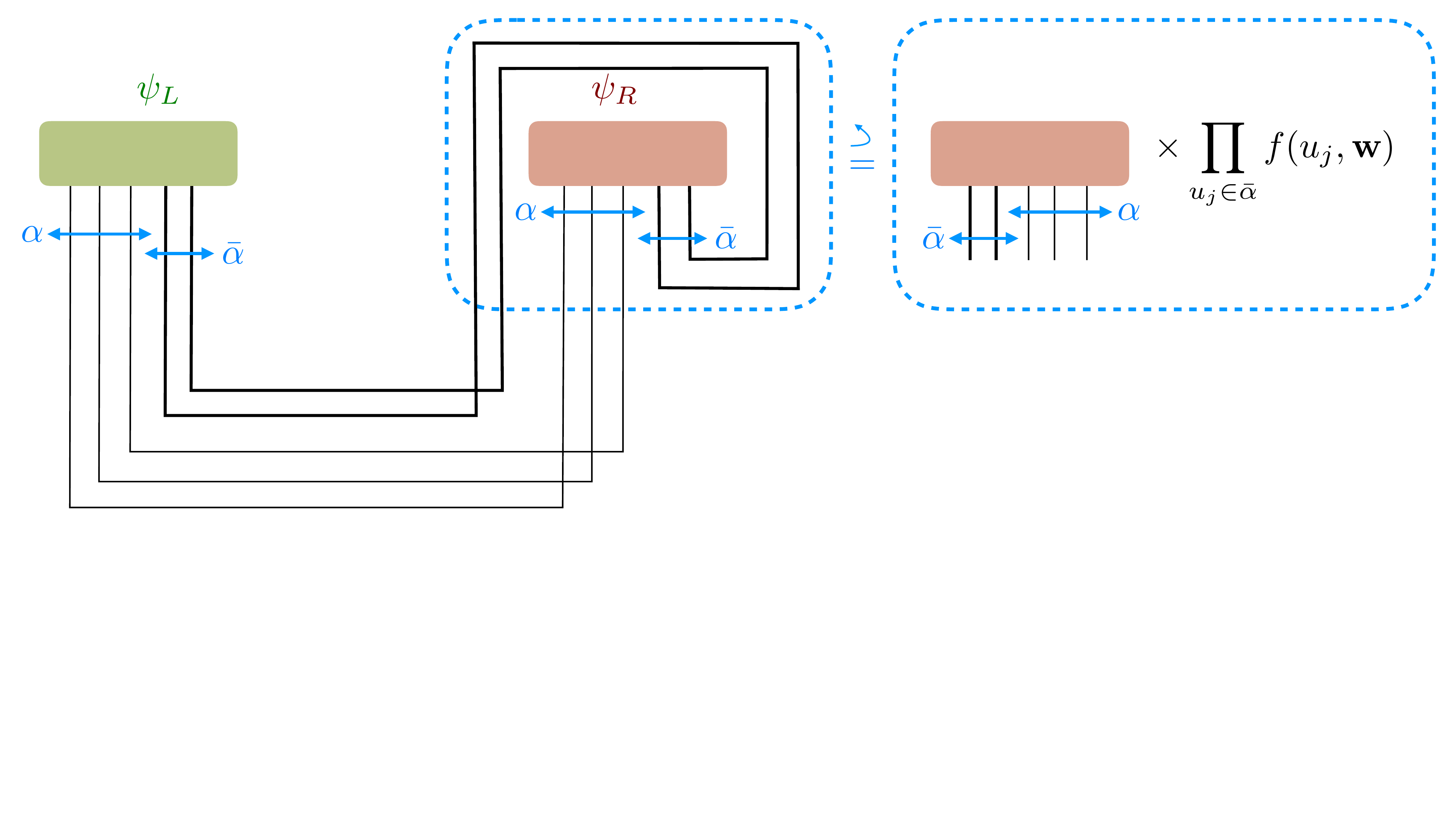} 
\end{center}
\vspace{-3.5 cm}
\caption{Matrix part for $\bar{\alpha}\neq \varnothing$: The magnons for the right wing are in a different order from those in the wave function. To contract the wave function with the hexagon, we have to rewrite it using the ``nested periodicity'' \eqref{nestedperiodicity}.} \la{Rotating}
\end{figure}

By contrast, when $\bar{\alpha}$ is not an empty set, things are a little bit more involved. In the diagram that computes the matrix part, the magnons for the right wing are not in the same order as those in the wave function $\dot{\Psi}$ (see figure \ref{Rotating}). Thus we first have to rewrite the wave function $\dot{\Psi}$ in a different order. This can be done by using the nested periodicity \eqref{nestedperiodicity}. It produces a product of phase factors $f$, which can be read off from the asymptotic Bethe equation for the momentum-carrying root:
\beq\label{fofu}
f(u)=\prod_{v_i^{(1)} \in{\bf v}^{(1)},v_k^{(3)}\in{\bf v}^{(3)}}\frac{x^{-}(u)-x(v_i^{(1)})}{x^{+}(u)-x(v_i^{(1)})}\frac{1-1/x^{-}(u)x(v_k^{(3)})}{1-1/x^{+}(u)x(v_k^{(3)})}\,.
\eeq
Here $x(u)$ is the Zhukowski variable $u=g (x+1/x)$ with $g=\sqrt{\lambda}/4\pi$, and $f^{\pm}(u)\equiv f(u\pm i/2)$.
Note that we should only take factors which depend on $v$'s since $f$ is the phase factor coming just from the right wing.
After doing so, we can straightforwardly contract the wave functions with the hexagons and act the S-matrices on the wave functions. This leads to an expression
\beq
\mathcal{H}_{\alpha,\bar{\alpha}}=\left(\prod_{u_i\in \bar{\alpha}}f(u_i)\right)\left(\prod_{\substack{i<j\\u_i,u_j\in\alpha}}h(u_i,u_{j})\right)
\left(\prod_{\substack{i<j\\u_i,u_j\in\bar{\alpha}}}h(u_i,u_{j})\right)\langle \overleftarrow{\dot{\Psi}}_{\alpha\bar{\alpha}}|\Psi_{\alpha\bar{\alpha}}\rangle
\eeq
It turns out\footnote{Using the nested level Bethe equations we can reverse the nested-wavefunction and show $\langle\overleftarrow{\dot{\Psi}}| = \langle \dot{\Psi}|\,$, taking special care with complex roots. A detailed explanation for the $\mathfrak{psu}(1,1|2)$ subsector is given in Appendix \ref{su112}, and we restrain from presenting the analytic proof for more general cases. For the $\mathfrak{so}(6)$ sector at tree level, we derived the formula \eqref{Main} from scratch, namely without ever resorting to the hexagon formalism, by developing the algebraic Bethe ansatz for that sector. This provides another independent support for the formula \eqref{Main}. See Appendix \ref{ap:so(6)} for details. } 
that the contraction $\langle \overleftarrow{\dot{\Psi}}|\Psi\rangle$ coincides with the usual scalar product in the $\mathfrak{psu}(2|2)$ spin chain $\langle \dot{\Psi}|\Psi\rangle$. It is thus independent of the order of momentum-carrying roots and becomes a partition-independent prefactor. Furthermore, because of the orthogonality of two different on-shell states, it vanishes unless all the roots in two wings are equal, namely ${\bf v}^{(i)}={\bf w}^{(i)}$. This suggests the existence of a hidden symmetry, which forces infinitely many structure constants to vanish. In the next subsection, we explicitly construct such a symmetry using the transfer matrix of $\mathfrak{psu}(2|2)$. 
\begin{figure}[t]
\begin{center}
\includegraphics[scale=0.55]{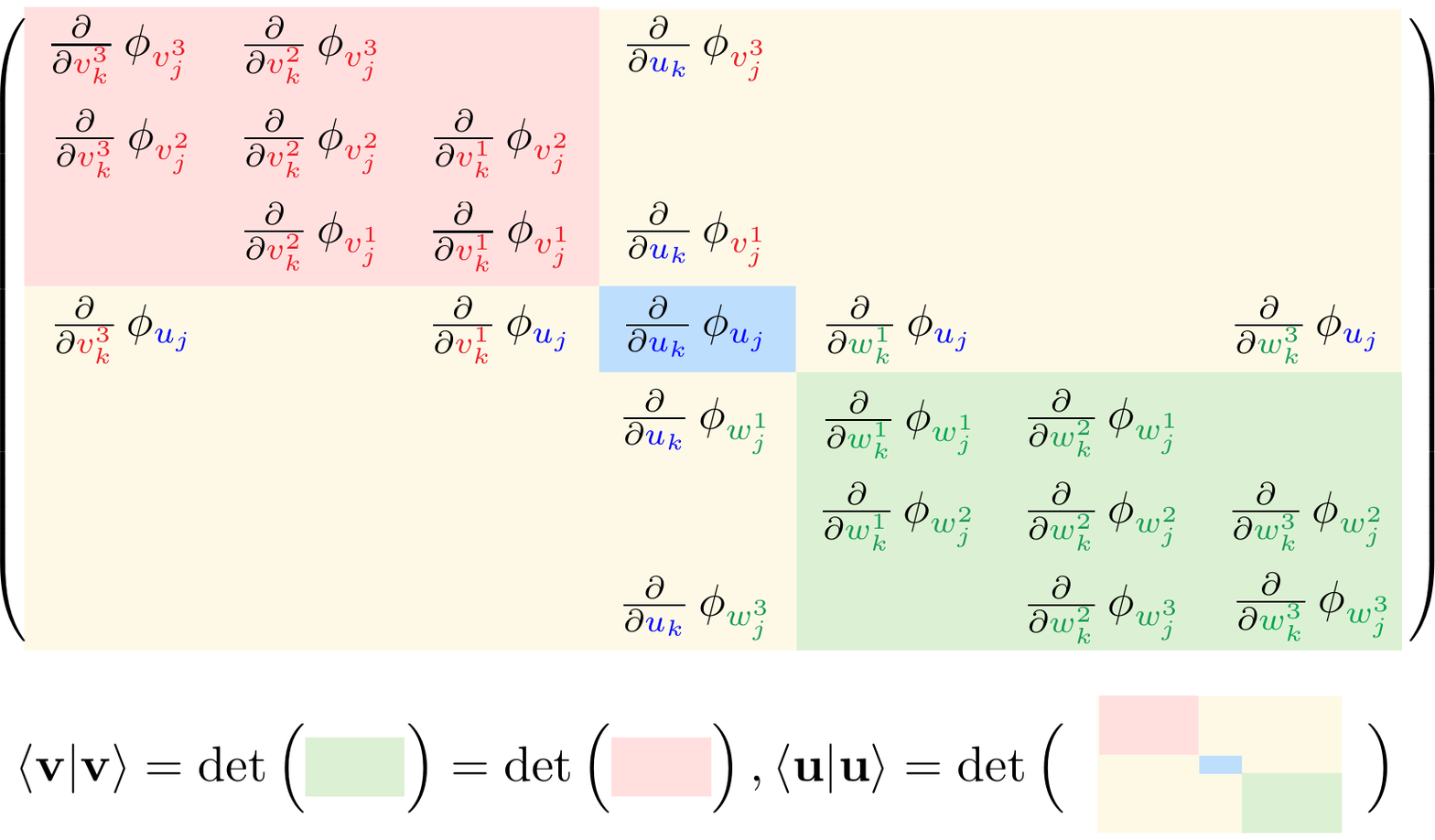} 
\end{center}
\vspace{-3 cm}
\caption{Definitions of the Gaudin norms $\langle{\bf u}|{\bf u} \rangle$ and $\langle{\bf v} |{\bf v}\rangle$: $\phi_{u,v,w}$ is a logarithm of the nested Bethe equation, $e^{i\phi_{u,v,w}}=1$. $\langle{\bf u}|{\bf u} \rangle$ is a determinant of the full matrix shown above whereas $\langle{\bf v}|{\bf v} \rangle$ is a determinant of the upper (or equivalently lower) diagonal matrix shown in red (green). The missing matrix elements are vanishing, as a result of the structure of the $\mathfrak{psu}(2,2|4)$ BAEs (e.g., there is no interaction between auxiliary roots $w$ and $v$ lying on different wings).}\la{Gaudin} 
\end{figure}

Putting together all the elements, we obtain our main formula for the structure constant in higher rank sectors,
\beq\la{Main}
\left(\frac{C^{\circ\circ\bullet}_{123}}{C_{123}}\right)^2=\frac{\langle{\bf v} |{\bf v}\rangle^2\prod_{k=1}^{K}\mu (u_k)}{\langle{\bf u}|{\bf u} \rangle\prod_{i<j}S(u_i,u_j)}\mathcal{A}^2\,.
\eeq
Here $\mu(u)$ is the measure \cite{C123Paper} and $\mathcal{A}$ is a higher-rank generalization of the sum over partitions, which reads
\beq\label{higherrankA}
\mathcal{A}=\prod_{i<j}h(u_i,u_j)\sum_{\alpha\cup\bar{\alpha}= {\bf u}}(-1)^{|\bar{\alpha}|}\prod_{j\in \bar{\alpha}}f(u_j)e^{i p(u_j) \ell_{31}} \prod_{i\in\alpha,j\in\bar{\alpha}}\frac{1}{h(u_i,u_j)}\,.
\eeq
The factor $\langle{\bf u}|{\bf u} \rangle$ denotes the Gaudin norm of the full $\mathfrak{psu}(2,2|4)$ spin chain whereas $\langle{\bf v} |{\bf v}\rangle$ is the norm for the (right) wing. Their precise definitions are given in figure \ref{Gaudin}.

As shown in Appendix \ref{ap:rewriting}, the ratio $\langle{\bf v} |{\bf v}\rangle^2/\langle{\bf u}|{\bf u} \rangle$ can be rewritten as a single determinant by eliminating the dependence on the roots at higher levels. It is also possible to express the sum over partition $\mathcal{A}$ as a Pfaffian of a $2K\times 2K$ matrix. Combining these two expressions, we can express the square of the structure constant simply as a ratio of two determinants. See Appendix \ref{ap:pfaffian} for details.

In section \ref{sec:comparison}, we will use this formula to reproduce the data obtained by the OPE decomposition of the four-point functions.
\subsection{``Yangian'' Symmetry}\label{sec:yangian}
As we saw above, the structure constant vanishes unless the roots in the two wings are identical. Below we uncover the underlying symmetry responsible for such a super-selection rule.

 Let us take a look again at the matrix part of $C_{123}^{\circ\circ\bullet}$. Using the Yang-Baxter relation, we can show that the difference of transfer matrices acting on two wings must always vanish if the state is contracted with the hexagon (see figure \ref{YangianFig}):
 \beq\label{yangian}
 \left(\langle\mathfrak{h}|\otimes \langle \mathfrak{h}|\right) (T_{r}(u)-\overleftarrow{\dot{T}}_{r}(u))=0\,.
  \eeq   
Here $r$ can be any representation of $\mathfrak{psu}(2|2)$ and $T$ and $\overleftarrow{\dot{T}}$ denote the forward and the backward transfer matrices acting on the left and the right $\mathfrak{psu}(2|2)$ respectively. This property turns out to be true even more generally: It is in fact easy to see that this also holds for correlators with more than one non-BPS operators, and it even holds in the presence of wrapping corrections if we ignore the subtleties coming from double-pole singularities \cite{4loop}.
\begin{figure}[t]
\begin{center}
\includegraphics[scale=0.25]{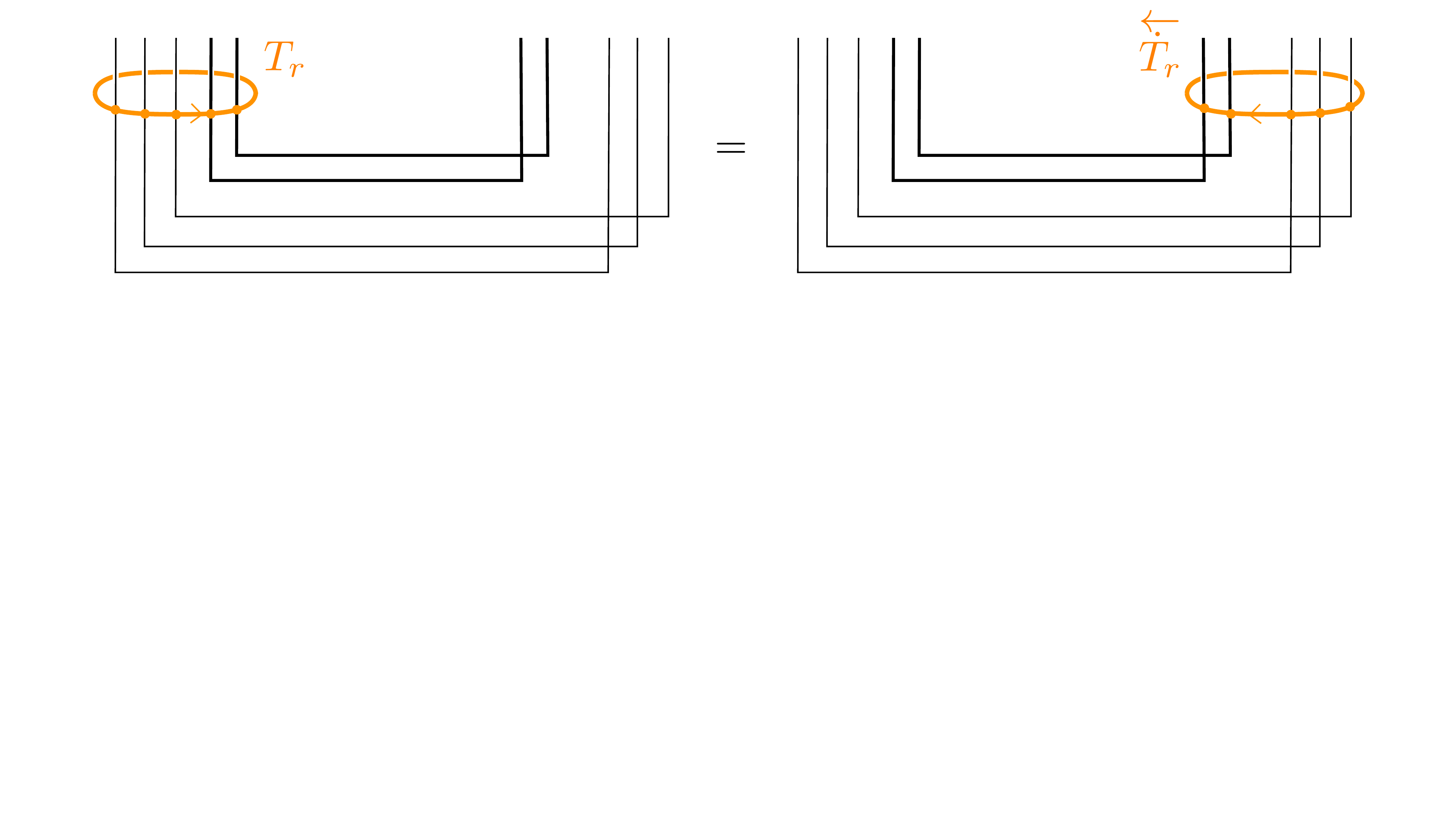} 
\end{center}
\vspace{-6.2 cm}
\caption{``Yangian'' symmetry for three-point functions. The thin black lines denote the magnons in the first hexagon $(\alpha)$ whereas the thick bold lines denote the magnons in the second hexagon $\bar{\alpha}$. Using the Yang-Baxter equation, one can move the transfer matrix from the left to the right.} \la{YangianFig}
\end{figure}

As is well-known, the expansion of transfer matrices yields mutually commuting charges. Thus, the relation \eqref{yangian} is manifestation of infinitely many conservation laws hidden in the three-point function. With a slight abuse of the word, we call it ``Yangian symmetry'' in this paper.

When the state we contract is the on-shell nested Bethe state, we can replace the symmetry generator $T_{r}(u)-\overleftarrow{\dot{T}}_{r}(u)$ by its eigenvalue. Then it follows from \eqref{yangian} that, unless
\beq\label{eigenvalue}
T_{r}(u)-\overleftarrow{\dot{T}}_{r}(u)=0
\eeq
is satisfied as an eigenvalue equation, the structure constant must vanish. The eigenvalues of these transfer matrices are expressed in terms of nested roots \cite{Beisert:2006qh} and the only possible way to satisfy \eqref{eigenvalue} is to set the rapidities of two wings to be identical. This is the symmetry origin\footnote{The symmetry we constructed here is reminiscent of the ``monodromy relations'' studied at weak coupling in \cite{monodromy}. It would be interesting to understand the relation between the two.} of our super-selection rule.

In integrable systems, an infinite number of commuting charges are often accompanied by a real Yangian symmetry, namely a set of non-commutative non-local charges. An explicit construction of such a symmetry for our case is an interesting open problem for the future.

\section{Comparison with Data\la{sec:comparison}}
In this section we combine the previous two sections. Namely, we put the integrability predictions of section \ref{sec3} to test by comparing them with the OPE expansion described in section \ref{sec2} of perturbative four point functions.

Integrability yields predictions for individual structure constants given a set of Bethe roots corresponding to the non-BPS operator at hand. 
These operators have different quantum dimensions $\Delta$ as read off from their Bethe roots  but classically there is a large number of operators with the same classical dimension $\Delta_\text{classical}=\Delta-\delta \Delta$ -- the right hand side of (\ref{R1}).  In perturbation theory, what shows up in the OPE of a four-point function are sum rules over these degenerate operator spaces. The summand is the square of the structure constants weighted by powers of the quantum anomalous dimension $\delta \Delta$.  

This was illustrated in detail for the simplest $\mathfrak{sl}(2)$ sector in \cite{SL2Tailoring}, see for example formulae (56--65) therein summarizing some of those sum rules. Here we are dealing with the full nested space so the sum rules are a bit more involved; they are sums over all finite solutions to Bethe equations whose occupation numbers $K,K_j$ and length $L$ yield the same $\mathfrak{psu}(2,2|4)$ classical charges appearing in the right hand side of (\ref{R1}--\ref{R4}).

Up to one loop, for instance, we can easily use the perturbative results of \cite{DolanData} to extract predictions for the sum rules $\mathcal{P}^{(0,0)}$, $\mathcal{P}^{(0,1)}$ and $\mathcal{P}^{(1,1)}$ defined as\footnote{In other words, 
$\mathcal{P}^{(0,0)}\equiv \sum_{\{u\}} (C^{(0)}_{\mathbf{u}} )^{2}$, $\mathcal{P}^{(1,0)}\equiv \sum_{\{u\}}\,2\,C^{(0)}_{\mathbf{u}}\,C^{(1)}_{\mathbf{u}}$ and $\mathcal{P}^{(1,1)}\, \equiv \, \sum_{\{u\}}\, \gamma^{(1)}_{\mathbf{u}}\,  (C^{(0)}_{\mathbf{u}} )^{2}
$
where
$
\delta\Delta = g^{2}\,\gamma^{(1)}_{\mathbf{u}} + g^{4}\,\gamma^{(2)}_{\mathbf{u}}\, + \, \cdots$ and $C^{\circ\circ\bullet}_{\mathbf{u}} = C^{(0)}_{\mathbf{u}} + g^{2}\,C^{(1)}_{\mathbf{u}} + g^{4}\, C^{(2)}_{\mathbf{u}}\, + \, \cdots$.} 
\begin{equation}
\la{sumEq}\sum_{\text{Bethe solutions with fixed r.h.s. of (\ref{R1}--\ref{R4})}}  \left(\,C^{\circ\circ\bullet}\,\right)^{2} e^{{\color{blue}y}\, \delta \Delta}  \, \equiv \, \mathcal{P}^{(0,0)} +g^{2}\, \mathcal{P}^{(1,0)} + g^{2}\,{\color{blue}y} \mathcal{P}^{(1,1)} + \mathcal{O}(g^4)
\end{equation} 
These predictions are summarized in table \ref{table} for one loop OPE data extracted from the four point function of $1/2$ BPS operators of length $p=4$. We provide the sum rules corresponding to non-BPS operators with $\mathfrak{so}(6)$ charges $0\leq m \leq n \leq 2\,$ and twist $\tau\leq 2\,p-2$. At twist~$\tau =2p$ the OPE would be contaminated by double trace contributions and we could no longer cleanly test the single trace integrability predictions against it.\footnote{Of course, we could simply consider larger external operators to delay the double trace contribution as much as we want. It would also be interesting to play with the OPE analysis varying the external dimensions in order to isolate the double trace contribution from the extremal one. This would provide valuable data for guiding an integrability based approach towards studying extremal or double trace correlation functions.} To generate this table we used a four point function with external operators of length $p=4$ so that at one loop we can text integrability predictions in the OPE involving operators of twist $2,4$ and $6$. Had we used a larger $p$ and we could have tested those twists (the result would be the same at this loop order) and more. 

Having predictions for the right hand side of (\ref{sumEq}) we turn to the left hand which we will now generate using integrability.

\begin{landscape}
\begin{table}
\centering
\caption{The sum of $\mathcal{P}^{(a,b)}$ from superconformal block expansion $\mathbb{P}^{[n,m]}_{\tau,s}=\mathcal{P}^{(0,0)}+g^2\mathcal{P}^{(1,0)}+g^2y\mathcal{P}^{(1,1)}$ of operators with twist $\tau=\Delta_{classical}-s$ and spin $s$. The data in color has been checked against integrability and the data in red is for the $\mathfrak{sl}(2)$ sector.}
\renewcommand{\arraystretch}{1.8}
\begin{tabular}{l|llll}
\multicolumn{1}{l}{\ \ $s$} & \multicolumn{1}{c}{$0$} & \multicolumn{1}{c}{$2$} & \multicolumn{1}{c}{$4$} & \multicolumn{1}{c}{$6$}\\
\hline
$\mathbb{P}^{[0,0]}_{2,s}$ & \textcolor{red}{$\frac{1}{3}-4g^2+4g^2y$} & \textcolor{red}{$\frac{1}{35}-\frac{205}{441}g^2+\frac{10}{21}g^2y$} & \textcolor{red}{$\frac{1}{462}-\frac{1106}{27225}g^2+\frac{7}{165}g^2y$} & $\frac{1}{6435}-\frac{14380057}{4509004500}g^2+\frac{
761}{225225}g^2y$\\

$\mathbb{P}^{[0,0]}_{4,s}$ & \textcolor{blue}{$\frac{2}{15}-\frac{79}{75}g^2+\frac{14}{15} g^2y$} & \textcolor{blue}{$\frac{13}{378}-\frac{9223}{23814}g^2+\frac{71}{189} g^2y$} & \textcolor{blue}{$\frac{1}{198}-\frac{137053}{2044900}g^2+\frac{431}{6435} g^2y$} & $\frac{43}{72930}-\frac{1514205197}{173746973400}g^2+\frac{905}{102102} g^2y$\\

$\mathbb{P}^{[1,1]}_{4,s}$ & \textcolor{red}{$\frac{7}{5}-8g^2+8 g^2y$} & \textcolor{red}{$\frac{16}{63}-\frac{196}{81}g^2+\frac{22}{9} g^2y$} & \textcolor{red}{$\frac{29}{858}-\frac{1873528}{4601025}g^2+\frac{892}{2145} g^2y$} & $\frac{46}{12155}-\frac{12573551}{239320900}g^2+\frac{419}{7735} g^2y$\\

$\mathbb{P}^{[0,0]}_{6,s}$ & \textcolor{blue}{$\frac{13}{210}-\frac{31}{63}g^2+\frac{7}{15} g^2y$} & $\frac{23}{660}-\frac{566107}{1633500}g^2+\frac{557}{1650} g^2y$ & $\frac{7}{825}-\frac{1448972527}{15030015000}g^2+\frac{35891}{375375} g^2y$ & $\frac{37}{25194}-\frac{19377905081}{1052328186000}g^2+\frac{184}{9945} g^2y$\\

$\mathbb{P}^{[1,1]}_{6,s}$ & \textcolor{blue}{$\frac{4}{5}-\frac{1264}{245}g^2+\frac{172}{35} g^2y$} & $\frac{47}{110}-\frac{20891}{5445}g^2+\frac{208}{55} g^2y$ & $\frac{368}{3575}-\frac{5619351}{5112250}g^2+\frac{3932}{3575} g^2y$ & $\frac{149}{8398}-\frac{9215599347}{43197422450}g^2+\frac{31709}{146965} g^2y$\\

$\mathbb{P}^{[2,0]}_{6,s}$ & \textcolor{blue}{$\frac{4}{15}-2g^2+2 g^2y$} & \textcolor{blue}{$\frac{15}{77}-\frac{1973}{1089}g^2+\frac{20}{11} g^2y$} & $\frac{2}{39}-\frac{84593}{152100}g^2+\frac{329}{585} g^2y$ & $\frac{98}{10659}-\frac{57003741}{511212100}g^2+\frac{1284}{11305} g^2y$\\

$\mathbb{P}^{[2,2]}_{6,s}$ & \textcolor{red}{$\frac{26}{7}-12g^2+12 g^2y$} & \textcolor{red}{$\frac{12}{11}-\frac{2494}{363}g^2+\frac{76}{11} g^2y$} & \textcolor{red}{$\frac{12}{55}-\frac{15917}{8450}g^2+\frac{124}{65} g^2y$} & $\frac{145}{4199}-\frac{9212823}{25560605}g^2+\frac{831}{2261}g^2y$\\

\multicolumn{1}{l}{}\\

\multicolumn{1}{l}{\ \ $s$} & \multicolumn{1}{c}{$1$} & \multicolumn{1}{c}{$3$} & \multicolumn{1}{c}{$5$} & \multicolumn{1}{c}{$7$}\\
\hline

$\mathbb{P}^{[1,0]}_{4,s}$ & \textcolor{blue}{$\frac{1}{5}-2g^2+2 g^2y$} & \textcolor{blue}{$\frac{3}{77}-\frac{521}{1089}g^2+\frac{16}{33} g^2y$} & \textcolor{blue}{$\frac{1}{195}-\frac{10909}{152100}g^2+\frac{43}{585} g^2y$} & $\frac{2}{3553}-\frac{4415079}{511212100}g^2+\frac{101}{11305} g^2y$\\

$\mathbb{P}^{[1,0]}_{6,s}$ & \textcolor{blue}{$\frac{1}{4}-\frac{20}{9}g^2+\frac{13}{6} g^2y$} & $\frac{9}{104}-\frac{1179}{1300}+\frac{9}{10} g^2y$ & $\frac{3}{170}-\frac{1179497}{5664400}g^2+\frac{249}{1190} g^2y$ & $\frac{25}{9044}-\frac{2070147901}{57971452140}g^2+\frac{41291}{1139544} g^2y$\\

$\mathbb{P}^{[2,1]}_{6,s}$ & \textcolor{blue}{$\frac{8}{7}-8g^2+8 g^2y$} & \textcolor{blue}{$\frac{29}{78}-\frac{5186}{1521}g^2+\frac{134}{39} g^2y$} & $\frac{69}{935}-\frac{11553}{14450}g^2+\frac{69}{85} g^2y$ & $\frac{335}{29393}-\frac{30397561}{219090900}g^2+\frac{3207}{22610} g^2y$ 
\end{tabular} \la{table}
\end{table}
\end{landscape}

To reproduce this OPE data from integrability we first find all (wing-symmetric) solutions\footnote{Bethe solutions with asymmetric wings give a vanishing structure constant $C^{\circ\circ\bullet}$. We exclude as well symmetric solutions with $w^{1}=v^{(1)}=0$, as they do not render highest weights.} of Beisert-Staudacher Bethe equations with finite Bethe roots. First we set $g=0$ and solve these equations to leading order at weak coupling. This part is hard. Then to get the quantum corrections to the Bethe roots we simply correct the Bethe roots perturbatively by linearizing the $\mathcal{O}(g^{2})$ Bethe equations around each tree level seed solution. This part is straightforward. Once we get the Bethe roots we plug them into the Hexagon prediction~(\ref{Main}) with $l=L/2$, sum over all solutions as indicated in the left hand side of~(\ref{sumEq}). The result is then compared with the OPE predictions for the right hand side of~(\ref{sumEq}) which we extracted from perturbative data and summarized in table \ref{table}. 

To find all the Bethe ansatz solutions at tree level  we resorted to various pieces of technology. The simplest Bethe equations correspond to the $\mathfrak{sl}(2)$  sector where we only excite the middle node and Bethe solutions for operators of spin $s$ are given by sets of real roots $\{u_{1},\cdots , u_{s}\}$. Solving Bethe equations in the $\mathfrak{sl}(2)$ sector in Mathematica is absolutely straightforward, see for example \cite{NorditaPedroLectures}. Checks of OPE data against integrability conjectures  were already performed in \cite{C123Paper} and earlier in \cite{SL2Tailoring}. The sums $\mathcal{P}^{(a,b)}$  for this sector are highlight in red in table~\ref{table}. 
 
For other sectors such as higher rank sectors with excited wing nodes, Bethe equations become more complicated and also admit complex solutions including at times so called exceptional solutions \cite{SingularNepomechie,SingularArutyunov}. It is the existence of complex solutions which renders the problem of finding all solutions to the Bethe equations much more challenging in this case. One way to proceed which we found quite useful is to use a Baxter formulation of Bethe equations and solve directly for the Baxter polynomials and the transfer matrix eigenvalues rather than individual Bethe roots. Another useful numerical method is the so-called Homotopy continuation method \cite{Homotopy} where one starts from some simpler equations and adiabatically deform them until they become the Beisert-Staudacher equations. For $\mathfrak{su}(2)$ solutions this was proposed in \cite{Homotopy} and its generalization to the nested case also works very well. The third method -- and the one we found to be the most convenient -- is however to use the very powerful recently proposed analytic solver of \cite{Qsolver,Qsolver2} based on the $Q$-system\footnote{We are very grateful to C.~Marboe and D.~Volin for sharing a working code of the fast analytic solver for $\mathfrak{psu}(2,2|4)$.}. This provided us with the complete set of Bethe solutions needed to reproduce all the number in blue in tables \ref{table} using the hexagon conjecture~(\ref{Main}). 
 
To illustrate what goes into these computations consider the following example. For global charges $\Delta-\delta \Delta= 8$, $s=2\,$, $n=2$ and $m=0$ there are  20 wing-symmetric Bethe solutions, each of them with $6$ middle node roots $\{u\}$ and $2$ roots in the first left and right fermionic nodes $\{w^{(1)}\}$  and $\{v^{(1)}\}$. Performing the sum over Bethe solutions with a high numerical precision we obtain:
\begin{align}
\sum_{\text{20 solutions}}\left(C^{\circ\circ\bullet}\right)^{2}\,& = 0.194805\mathbf{194805}194805\mathbf{194805}194805\mathbf{194805}\nonumber\\
&\qquad - g^{2}\;
 1.8117539026629935720844\mathbf{8117539026629935720844}
\end{align}
we then recognize this renders the rational numbers:
\begin{equation}
\sum_{\text{20 solutions}}\left(C^{\circ\circ\bullet}\right)^{2}\, = \, \frac{15}{77} \, - \, g^{2}\,\frac{1973}{1089}
\end{equation}
In an attached \texttt{Mathematica} notebook the reader can find our conjecture (\ref{Main}) coded up up to one loop order and how the twenty solutions beautifully add to this nice rational number which perfectly reproduces the perturbative OPE data. All other blue numbers in table \ref{table} were checked in the same way. 

Note in particular that there is no data in table \ref{table} when $s+n-m$ is odd despite the fact that there are definitely Bethe solutions yielding these quantum numbers. The point is that operators are absent in the OPE of identical operators for symmetry reasons. It is nice to see how that comes about from our integrability construction. The sum over partitions (\ref{higherrankA}) is written in terms of the bridge length $\ell_{31}$. However, nothing in the original problem singles out this particular adjacent channel. We can find an equivalent formula expressed in terms of the complementary bridge length $\ell_{23}$ when the Bethe state is on shell and cyclic. Namely, using the ABA equations, $e^{ip_{\bar{\alpha}}L_{3}}f(u_{\bar{\alpha}})^2S_{\bar{\alpha}\alpha} = 1$ and $f(u_{\alpha})f(u_{\bar{\alpha}}) = 1$, the permutation property of the hexagon form factor $h_{\alpha\bar{\alpha}}S_{\bar{\alpha}\alpha} = h_{\bar{\alpha}\alpha}$, and the zero momentum condition $e^{-ip_{\bar{\alpha}}} = e^{ip_{\alpha}}$, one easily derives that
\beq
\sum_{\alpha}(-1)^{|\bar{\alpha}|} f(u_{\bar{\alpha}})e^{ip_{\bar{\alpha}}\ell_{31}} \frac{1}{h_{\alpha\bar{\alpha}}} = (-1)^{K}\sum_{\alpha}(-1)^{|\bar{\alpha}|} f(u_{\bar{\alpha}})e^{ip_{\bar{\alpha}}\ell_{23}} \frac{1}{h_{\alpha\bar{\alpha}}}\, ,
\eeq
where $K = |\alpha| + |\bar{\alpha}|$ is the total number of magnons. For two identical operators, the spin chain is split into two equal parts of length $\ell_{13} = \ell_{23} = L_{3}/2$ and the previous relation turns into a selection rule : $\mathcal{A} = 0$ for $K$ odd. In terms of the quantum numbers of the superconformal primary, see equation (\ref{R1}), it happens whenever $(-1)^{K} = (-1)^{s+n-m} = -1$, in agreement with the symmetry property of the 4-point function.

Finally, it is worth stressing that while all the checks we performed worked like a charm, they do not exhaust the available perturbative data by any stretch. Even at tree level and one loop we only confirmed the predictions in blue in table \ref{table}.
From a Bethe ansatz point of view, most of these solutions are not general enough as they do not excite roots associated with all $\mathfrak{psu}(2,2|4)$ Dynkin nodes. The only solutions which contain roots of type $v^{(3)}$ and $w^{(3)}$ are the ones which contribute to $\mathbb{P}_{\tau= 6 , s=0}^{n=0,m=0}$ in table~\ref{table}. These solutions have some peculiarities, such as the appearance of odd powers of $g$ in the rapidities, which are explained in Appendix \ref{ap:special case}. It would be very interesting -- even at this low loop order -- to perform a few higher twist checks and probe various Bethe solutions in full generality. 
It would also be very interesting to expand Bethe ansatz further and compare the integrability predictions with the available data at two \cite{Ref2Loops}, three \cite{Ref3Loops} or even four loops \cite{Ref4Loops}. When going to higher loops we should either start including finite size corrections to the three-point functions \cite{appeared,3loop,4loop} (hard) or increase the length $p$ of the external operators as explained in section \ref{sec2} (easy). 
 
 \newpage

\section{Conclusions and Outlook} \la{sec5}
We initiated the study of four-point functions of large BPS operators by combining the operator product expansion and integrability. We found a compact formula for structure constants in higher rank sectors and checked it against OPE data at tree level and 1 loop. 

There are numerous physically interesting questions which can be addressed with the methods and the results in this paper. The most important among them\footnote{Other interesting regimes to study would be the Regge limit and the double light-cone limit.} is, perhaps, to study the strong coupling limit and understand the emergence of the bulk locality. Since we are discarding the double-trace contributions, we need to take a careful double-scaling limit in which both the coupling $g$ and the lengths of the external operators $p_i$ tend to infinity. In such a limit, the solutions to the Bethe equation become dense and the summation over the solutions would be effectively replaced by integrals. It would be extremely interesting to see if such integrals, combined with the Pfaffian/determinant representations presented in this paper, give us analytic control over the local physics in AdS.

Right now, we are solving the Bethe equation and computing individual structure constants. However, what we really want to know is the full four-point function, not the individual structure constants. Furthermore, solving the Bethe equation would become horrendously complicated when the length of the operators and the number of magnons are large. It is thus important to develop a method which allows us to compute the sum over the states without explicitly solving the Bethe equations\footnote{In the context of the scattering amplitudes, there are representations of amplitudes given by summation over algebraic equations known as scattering equations \cite{SEqs} which are formally not so dissimilar from the Bethe equations encountered here. There, beautiful methods have been developed to perform to sum over the solutions to the scattering equation without even explicitly solving those equations, see e.g. \cite{Chrys}. Can we do something similar here?}.

Quite recently, there appeared an alternative approach to study higher-point functions called hexagonalization\footnote{For tree-level four-point functions involving non-BPS operators, an extensive study in the SU(2) sector was performed in \cite{CaetanoEscobedo}, and expressions in terms of sums over partitions, akin to the ones that arise from the hexagon formalism, were derived.
A similar correlator in special kinematics was revisited recently \cite{cushion} and it was observed that the result can be expressed in terms of hexagon form factors, based on the idea akin to hexagonalization. } \cite{hexagonalization}. In that approach, the hexagons are glued always along the mirror edges and the cross ratios appear as the weight factors for mirror particles. By contrast, in our approach, we glue physical edges and the cross-ratio dependence comes entirely from the conformal blocks. 

Both approaches have its own advantages and drawbacks. For example, for large operators and for the leading OPE contributions, the approach proposed here extends to higher loops in a trivial way since after solving Bethe equations at tree level it is trivial to correct them perturbatively to any loop order. The hexagonalization approach, on the other hand leads directly to beautiful resummed expressions for the full four point function without ever solving any Bethe equations but the number of mirror particle integrals grows with the loop order. The approach here struggles when we reach maximal twist and start becoming contaminated by double trace operators while the hexagonalization method automatically incorporates these effects. Super-conformal are manifest here since we sum over super-conformal primaries with the help of super-conformal blocks but crossing symmetry is far from obvious while the converse is true in the hexagonalization approach. The list could go on and on. Clearly, understanding the relation between two approaches is a very interesting future problem. If we could take the best out of each of them we would call it a win! 
 
\section*{Acknowledgments}

We are grateful to Christian Marboe and Dmytro Volin for discussions and sharing a useful Mathematica code with us. We thank Carlo Meneghelli for comments on the draft. Research at the Perimeter Institute is supported in part by the Government of Canada through NSERC and by the Province of Ontario through MRI. The work of De-liang Zhong was supported by the European Research Council (Programme ERC-2012- AdG 320769 AdS-CFT-solvable).

\appendix
\addtocontents{toc}{\setcounter{tocdepth}{1}}

\section{Superconformal Blocks} \la{CBAppendix}

The super-OPE relies on the use of a superconformal block to resum the contributions of all the superconformal descendants of a given superconformal primary with weights $\Delta, s, m, n$. In this appendix we record its expression when the superconformal primary that is flowing is either long or half-BPS and when the external operators are identical half-BPS superconformal primaries.

The non-BPS block presented in (\ref{long}) can be read off from various references, e.g.~\cite{Nirschl:2004pa,Dolan:2004mu,Dolan:2004iy,Bissi}. In contrast, less is written explicitly about the BPS block. Following~\cite{Nirschl:2004pa,Dolan:2004mu} it has become conventional to decompose the protected part of the four point function over an OPE-like basis of (single variable hypergeometric) functions \cite{Dolan:2004iy,Nirschl:2004pa}, which solve a SUSY version \cite{Bissi} of the Casimir eigenvalue equation \cite{Dolan:2003hv}. These functions are however \textit{not} identical to the BPS blocks we are after, as one can easily see by checking their content in usual conformal waves. We give in (\ref{BPSb}) the expression we have found for the BPS block by adding enough of the former functions together until we got the appropriate OPE content for an half-BPS multiplet. A general formula for all the superconformal blocks of interest can also be found in \cite{Doobary:2015gia}.

\subsubsection*{Non-BPS blocks}

These are given concisely by
\beqa\label{long}
&&\mathcal{F}_{\Delta,s,n,m}(z,\bar z,\alpha,\bar \alpha) = (z-\alpha)(z-\bar\alpha)(\bar z-\alpha)(\bar z-\bar\alpha) \times \\
&&\!\!\!\!\!\!\!\!\!\,\,\,\,\times \(F_{\Delta,s}(z,\bar z) =(-1)^s \, \frac{h_{\frac{\Delta +s}{2}}(z) h_{\frac{\Delta -s-2}{2}}\left(\bar{z}\right)-h_{\frac{\Delta -s-2}{2} }(z) h_{\frac{\Delta +s}{2}}\left(\bar{z}\right)}{(z-\bar z)/z\bar z } \) \nn
   \\
   &&\!\!\!\!\!\!\!\!\!\,\,\,\,\times\( Y_{n,m}(\alpha,\bar\alpha) = \frac{ (m!)^2 ((n+1)!)^2}{(2 m)!  (2 n+2)!}\times   \frac{P_m (\frac{2}{\alpha }-1 ) P_{n+1} (\frac{2}{\bar{\alpha }}-1 )-P_{n+1} (\frac{2}{\alpha }-1 ) P_m (\frac{2}{\bar{\alpha }}-1 )}{   \left(\alpha
   -\bar{\alpha }\right)\alpha \bar{\alpha }}\)\,,\nn
\eeqa
where $P_n$ are Legendre polynomials and where $h_\lambda(z)=z^\lambda \, _2F_1(\lambda {\color{red}+2},\lambda {\color{red}+2};2 \lambda {\color{red}+4};z)$, with the shifts in red being an $\mathcal{N}=4$ SUSY shift. In AdS/CFT jargon, we can say that the first line is SUSY, the second is AdS and the third accounts for the sphere. The slightly unconventional normalization factor in the last line ensures that the R-charge blocks behave as~$Y_{n,m}(\alpha,\bar\alpha)  = 1\times \alpha^{-n-2} \bar\alpha^{-m-2}(1+\mathcal{O}(\alpha,\bar \alpha))$. The bosonic blocks are normalized so that~$F_{\Delta,s}(z,\bar z)\simeq   (\bar z z)^{\Delta /2} (-1)^s\big({ \left(\frac{z}{\bar z}\right)^{\frac{s}{2}+\frac{1}{2}}-\left(\frac{\bar z}{z}\right)^{\frac{s}{2}+\frac{1}{2}} }\big)/\big({(\tfrac{z}{\bar{z}})^{{1}/{2}}-(\tfrac{\bar{z}}{z})^{{1}/{2}}}\big)$, in the OPE limit where $z,\bar z\to 0$ with $z/\bar z$ fixed. 

By sending $z, \bar{z}\rightarrow 0$, one reads out the $\mathfrak{su}(4)$ block of the superconformal primary, with Dynkin labels $[n-m, 2m, n-m]$,
\beq
Z_{n, m}(\alpha, \bar{\alpha}) = (\alpha\bar{\alpha})^2 Y_{n, m}(\alpha, \bar{\alpha})\, ,
\eeq
while by sending $\alpha, \bar{\alpha}\rightarrow 0$ one recovers the conformal block of a SUSY descendent with dimension $\Delta+4$ and spin $s$,%
\footnote{This descendent has shifted labels $n+2, m+2$ compared to those of the superconformal primary, see e.g. table (8.1) in \cite{Dolan:2001tt}.}
\beq
G_{\Delta+4, s}(z, \bar{z}) = (z\bar{z})^2 F_{\Delta, s}(z, \bar{z})\, .
\eeq
Equivalently, the $\mathfrak{so}(2,4)$ block $G_{\Delta, s}$ for a conformal primary with dimension $\Delta$ and spin $s$ is given~\cite{Dolan:2000ut} by $F_{\Delta, s}$ \textit{without} the shifts in red in the arguments of the hypergeometric function $h$ below (\ref{long}).

\subsubsection*{BPS blocks}

These can be written as linear combinations of six bosonic blocks,
\beqa\label{BPSb}
\mathcal{F}_{\Delta}&=&{\color{blue}G_{\Delta ,0} Z_{\frac{\Delta }{2},\frac{\Delta }{2}}}+\frac{\Delta ^2 {\color{blue} G_{\Delta +1,1}
   Z_{\frac{\Delta }{2},\frac{\Delta }{2}-1}}}{2^4 (\Delta -1) (\Delta
   +1)} 
   +\frac{(\Delta +2)^2 \Delta ^2 {\color{blue} G_{\Delta +2,2}
   Z_{\frac{\Delta }{2}-1,\frac{\Delta }{2}-1}} }{2^8 (\Delta -1)
   (\Delta +1)^2 (\Delta +3)} \nn \\
   &+&
  \frac{(\Delta -2)^2 \Delta ^2{\color{blue}G_{\Delta
   +2,0} Z_{\frac{\Delta }{2},\frac{\Delta }{2}-2}}}{2^8 (\Delta -3)
   (\Delta -1)^2 (\Delta +1)}
   +
   \frac{(\Delta -2)^2
   (\Delta +2)^2 \Delta ^2 {\color{blue}G_{\Delta +3,1} Z_{\frac{\Delta
   }{2}-1,\frac{\Delta }{2}-2}}}{2^{12} (\Delta -3) (\Delta -1)^2 (\Delta
   +1)^2 (\Delta +3)} \nn
 \\
   &+&\frac{(\Delta -2)^2 (\Delta +2)^2 \Delta ^4{\color{blue}G_{\Delta +4,0}
   Z_{\frac{\Delta }{2}-2,\frac{\Delta }{2}-2} } }{2^{16} (\Delta -3)
   (\Delta -1)^3 (\Delta +1)^3 (\Delta +3)}\, .
\eeqa
They are normalized by their OPE behaviour~$\mathcal{F}_{n} = (z\bar z)^n Z_{n,n}(\alpha,\bar \alpha) (1+\mathcal{O}(z,\bar z))$.

The formula (\ref{BPSb}) agrees with those given in \cite{Dolan:2001tt} for $\Delta=2$ (irrep $\textbf{20} = [0, 2, 0]$) and $\Delta = 4$ (irrep $\textbf{105} = [0, 4, 0]$), see equations (8.17) and (8.24) in \cite{Dolan:2001tt}. Each of the 6 conformal waves in (\ref{BPSb}) corresponds to one bosonic conformal primary, with zero hypercharge $Y=0$ and in a left-right symmetric irrep of $\mathfrak{so}(3,1)$ and $\mathfrak{su}(4)$, on the middle line of the half-BPS supermultiplet given in table (B.1) of \cite{Dolan:2001tt}. For the stress tensor multiplet $\Delta = 2$, only the first three terms survive in (\ref{BPSb}), corresponding to the dimension 2 chiral primary, the dimension 3 R-symmetry current and the dimension 4 stress energy tensor, in agreement with the bosonic (hypercharge zero) components in the $\Delta = 2$ supermultiplet reviewed in table (2.15) of \cite{Dolan:2001tt}. Notice that the dual and self-dual parts of the dilaton carry non zero hypercharges and thus decouple, in accord with the non-renormalization property of the BPS structure constant.

\section{Dualization, Gaudin Norm and Diagonal Symmetry}\la{ap:rewriting}

The considerations in Section \ref{sec3} were based on the description of the Bethe states in the $\mathfrak{sl}(2)$ grading, corresponding to $\eta_{1} = \eta_{2} = -1$ in the notations of \cite{BS}. This choice was motivated by the comparison with data. Conformal primaries in the $\mathfrak{sl}(2)$ grading are indeed closer to the superconformal primaries (they share the same length in the spin chain description for instance). We would nonetheless get a fully equivalent description in the $\mathfrak{su}(2)$ grading, corresponding to the choice $\eta_{1} = \eta_{2} = +1$, as shown in appendix \ref{su112} in a particular subsector. As well known, the two gradings are related by the (simultaneous) dualizations of the fermionic nodes in the two wings of the $\mathfrak{psu}(2, 2|4)$ Dynkin diagram \cite{BS}. In this appendix, we will show that our main formula (\ref{Main}) transforms properly under this diagonal dualization. We shall also demonstrate that it is invariant under diagonal $\mathfrak{su}(2|2)_{D}$ transformations, including the length changing effect accompanying them \cite{BS}. To prove both properties, we shall find convenient to first show that the ratio of determinants appearing in (\ref{Main}) can be written as a single Gaudin determinant for the effective BAEs for the main roots $\textbf{u}$. The latter are obtained after implicitly integrating out the auxiliary rapidities along the wings, at given wing mode numbers $m_{\textbf{v}} = m_{\textbf{w}}$.

\subsubsection*{Cosmetic rewriting}

To simplify the discussion, we assume that we can unite the fermionic roots of type 1 and type 3, on each wing of the diagram \ref{Dynkin}, by applying the dynamical transformation of \cite{BS}. It amounts to inverting the Zhukowski roots of type 3 such that they appear as roots of type $1$, while simultaneously redefining the length of the operator, $L\rightarrow L-2K^{(3)}$, in order to absorb the momentum factor spit out during the inversion. It is clear from the structure of the wing dependent factor (\ref{fofu}) why we can do that in the sum over partitions (\ref{higherrankA}). It will become clear, at the end of the appendix, why we can also do it at the level of the Gaudin determinants (once these ones are properly projected down to the subspace of cyclic states).

The formula for the higher rank structure constant (\ref{Main}) factorizes into two main blocks. The first one is the ratio of determinants $r$,
\beq\label{rr}
r^2 = \frac{\mathcal{h}\textbf{v}|\textbf{w}\mathcal{i}^2}{G} = \frac{\mathcal{h}\textbf{v}|\textbf{v}\mathcal{i}\mathcal{h}\textbf{w}|\textbf{w}\mathcal{i}}{G}\, ,
\eeq
with $\textbf{v} = \Psi$ and $\textbf{w} = \dot{\Psi}$ the on-shell left and right wing wave functions and with $G = \mathcal{h}\textbf{u}|\textbf{u}\mathcal{i}$ the Gaudin determinant of the full Bethe state, see figure \ref{Gaudin}. The next one is the partition dependent factor
\beq\label{aa}
a_{\alpha\bar{\alpha}} = \frac{(-1)^{|\bar{\alpha}|}}{h_{\alpha\bar{\alpha}}}\, e^{ip_{\bar{\alpha}} \ell_{31}}\, T(\bar{\alpha}) = \frac{(-1)^{|\bar{\alpha}|}}{h_{\alpha\bar{\alpha}}}\,e^{\frac{i}{2}p_{\bar{\alpha}} (L+L_{1}-L_{2})}\, T(\bar{\alpha})\, ,
\eeq
with implicit products over the elements of the various sets and with $T =$ $\mathcal{h}\Psi|\, \mathbb{T}\,|\dot{\Psi}\mathcal{i}/\mathcal{h}\Psi|\dot{\Psi}\mathcal{i}$ the eigenvalue of the $SU(2|2)$ transfer matrix in the diagonal state $\Psi = \dot{\Psi}$. The latter transfer matrix is evaluated in (\ref{aa}) on the Bethe roots $\bar{\alpha} \subset \textbf{u}$,
\beq\label{osT}
T(\bar{\alpha}) = \prod_{j\in \bar{\alpha}}T(u_{j})\, .
\eeq
We normalize it such that $T(u_i) = 1$ for ${\textbf u} = \{u_i\}$ in the $\mathfrak{sl}(2)$ sector. For a general Bethe state, one finds that $T(u_{i}) = f(u_{i})$ with the $f$  factor (\ref{fofu}), see e.g.~\cite{Beisert:2006qh}. There is no need to carry out the sum over the partitions, since our discussion will apply to each partition independently. We also ignore the remaining overall factors in (\ref{Main}) and (\ref{higherrankA}), like the product over the measures, etc., which also play no role at all. 

The transfer matrix (\ref{osT}) responds promptly to all the manipulations we want to perform. Changing ``its" grading, for instance, is straightforward, using the general expression given in \cite{Beisert:2006qh}: it boils down to replacing the auxiliary roots $\textbf{y}$ and their S-matrices $S$ by their dual versions $\tilde{\textbf{y}}$ and $\tilde{S}$. (This is so up to an overall factor $A_{\bar{\alpha}\alpha}$ that is used to convert $h_{\alpha\bar{\alpha}}$ in (\ref{aa}) from its $\mathfrak{sl}(2)$ to its $\mathfrak{su}(2)$ value \cite{C123Paper}.) What is less obvious is that the ratio (\ref{rr}) responds in the exact same way. The problem being that in (\ref{rr}) one needs to take derivatives of the BAEs, before dualizing. As we shall see, for the ratio (\ref{rr}) one can equivalently proceeds in the reverse order, that is dualize before taking the derivatives.

\subsubsection*{Induced Gaudin determinant}

The separation between main roots $\textbf{u}$ and auxiliary roots $\textbf{v}$ is suggestive of a factorization into two determinants for the two subsystems of equations $\phi_{\textbf{u}} = 2\pi m_{\textbf{u}}$ and $\phi_{\textbf{v}} = 2\pi m_{\textbf{v}}$. Were these two subsystems independent, we would of course immediately conclude that
\beq\label{fact}
G\big|_{\textrm{no interaction}} = G_{\textbf{u}} \times G_{\textbf{v}}\, .
\eeq
(Note that we will not need to distinguish between the two types of auxiliary roots that we have at our disposal. This is why we unite them into a single set, $\textbf{v}\cup \textbf{w} \rightarrow \textbf{v}$. The cut off between real and auxiliary roots is actually immaterial and our discussion applies to a general decomposition into two or more non-overlapping subsystems.) The factorization (\ref{fact}) would also apply to triangular systems, that are such that the dynamics of one subsystem does not depend on the complementary subset of roots. If, for instance, the $\textbf{u}$ system of equations $\phi_{\textbf{u}} = 2\pi m_{\textbf{u}}$ is such that $\partial_{\textbf{v}}\phi_{\textbf{u}} = 0$, then we would still have (\ref{fact}) except that $G_{\textbf{v}}\rightarrow G_{\textbf{v}|\textbf{u}}$, with $G_{\textbf{v}|\textbf{u}} = \textrm{det}\, \partial \phi_{v_{i}}/\partial v_{j}$ the determinant of the $\textbf{v}$ system of equations $\phi_{\textbf{v}} = 2\pi m_{\textbf{v}}$ (with the roots $\textbf{u}$ entering as external parameters). The point is that we can always bring the full system to a triangular form if we treat the $\textbf{v}$'s as being the slaves of the $\textbf{u}$'s. What we get in the more general case is that the factor $G_{\textbf{u}}$ is replaced by the determinant $G_{\textbf{u}|\phi_{\textbf{v}}} = \textrm{det}\, d\phi_{u_{i}}/u_{j}$, with the derivative $d/du_{j}$ being taken at fixed mode numbers $\phi_{\textbf{v}} = 2\pi m_{\textbf{v}}$ instead of fixed rapidities $\textbf{v}$.

The proof of this factorization is elementary. We simply need to recall the interpretation of the determinant $G$ as the Jacobian for the mapping between rapidities and mode numbers, 
\beq
G\,  d\textbf{u}\wedge d\textbf{v} = d\phi_{\textbf{u}}\wedge d\phi_{\textbf{v}}\, ,
\eeq
and evaluate it in two steps,
\beq
G\,  d\textbf{u}\wedge d\textbf{v} = \frac{G}{G_{\textbf{v}|\textbf{u}}}\,  d\textbf{u}\wedge d\phi_{\textbf{v}} = \frac{G}{G_{\textbf{u}|\phi_{\textbf{v}}}G_{\textbf{v}|\textbf{u}}}\,  d\phi_{\textbf{u}}\wedge d\phi_{\textbf{v}}\, ,
\eeq
that is
\beq
G = G_{\textbf{u}|\phi_{\textbf{v}}}G_{\textbf{v}|\textbf{u}}\, .
\eeq
We would arrive at the same conclusion starting with 
\beq
\frac{1}{G} = \int d\textbf{u}d\textbf{v}\,  \delta(\phi_{\textbf{u}}-2\pi m_{\textbf{u}})\delta(\phi_{\textbf{v}}-2\pi m_{\textbf{v}})\, ,
\eeq
integrating over $\textbf{v}$, at fixed $\textbf{u}$, and then integrating over $\textbf{u}$,
\beq
\frac{1}{G} = \int d\textbf{u}\,  \frac{1}{G_{\textbf{v}|\textbf{u}}} \delta(\phi_{\textbf{u}}\big|_{\textbf{v} = \phi_{\textbf{v}}^{-1}(m_{\textbf{v}}, \textbf{u})}-2\pi m_{\textbf{u}}) = \frac{1}{G_{\textbf{u}|\phi_{\textbf{v}}}G_{\textbf{v}|\textbf{u}}}\, .
\eeq
As alluded to before, $G_{\textbf{v}|\textbf{u}}$ is the minor of $G$, obtained by deleting the $\phi_{\textbf{u}}$-rows and $\textbf{u}$-columns, while $G_{\textbf{u}|\phi_{\textbf{v}}}$ is the induced Gaudin determinant for the set of effective equations for the $\textbf{u}$'s. The latter are obtained by 1) solving the equations for $\textbf{v}$ at given mode numbers $m_{\textbf{v}}$ and for a given set of rapidities $\textbf{u}$ (assuming that solutions exist) and 2) plugging the solution $\textbf{v} = \phi^{-1}_{\textbf{v}}(m_{\textbf{v}}, \textbf{u})$ into the equations for $\textbf{u}$ (assuming that it is unique). The nice thing about $G_{\textbf{u}|\phi_{\textbf{v}}}$ is that it eliminates the dependence on the variables used to parameterize the directions transverse to the subspace $\phi_{\textbf{v}} = 2\pi m_{\textbf{v}}$. It indeed measures the density of states (per volume $d\textbf{u}$) on a given ``physical subspace" $\phi_{\textbf{v}} = 2\pi m_{\textbf{v}}$.

In the case of interest we have a tripartite decomposition $\textbf{u}\cup \textbf{v}\cup \textbf{w}$ where $\textbf{v}$ and $\textbf{w}$ do not interact with each other, see figure \ref{Gaudin}. Hence we can write
\beq
G = G_{\textbf{u}|\phi_{\textbf{v}, \textbf{w}}} G_{\textbf{v}, \textbf{w}|\textbf{u}} = G_{\textbf{u}|\phi_{\textbf{v}, \textbf{w}}} G_{\textbf{v}|\textbf{u}} G_{\textbf{w}|\textbf{u}}\, ,
\eeq
and since $\textbf{v} = \textbf{w}$, as a result of the on-shell super selection rules, we also have $G_{\textbf{v}|\textbf{u}}  = G_{\textbf{w}|\textbf{u}}$. Therefore
\beq\label{inducedformula}
\frac{1}{r^2} =  G/G_{\textbf{v}|\textbf{u}}^2 = G_{\textbf{u}|\phi_{\textbf{v}, \textbf{w}}}\big|_{\textbf{v} = \textbf{w}}\, .
\eeq
It shows that $1/r^2$ is the induced Gaudin determinant for the roots $\textbf{u}$ on the subspace $\phi_{\textbf{v}} = \phi_{\textbf{w}} = 2\pi m_{\textbf{v}}$. As such, it should be clear that it does not depend on how we parameterize the higher levels of the wave function and, in particular, on which grading we choose; it is invariant under dualization of the auxiliary roots $\textbf{v}\rightarrow \tilde{\textbf{v}}$,
\beq
\frac{d}{du_{j}} \phi_{u_{j}}(\textbf{u}, \textbf{v}) = \frac{d}{du_{j}} \tilde{\phi}_{u_{j}}(\textbf{u}, \tilde{\textbf{v}})\, ,
\eeq
since, on both sides of this equation, the total derivatives are taken along the same physical subspace. Put differently, when computing $G_{\textbf{u}|\phi_{\textbf{v}, \textbf{w}}}$ we are allowed to dualize (or equivalently use the on-shell conditions for the $\textbf{v}\cup \textbf{w}$'s) prior to take the derivatives w.r.t.~the $\textbf{u}$'s.

\subsubsection*{Diagonal symmetry}

Let us comment finally on the diagonal $\mathfrak{su}(2|2)_{D}$ symmetry. This is a symmetry of the structure constant \cite{C123Paper}. It should thus be reflected in our final expression (\ref{Main}). As well known, see e.g. \cite{Beisert:2006qh}, this symmetry can be phrased as the invariance under addition of auxiliary roots at the special points $y = 0, \infty$.%
\footnote{One must also include $w=\infty$, with $w=v^{(2)}$ an $\mathfrak{su}(2)$ root, as well as $\tilde{w} = \infty$ in the dual frame for the second $\mathfrak{su}(2)\subset \mathfrak{su}(2|2)$. Combinations of the type $(y, w) = (0, \infty)$ and $(y, w) = (\infty, \infty)$, with $w-g(y+1/y)$ held fixed, should also be considered to get all the supercharges.} For $y=0$ the symmetry is dynamical and related to the joining/splitting of free multiplets at the unitarity bound \cite{BS}. (This is literally true for states in the $\mathfrak{su}(1,1|2)$ sector.) It holds for cyclic states only and comes along with a redefinition of the length of the spin chain \cite{BS}.%
\footnote{Spin chain states form proper $\mathfrak{psu}(2,2|4)$ supermultiplets only if they are cyclic. Non cyclic states have no counterparts in the gauge theory.}
Roughly speaking, adding or removing a root at $y = \dot{y} = 0$ soaks up or spits out two units of spin chain length. For example, a length $L$ two derivative BMN operator and a length $L+2$ two scalar BMN operator, sharing the same roots $u_{1} = -u_{2}$, fall in the same supermultiplet and differ by the adjunction of a root at $y = \dot{y} = 0$ (+ some at infinity).

The diagonal symmetry is manifest in (\ref{aa}) since the transfer matrix is invariant under sending roots to infinity. It also transforms multiplicatively, in the spin chain frame, under addition of a root at $y=0$,
\beq
T(u; y=0, \textrm{rest}) = e^{-ip(u)}\times T(u; \textrm{rest})\, ,
\eeq
hence implementing the length changing effect $L\rightarrow L-2$ for the removal of the fermionic roots $y=0$ and $\dot{y} = 0$ from the state. (We work here in the non compact grading.) It is also not difficult to prove that the ratio (\ref{rr}) remains unchanged when sending auxiliary roots to infinity. It does not transform correctly when setting roots at $y=\dot{y} = 0$ however. The problem is that the Gaudin determinant requires to take derivatives of the BAEs 
\textit{prior} to impose cyclicity, while we would need the reverse to make our point. The latter two operations do not quite commute, which is why the ratio (\ref{rr}) is not per se a diagonal invariant. The mismatch is not that big however and, as shown below, drops out of the full structure constant.

Given our earlier discussion, it should be clear that the quantity that is invariant is the induced norm on the subspace of cyclic states. This is not quite the same as (\ref{rr}). The Gaudin determinant $G_{\textbf{u}|\phi_{\textbf{v}, \textbf{w}}}|_{\textbf{v} = \textbf{w}}$ measures the density of unconstrained spin chain states on the subspace $\phi_{\textbf{v}} = \phi_{\textbf{w}} = 2\pi m_{\textbf{v}}$. The latter counts $L$ too many states, as compared to the gauge theory, because of the $L-1$ subspaces of unrealized solutions to the equation $e^{ip_{\textbf{u}}L} = 1$ (which itself is a consequence of the BAEs). Hence, imposing the cyclic condition $e^{ip_{\textbf{u}}} = 1$ at the level of the determinant amounts to rescaling it by $L$, and the invariant quantity is
\beq
L r^2\, ,
\eeq 
and not $r$ alone. This small extra factor makes a difference since the length $L$ transforms in the multiplet splitting/joining. The product $Lr^2$ should not. In the hexagon construction, the factor $\sqrt{L}$ multiplying $r$ is provided by the vacuum structure constant $C^{\circ\circ\circ} = \sqrt{L_{1}L_{2}L}/N$ where $N$ is the rank of the gauge group. The (properly normalized) structure constant is thus an $\mathfrak{su}(2|2)_{D}$ invariant, as expected.

Technically, to prove the invariance of $Lr^2$, we write
\beq
\begin{aligned}
G_{\textbf{u}|\phi_{\textbf{v}, \textbf{w}}}\big|_{\textbf{w} = \textbf{v}} &= \textrm{det}\,\bigg[L\frac{dp_{j}}{du_{k}} + \frac{d}{idu_{k}}\log{\frac{x^{-}_{j}-y}{x^{+}_{j}-y}} + \frac{d}{idu_{k}}\log{\frac{x^{-}_{j}-\dot{y}}{x^{+}_{j}-\dot{y}}} + \ldots \bigg]_{y = \dot{y}=0} \\
&= \frac{L}{L-2} \times \textrm{det}\,\bigg[(L-2)\frac{dp_{j}}{du_{k}} + \ldots \bigg]\, ,
\end{aligned}
\eeq
where $L/(L-2)$ emerges as the ratio of two Jacobians : $dLP/dP = L$, obtained by replacing the equation $LP = 2\pi m$ by the cyclic constraint $P = 2\pi m$ in the determinant for the initial spin chain of length $L$,%
\footnote{We can always substitute $LP = 2\pi m$ to one of the equations defining the determinant, since this equation is just the sum of the rows of the Gaudin matrix.} and $dP/d(L-2)P = 1/(L-2)$, obtained by reversing the procedure for the determinant of the final spin chain of length $L-2$. Inbetween we have applied
\beq
\bigg\{\frac{d}{idu_{k}}\log{\left[\frac{x^{-}_{j}-y}{x^{+}_{j}-y}\right]}\bigg\}_{y= 0} = \frac{d}{idu_{k}}\log{\bigg[\frac{x^{-}_{j}-y}{x^{+}_{j}-y}}\bigg]_{y=0} = -\frac{dp_{j}}{du_{k}} \, ,
\eeq
and similarly for $\dot{y}$, which are valid as soon as the derivative $d/du_{k}$ is taken along the cyclic subspace $y = \dot{y} =0$.

By the same token, one can demonstrate that the dynamic transformation mentioned at the very beginning is correctly implemented in our expressions.

\section{Comparison with Data : A Special Case}\label{ap:special case}

In this appendix we analyze in detail how to obtain the OPE data sum $\mathbb{P}_{\tau= 6 , s=0}^{n=0,m=0}$ in table~\ref{table} using the conjecture  \eqref{Main}. Unlike the rest of the data that we have checked, this sum receives contributions from operators of different lengths. For these quantum numbers we have a total of $7$ wing-symmetric Bethe solutions:
\begin{equation}\label{eq:sol7}
\begin{matrix}
\text{Lenght}\,L &\text{Field content at } \mathcal{O}(g^{0})  &  \#  \,\text{Roots in }\mathfrak{sl}(2)\text{-grading}  & \text{\#\,Wing-symmetric sols}  \\
\\
4 & \qquad Tr(D\bar{D}Z\bar{Z}Z\bar{Z})+\cdots  & \qquad \{1,2,3,6,3,2,1\} & \qquad 2 \\
\\
6 & \qquad Tr(Z\bar{Z}Z\bar{Z}Z\bar{Z})+\cdots  & \qquad \{0,2,4,6,4,2,0\} & \qquad 5
\end{matrix}
\end{equation}

The five solutions with $L=6$ correspond to operators in the $so(6)$ sector at $\mathcal{O}(g^{0})$. At loop level, their roots receive corrections in even powers of the coupling $g$ and  can be used straightforwardly in \eqref{Main} to obtain the corresponding structure constants. Theses solutions  behave in a standard way so we do not review them in this appendix.

The two solutions with $L=4$ have roots in all the $7$ nodes of the $\mathfrak{psu}(2,2|4)$ Dynkin diagram. Hence it constitutes an interesting case that proves all the components of the conjecture \eqref{Main}. In the following we analyze in more detailed these solutions.

At $g=0$ these solutions contain two vanishing fermionic roots $v^{(3)}_{1}=v^{(1)}_{1}=0$ (similary $w^{(3)}_{1}=w^{(1)}_{1}=0$). These however are not associated to the action of supercharges. As we show in table \ref{table:roots} these zeros receive corrections at loop order and are lifted to take opposite non-zero values. Their corrections start at $\mathcal{O}(g^{1})$, unlike the rest of roots that start at $\mathcal{O}(g^{2})$, and have an unusual expansion in odd powers\footnote{The existence of these types of Bethe roots was first observed in \cite{BS}, see section 5.2 therein.} of $g$. In terms of the Zhukovski variables the relation between the fermionic roots translates into:
\begin{equation}\label{eq:Xfermion}
v^{(1)}_{1}=-v^{(3)}_{1}\,\to \, x(u^{(3)}_{1})=-\frac{1}{x(u^{(1)}_{1})}\quad\text{with: } \frac{v}{g}\,= x(v)+\frac{1}{x(v)}\, 
\end{equation}
At loop level we can perform a dynamical  transformation of the fermionic roots, as explained in appendix \ref{ap:rewriting}. We can treat the roots of type $(3)$ as of type $(1)$, in both wings, by going through the cut ($x\to \frac{1}{x}$) and increasing the length of the operator. In this way, at loop order, the operator with length $L=4$ and excitations  $\{1,2,3,6,3,2,1\}$ has an alternative description with length $L=6$ and excitations $\{0,2,4,6,4,2,0\}$. In this latter description the new fourth wing root of type $(1)$ is simply identified with the root of type $(3)$ in the former description and the correspondent Zhukovski variable changes as:
\begin{align}\label{eq:dynamicalT}
& \text{Zhukovsky:} & x(u^{(3)}_{1}) &\,\to\, x(u^{(1)}_{4})=\frac{1} {x(u^{(3)}_{1})}   &\nonumber\\
& \text{Wing root:} & v^{(3)}_{1} &\,\to\, v^{(1)}_{4} = v^{(3)}_{1} &\nonumber\\
& \text{Excitations:} & \{1,2,3,6,3,2,1\} &\,\to\, \{0,2,4,6,4,2,0\} &\nonumber\\
& \text{Length:} & L=4 &\,\to\, L=6 &
\end{align}

\begin{table}[H]
\renewcommand{\arraystretch}{1.8}
\begin{tabular}{l|l|l}
  & \text{First solution up to $\mathcal{O}(g^{4})$} & \text{Second solution up to $\mathcal{O}(g^{4})$}  \\ \hline
${\color{red}v^{(3)}_{1}}$ \!\! & ${\color{red}-2.1110824\, g + 5.3821312\, g^3 + \mathcal{O}(g)^{4}}$ & ${\color{red}-0.3000041 \,i\, g-5.082979 i\, g^3 } $ \\ \hline
 $v^{(2)}_{1}$ & $-0.37453099-3.8678404\, g^2$ & $-0.5540218\, i-1.802503 \,i\, g^2 $ \\ \hline
$v^{(2)}_{2}$ & $+0.37453099+3.8678404\, g^2$ & $+0.5540218\, i + 1.802503 \,i \, g^2$ \\ \hline
 ${\color{red} v^{(1)}_{1}}$ & ${\color{red}+2.1110824 \, g-5.3821312\, g^3}$ & ${\color{red}+0.3000041 \,i\, g + 5.082979 \,i\, g^3}$ \\ \hline
$v^{(1)}_{2}$ & $-0.41330424-2.7636175\, g^2$ & $1.0820445 \,i + 1.8719637\, i\, g^2$ \\ \hline
 $v^{(1)}_{3}$ & $+0.41330424 + 2.7636175 \,g^2$ & $-1.0820445 \,i-1.8719637\, i\, g^2$ \\ \hline
$u_{1}$ & $-0.074924705 g^2-0.43054180$ & $-5.3834596 g^2-1.2029572$ \\ \hline
$u_{2}$ & $-0.4259447-0.5088469 i-(2.949711+0.111629 i) g^2$ & $-3.9652234 g^2-0.53383287$ \\ \hline
$u_{3}$ & $-0.4259447+0.5088469 i-(2.949711-0.111629 i) g^2$ & $-1.7006510 g^2-0.15250255$ \\ \hline
$u_{4}$ & $+0.4259447-0.5088469 i+ (2.949711-0.111629 i) g^2$ & $0.15250255+1.7006510 \, g^2$ \\ \hline
 $u_{5}$ & $+0.4259447+0.5088469 i+(2.949711+0.111629 i) g^2$ & $0.53383287+3.9652234 \, g^2$ \\ \hline
 $u_{6}$ & $+0.43054180 + 0.074924705\, g^2$ & $1.2029572+5.3834596 \,g^2$ 
\end{tabular}
\caption{The two wing-symmetric Bethe solutions for $L=4$ , $\Delta_{0}=6$, $s=0$, $n=0$ and $m=0$  in the $\mathfrak{sl}(2)$-grading. The roots in red vanish at $\mathcal{O}(g^{0})$ and have an unusual expansion in odd powers $g$, with $v^{(3)}_{1}=-v^{(1)}_{1}$. The next correction is of $\mathcal{O}(g^{4})$ order for all the roots displayed.}
\label{table:roots}
\end{table}
When computing the correspondent normalized structure constants using the conjecture \eqref{Main} we can use any of the two equivalent descriptions in \eqref{eq:dynamicalT}. Using the roots in table \ref{table:roots}, the components of the conjecture behave according to the discussion in appendix \ref{ap:rewriting} :
\begin{equation}\label{eq:Addbar}
\mathcal{A}\big{|}_{L=4,\,\{1,2,3,6,3,2,1\}} \,=\,  \mathcal{A}\big{|}_{L=6,\,\{0,2,4,6,4,2,0\}} 
\end{equation}
and:
\begin{equation}\label{eq:ratioGddbar}
L\times \frac{\langle\mathbf{v}|\mathbf{v} \rangle^{2}}{\langle \mathbf{u} | \mathbf{u}\rangle}\bigg{|}_{L=4,\,\{1,2,3,6,3,2,1\}}  = L\times \frac{\langle\mathbf{v}|\mathbf{v} \rangle^{2}}{\langle \mathbf{u} | \mathbf{u}\rangle}\bigg{|}_{L=6,\,\{0,2,4,6,4,2,0\}} 
\end{equation}
Although a couple of fermionic roots in table \ref{table:roots} have an expansion in odd powers of $g$, the resulting components \eqref{eq:Addbar} and \eqref{eq:ratioGddbar} of the hexagon conjecture have the usual expansion in even powers of $g$. By a simple inspection of \eqref{fofu} we can check the Zhukovski variables in \eqref{eq:Xfermion} fuse to give an expansion in even powers of $g$ for the sum over partitions $\mathcal{A}$.

Finally to reproduce the correspondent OPE data in table  \ref{table} we plug the $\mathcal{O}(g^{4})$ roots of the Bethe solutions \eqref{eq:sol7} in formula \eqref{Main} and sum over all $7$ contributions obtaining:
\begin{equation}
\mathbb{P}_{\tau= 6 , s=0}^{n=0,m=0}\big{|}_{y=0} = \sum_{\text{2 solutions with}\atop L=4,\{1,2,3,6,3,2,1\}}  \,\left(C^{\circ\circ\bullet}\right)^{2}+\sum_{\text{5 solutions with}\atop L=6,\{0,2,4,6,4,2,0\}}  \,\left(C^{\circ\circ\bullet}\right)^{2}\, = \, \frac{13}{210}-\frac{31}{63}\,g^{2}+\mathcal{O}\left(g^4\right)\,. \nn
\end{equation}

\section{Pfaffian Representations\la{ap:pfaffian}}
In this Appendix, we show that the sum over partition \eqref{higherrankA} at finite coupling can be recast as a Pfaffian of a finite-dimensional matrix. This rewriting has two virtues: It reduces the cost of numerical computation, which is extremely heavy when the number of roots is large. In addition, the argument is potentially applicable to the mirror corrections as well.

The first step is to rewrite \eqref{higherrankA} as
\beq
\mathcal{A}=\prod_{i<j}h(u_i,u_j)\sum_{\bar{\alpha}\subset  {\bf u}}(-1)^{|\bar{\alpha}|}\prod_{\substack{k<l\\k,l\in \bar{\alpha}}}H(u_k,u_l)\prod_{j\in \bar{\alpha}} e_j
\eeq
with\footnote{Note that $h(u,v)$ is given by \cite{C123Paper}
\beq
h(u,v)=\frac{x^{-}-y^{-}}{x^{-}-y^{+}}\frac{1-1/x^{-}y^{+}}{1-1/x^{+}y^{+}}\frac{1}{\sigma (u,v)}\,.
\eeq
}
\beq
\begin{aligned}\la{interaction}
H(u,v)&\equiv h(u,v)h(v,u)=\frac{(x^{-}-y^{-})(x^{+}-y^{+})(1-1/x^{-}y^{+})(1-1/x^{+}y^{-})}{(x^{+}-y^{-})(x^{-}-y^{+})(1-1/x^{+}y^{+})(1-1/x^{-}y^{-})}\,,\\
e_j&\equiv \frac{f(u_j)e^{i p(u_j) \ell_{31}}}{\displaystyle{\prod_{i\in {\bf u},i\neq j}}h(u_i,u_j)}\,.
\end{aligned}
\eeq
The next step is to use the identity\footnote{We found this formula empirically using Mathematica. We then learned that it is a special case of the elliptic generalization of Schur's Pfaffian formula (see eq. (16) of \cite{ellipticpfaffian}).}, which holds for even $N$,
\beq
{\rm pf} \left(k(x_n,x_m) \right)_{1\leq n,m\leq N} =\prod_{i<j}k(x_i,x_j)\,,
\eeq
where ${\rm pf}$ denotes a Pfaffian of a matrix and $k(x,y)$ is given by
\beq
k(x,y)=\frac{x-y}{1-xy}\,.
\eeq
Using this identity, we can rewrite a product of $H(u,v)$ over a set of rapidities $s=\{u_1,u_2,\ldots\}$ as 
\beq
\prod_{\substack{ i<j\\i,j\in s}}H(u_i,u_j)=(-1)^{\frac{n(n-1)}{2}}\left[\prod_{i=1}^{n}k(x_i^{+},x_i^{-})\right]{\rm pf} K_{s} \,,
\eeq
with
\beq
\left( K_{s}\right)_{ij}=k(t_i,t_j)\,,\qquad\quad  t_i= \begin{cases}x_i^{+} &1\leq i\leq |s|\\1/x^{-}_{i-|s|}&|s|< i\leq 2|s|\end{cases}  \qquad \left(x_i \equiv x(u_i)\right) \,.
\eeq
We thus have
\beq\label{pfaffexp}
\mathcal{A}=\prod_{i<j}h(u_i,u_j)\underbrace{\sum_{\bar{\alpha}\subset {\bf u}}(-1)^{|\bar{\alpha}|}\left(\prod_{u_j\in\bar{\alpha}}g_j\right){\rm pf}K_{\bar{\alpha}}}_{(\ast)}\,,
\eeq
with $g_j\equiv k(x_j^{+},x_j^{-})e_j$. The series representation \eqref{pfaffexp} is akin to the expansion of the so-called Fredholm Pfaffian \cite{Rains}. In fact, using the expansion formula for the Pfaffian, we can recast it into a Pfaffian of a $2K\times 2K$ matrix as follows:
\beq\label{finitefred}
(\ast)={\rm pf }\left(J-EK_{{\bf u}}E\right)\,,
\eeq
Here $J$ and $E$ are given by
\beq
J=\left(\begin{array}{cc}0 &I\\-I&0\end{array}\right)\,,\qquad E=\left(\begin{array}{cc}{\rm diag}(g_1,\ldots,g_M) &0\\0&I\end{array} \right)\,,
\eeq
with  $I_{M}$ being the identity matrix of rank $K$.

 Using \eqref{finitefred} and the well-known fact that a square of a Pfaffian is a determinant, we can express $\mathcal{A}^2$ in \eqref{Main} as a determinant. After rewriting a bit, the result reads
 \beq
 \mathcal{A}^2 =\left(\prod_{i< j}h(u_i,u_j)\right)^2 \det \left(I-\mathcal{K}\right)_{2K\times 2K}
  \eeq 
  where $\mathcal{K}$ has a block structure
  \beq
  \mathcal{K}\equiv \left(\begin{array}{cc}\mathcal{K}_{11}&\mathcal{K}_{12}\\\mathcal{K}_{21}&\mathcal{K}_{22}
  \end{array}\right)\,,
  \eeq
  with
  \beq
  \begin{aligned}
  &\left(\mathcal{K}_{11}\right)_{ij}=g_i k(x_i^{+},1/x_j^{-})\,,&& \left(\mathcal{K}_{12}\right)_{ij}=g_i k(x_i^{+},x_j^{+})\,,\\
  &\left(\mathcal{K}_{21}\right)_{ij}=-g_i k(1/x_i^{-},x_j^{+})\,,&& \left(\mathcal{K}_{22}\right)_{ij}=-g_i k(1/x_i^{-},1/x_j^{-})\,.
  \end{aligned}
  \eeq
Combining this result with the result in Appendix \ref{ap:rewriting}, we can express the square of the structure constant simply as a ratio of two determinants as
\beq
\begin{aligned}
\left(\frac{C^{\circ\circ\bullet}_{123}}{C_{123}}\right)^2=\left(\prod_k \mu(u_k) \prod_{i\neq j} h(u_i,u_j)\right)\frac{\det \left(I-\mathcal{K}\right)}{G_{\textbf{u}|\phi_{\textbf{v}, \textbf{w}}}\big|_{\textbf{v} = \textbf{w}} }\,,
\end{aligned}
\eeq  
where $G_{\textbf{u}|\phi_{\textbf{v}, \textbf{w}}}$ is the induced Gaudin determinant defined in \eqref{inducedformula}.

  Let us finally mention the potential application to the mirror corrections. The structure of the interaction term in the mirror-particle integrand given in \cite{3loop} takes the same form as \eqref{interaction}. Therefore, one can use the generalized Schur's Pfaffian identity also for the mirror particles and recast each term as a Pfaffian. Furthermore, in the case of mirror particles, the full expansion coincides exactly with the expansion of the Fredholm Pfaffian. It is still to be seen if the sum over bound-state indices leads to a further simplification, but in any case, such an expression would certainly be useful for resumming the finite size correction \cite{clustering} at finite coupling.  
\section{The $\mathfrak{so}(6)$ Structure Constant at Tree Level}\label{ap:so(6)}
In this appendix we compute the tree-level three point function of operators in the $\mathfrak{so}(6)$ sector of planar $N=4$ SYM. We focus on the case of two $\frac{1}{2}$-BPS operators and one non-BPS single-trace operators:
\begin{equation}
\mathcal{O}_{1} = Tr(\tilde{Z}^{L_{1}})\,, \qquad \mathcal{O}_{2} = Tr(\bar{Z}^{L_{2}})\,,\qquad \mathcal{O}_{3} = Tr(\underbrace{ Z\bar{Z}X\bar{X}\cdots}_{L} )+\cdots \la{setup}
\end{equation}
The protected operators are given BMN vacua spanned by the elementary field $\bar{Z}$ and by the rotated field $\tilde{Z} \,\equiv\, Z+\bar{Z}+X-\bar{X}\,$. The non-protected operator is given by an eigenstate of the one-loop dilatation operator. Such operator can be obtained by diagonalizing the Hamiltonian of the dual integrable spin chain \cite{so6Zarembo}, using the Algebraic Bethe Ansatz (ABA). We develop on such construction in section \ref{sec:ABA}. 

\begin{figure}[H]
\begin{center}
\begin{tikzpicture}
\node(C123) at (-2,5.5){\huge $\mathcal{C}_{123}\, = $};
\node(O2) at (-.8,.6){$\mathcal{O}_{1}\equiv Tr(\color{red}\tilde{Z}\cdots \tilde{Z}\,\color{blue}\tilde{Z}\cdots \tilde{Z}\color{black})$};
\node(O3) at (9.2,.6){$\mathcal{O}_{2}\equiv Tr(\color{blue}\bar{Z}\cdots \bar{Z}\,\bar{Z}\cdots \bar{Z}\color{black} )$};
\node(O1) at (4.1,9.4){$\mathcal{O}_{3}\equiv Tr(\overset{\color{red}{magnons}}{\Phi\cdots \Phi}\,\color{blue}Z\cdots Z\color{black} )$};

\draw [thick,-,double,blue,bend left]  (1.2,1) to (8.8,1);
  \draw [thick,-,double,blue,bend left]  (1,1) to (9,1);
  \draw [thick,-,double,blue,bend left]  (.8,1) to (9.2,1);
  \draw [thick,-,double,blue,bend left]  (.6,1) to (9.4,1);
  \draw [thick,-,double,blue,bend left]  (.4,1) to (9.6,1);
  \draw [thick,-,double,blue,bend left]  (.2,1) to (9.8,1);
\draw [thick,-,double,blue,bend left]  (10,1) to (5,9);
\draw [thick,-,double,blue,bend left]  (10.2,1) to (5.2,9);
\draw [thick,-,double,blue,bend left]  (10.4,1) to (5.4,9);
\draw [thick,-,double,blue,bend left]  (10.6,1) to (5.6,9);
\draw [thick,-,double,blue,bend left]  (10.8,1) to (5.8,9);
\draw [thick,-,double,blue,bend left]  (11,1) to (6,9);
 \draw [thick,-,double,red,bend right]  (0,1) to (4.7,9);
 \draw [thick,-,double,red,bend right]  (-.2,1) to (4.5,9);
 \draw [thick,-,double,red,bend right]  (-.4,1) to (4.3,9);
 \draw [thick,-,double,red,bend right]  (-.6,1) to (4.1,9);
 \draw [thick,-,double,red,bend right]  (-.8,1) to (3.9,9);
 \draw [thick,-,double,red,bend right]  (-1,1) to (3.7,9);
 \draw [thick,<->,red] (2,6) coordinate() node[left]{``l" bridge} to (4,4);
 \draw [thick,<->,blue] (5.5,4)  to (7.5,6)  node[right]{``L-l" bridge};
 \draw [thick,<->,blue] (4.8,1.5)  node[right]{trivial bridge}  to (4.8,3.5) ;
\end{tikzpicture}
\end{center}
\caption{Wick contraction for planar tree level structure constant $\mathcal{C}_{\mathcal{O}_{1}\mathcal{O}_{2}\mathcal{O}_{3}}$}
\label{fig:wickcontraction}
\end{figure}
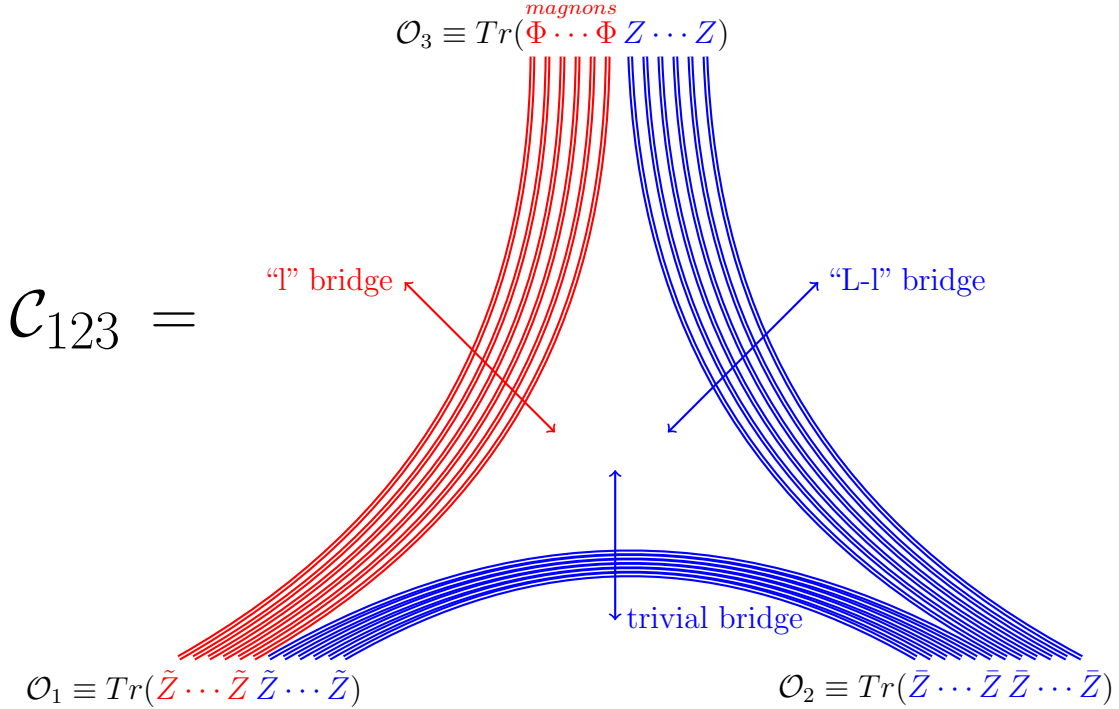

This three point function can be computed by Wick contractions as shown in figure \ref{fig:wickcontraction}. 
Following the tailoring procedure, introduced for the $\mathfrak{su}(2)$ sector in \cite{su2Tailoring}, we can express the wick contractions as scalar products of spin chain states dual to the single trace operators. In our configuration -- dubbed the reservoir picture in \cite{C123Paper} -- we have two trivial bridges which only feature propagators of the type $Z$-$\bar{Z}\,$ (blue lines). The only non-trivial wick contraction comes therefore from the bridge $l=(L+L_{1}-L_{2})/{2}$ between operators $\mathcal{O}_{1}$ and $\mathcal{O}_{3}$ and is given by the spin chain scalar product\footnote{This renders an unnormalized structure constant. The normalized version includes the norms of the three operators involved}
\begin{equation}\label{eq:c123brute}
\mathcal{C}_{123}\, = \, \langle \tilde{Z}^{l} \,|\, \Psi_{l}\rangle 
\end{equation}
where $\Psi_{l}$ is a sub-chain of length $l$ in the cyclic spin chain state $\Psi$, dual to the single trace operator i.e. $\mathcal{O}_{3}\equiv |\Psi\rangle$.
In this way the computation of the tree level three-point function is reduced to finding an inner product of states in the dual $\mathfrak{so}(6)$ spin chain. In the $\mathfrak{su}(2)$ sector the relevant scalar product was computed with ABA techniques \cite{su2Tailoring}, obtaining sum formulas or more compact determinant expressions for some special cases, see \cite{su2Slavnov} for a review. In \cite{su3Wheeler} $\mathfrak{su}(3)$ scalar products were computed and used in \cite{su3Foda} for the computation of structure constants in this sector. While for the $\mathfrak{so}(6)$ spin chain there are no available formulas for the scalar products\footnote{An attempt to conjecture the $\mathfrak{so}(6)$ scalar product, based on the results for $\mathfrak{su}(2)$, was given in \cite{so6wrong}} in the literature. One of the obstacles being the complexity of the ABA for $\mathfrak{so}(2n)$ models \cite{so6deVega}. To overcome this problem we construct an alternative version\footnote{$\mathfrak{so}(6)$ is special due to is isomorphism with $\mathfrak{su}(4)$} of the $\mathfrak{so}(6)$ ABA which allows for a simpler approach to the computation of scalar products\footnote{A very similar approach was developed independently by Carlo Meneghelli \cite{carloM}.}. In particular we use this machinery to find the scalar product \eqref{eq:c123brute} as a sum formula. This result makes direct contact with the conjecture \eqref{Main}, restricted to $\mathfrak{so}(6)$ at tree level, showing the structure of a sum over partitions and a matrix part depending on the nested levels of the Bethe Ansatz. It would be interesting to generalize this Bethe Ansatz to address operators in the full sector $\mathfrak{psu}(2,2|4)$ and reproduce the conjecture \eqref{Main} at tree level.

The rest of this appendix is organized as follows: in section \ref{sec:so6spinchain} we introduce the $\mathfrak{so}(6)$ integrable spin chain, the correspondent transfer matrix, as well as some notation.  In section \ref{sec:vertexmodel} we present a novel $\mathfrak{so}(6)$ vertex model, specifying the Boltzmann weights and the way to extract the Bethe states from the lattice.  In section \ref{sec:ABA} we develop a $\mathfrak{so}(6)$ ABA for the diagonalization of the spin chain Hamiltonian and transfer matrix. In section \ref{sec:YB} we present the Yang-Baxter algebra, showing how to use it to derive the so called wanted and unwanted terms of the ABA. In section \ref{sec:CBA} we present the coordinate Bethe Ansatz (CBA) which can be derived from our ABA and vertex model. Finally, in section \ref{sec:scalarproduct} we put in used the Yang-Baxter algebra to compute the tree level structure constant given by the scalar product \eqref{eq:c123brute}. We show how to simplify the result to obtain the $\mathfrak{so}(6)$ tree-level analog of the conjecture \eqref{Main}.

\subsection{The $\mathfrak{so}(6)$ spin chain}\label{sec:so6spinchain}
To obtain a basis of non-BPS operators we need to diagonalize the $\mathfrak{so}(6)$ integrable spin chain Hamiltonian:
\begin{equation}\label{eq:so6Hamiltonian}
\mathcal{H}_{\mathfrak{so}(6)} = \sum_{l=1}^{L} \left(I_{l,l+1}-P_{l,l+1}-\frac{1}{2}\,K_{l,l+1}\right)
\end{equation}
where $I$,$P$ and $K$ are identity, permutation and trace operators respectively.

This spin chain Hamiltonian is proportional to the   one loop dilatation operator in the $\mathfrak{so}(6)$ sector of $N=4$ SYM. The basis of eigenstates of this Hamiltonian constitutes a basis for non-BPS operator of the one loop $\mathfrak{so}(6)$ sector, which (partially) lifts the original tree level degeneracy. The single trace operators of this sector are mapped to cyclic states of the spin chain as:
\begin{equation}
Tr(Z\bar{Z}X\,Y\bar{X}) \longrightarrow |Z\bar{Z}X\,Y\bar{X}\rangle + \text{cyclic permutations}
\end{equation}
where the elementary fields are given by the complex scalars fields:
\begin{equation}\label{eq:Zletters}
Z \equiv \Phi_{12}\,,\qquad X \equiv \Phi_{23}\,,\qquad Y \equiv \Phi_{13}\,,\qquad \bar{Y} \equiv \Phi_{42}\,,\qquad \bar{X} \equiv \Phi_{14}\,,\qquad \bar{Z} \equiv \Phi_{34}\,.
\end{equation}
These scalar fields form a multiplet of the antisymmetric $\mathbf{6}$ representation of $\mathfrak{su}(4)$, isomorphic to the vector representation of $\mathfrak{so}(6)$. This isomorphism is realized by the transformation.
\begin{equation}\label{eq:Phimatrix}
\Phi_{a\,b} =\begin{pmatrix}
0 && \phi_{1}+i \phi_{4} && \phi_{2} + i \phi_{5} && \phi_{3} - i \phi_{6} \\
-\phi_{1} - i \phi_{4} && 0 && \phi_{3} + i \phi_{6} && -\phi_{2}+ i \phi_{5} \\
-\phi_{2}- i\,\phi_{5} && -\phi_{3}-i\, \phi_{6} && 0 && \phi_{1}- i \phi_{4} \\
-\phi_{3} + i\,\phi_{6} && \phi_{2}-i \phi_{5} && -\phi_{1} + i \phi_{4} && 0 \\
\end{pmatrix} 
\end{equation} 
In this appendix we stick to the basis of complex scalars \eqref{eq:Zletters}.

The ABA finds the spectrum of the $\mathfrak{so}(6)$ Hamiltonian by solving the eigenvalue problem of the transfer matrix, the trace of a monodromy operator. This  operator is constructed by ``scattering" a probe particle, in an auxiliary space $\mathbf{\Lambda}$ and spectral parameter (momentum)$u$, with all the spin chain sites. We can build various monodromies by choosing the auxiliary space to lie in any of the representations of the spin chain symmetry group:
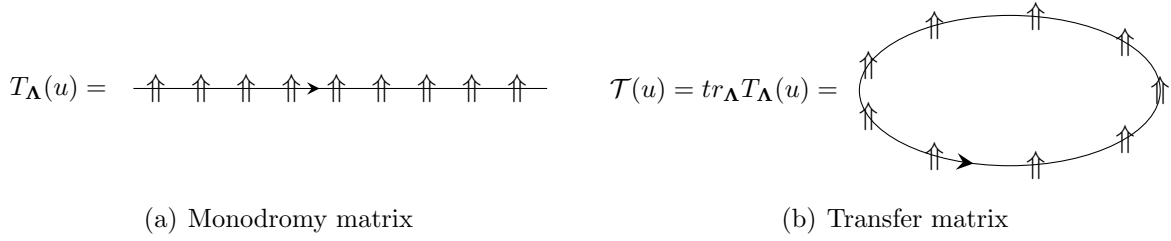
\begin{figure}[H]
\begin{center}
\subfigure[Monodromy matrix]
{
\begin{tikzpicture} 
\node[draw=none]  at  (0,-1.2) {};
\node[draw=none]  at  (-1.3,0) {\footnotesize $T_{\mathbf{\Lambda}}(u)=$};
\draw[decoration={markings,mark=at position .45 with
    {\arrow[scale=1.5,>=stealth]{>}}},postaction={decorate}] (-.3, 0) to (5.2,0);
\foreach \x in {0,.6,...,5}{
    \node[draw=none,minimum width=10pt] at ($(0,0)+(\x, 0)$) {$\Uparrow$};
    }
\end{tikzpicture}
}
\quad
\subfigure[Transfer matrix]
{
\begin{tikzpicture}
\node[draw=none]  at  (-3.8,0) {\footnotesize $\mathcal{T}(u)=tr_{\mathbf{\Lambda}}T_{\mathbf{\Lambda}}(u)=$};
\draw[decoration={markings,mark=at position .7 with
    {\arrow[scale=2,>=stealth]{>}}},postaction={decorate}] (0, 0) ellipse [x radius = 2cm, y radius = 1cm];
\foreach \x in {0,40,...,330}{
    \node[draw=none,minimum width=10pt] at ($(0,0)+(\x:2 and 1)$) {$\Uparrow$};
    }
\end{tikzpicture}
}
\end{center}
\caption{$\mathfrak{so}(6)$ Monodromy and Transfer matrices with $\mathbf{\Lambda}$ auxiliary space. The spin chain sites represent any of the elementary fields $\Uparrow \equiv \Phi_{ab}$ in \eqref{eq:Zletters}}.
\end{figure}
From all these possibilities one is distinguished and corresponds to the choice of auxiliary space in the same representation as the spin chain sites. The trace of this special choice, the transfer matrix, is a generating function of a family of local conserved charges including the nearest-neighbour Hamiltonian of the spin chain. For our $\mathfrak{so}(6)$ spin chain the distinguished monodromy is $T_{\mathbf{6}}$, its trace  generates local conserved charges such as the Hamiltonian \eqref{eq:so6Hamiltonian}. Other choices of auxiliary space do not generate the spin chain Hamiltonian, nevertheless their correspondent transfer matrices are in convolution with the distinguished one. This means that we can address the eigenvalue problem for the spin chain Hamiltonian and all the transfer matrices in convolution at once. So we can choose to solve the eigenvalue problem of the simplest transfer matrix. For our $\mathfrak{so}(6)$ spin chain the simplest choice  corresponds to $T_{\mathbf{4}}$ with auxiliary space in the $\mathbf{4}$ fundamental representation of $\mathfrak{su}(4)$ as:
\begin{equation}\label{eq:mono4}
\begin{tikzpicture}
\node(O1) at (-3,.9){$\left(T_{\mathbf{4}_{\,}}(u)\right)_{a\;;\,\Phi_{a_{1}b_{1}}\cdots \Phi_{a_{L}b_{L}}}^{b\;;\,\Phi_{c_{1}d_{1}}\cdots \Phi_{c_{L}d_{L}}}=$};

\node(ppp) at (5,.5){$\cdots$};
    \draw [thick,->-=.05,->-=.20,->-=.35,->-=.5,->-=.65,->-=.80,->-=0.95,black] (0,1.0) coordinate (a_1) node[left]{$a$} node [below]{\footnotesize $\quad u$}   to (7.0,1.0) coordinate (d_1) node[right]{$b$} ;
         
     \draw [double,->-=.166,->-=.833,black](1.0,0) coordinate (i_1) node[right]{\tiny $\theta_{1}$} node [below,black]{\footnotesize $\Phi_{a_{1}b_{1}}$}  to (1.0,2.0) coordinate (l_1) node [above]{\footnotesize $\Phi_{c_{1}d_{1}}$}  ;
      
     \draw [double,->-=.166,->-=.833,black] (2.0,0) coordinate (i_2) node[right]{\tiny $\theta_{2}$} node [below,black]{\footnotesize $\Phi_{a_{2}b_{2}}$}  to (2.0,2.0) coordinate (l_2) node [above]{\footnotesize $\Phi_{c_{2}d_{2}}$} ;
     
     \draw [double,->-=.166,->-=.833,black] (3.0,0) coordinate (i_3) node[right]{\tiny $\theta_{3}$} node [below,black]{\footnotesize $\Phi_{a_{3}b_{3}}$} to (3.0,2.0) coordinate (l_3) node [above]{\footnotesize $\Phi_{c_{3}d_{3}}$};
     
     \draw [double,->-=.166,->-=.833,black] (4.0,0) coordinate (i_4) node[right]{\tiny $\theta_{4}$} node [below,black]{\footnotesize $\Phi_{a_{4}b_{4}}$} to (4.0,2.0) coordinate (l_4)node [above]{\footnotesize $\Phi_{c_{4}d_{4}}$};

     \draw [double,->-=.166,->-=.833,black] (6.0,0) coordinate (i_6) node [below,black]{\footnotesize $\Phi_{a_{L}b_{L}}$} to (6.0,2.0) coordinate (l_6) node [above]{\footnotesize $\Phi_{c_{L}d_{L}}$};
     
       
    \coordinate (j_1) at (intersection of a_1--d_1 and i_1--l_1);
    \coordinate (j_2) at (intersection of a_1--d_1 and i_2--l_2);
    \coordinate (j_3) at (intersection of a_1--d_1 and i_3--l_3);
    
    \coordinate (k_1) at (intersection of a_1--d_1 and i_4--l_4);
  
    \coordinate (k_3) at (intersection of a_1--d_1 and i_6--l_6);
    \fill[black] (j_1) circle (2pt);
    \fill[black] (j_2) circle (2pt);
    \fill[black] (j_3) circle (2pt);
    \fill[black] (k_1) circle (2pt);
    
    \fill[black] (k_3) circle (2pt);
\end{tikzpicture}
\end{equation}
where $\Phi_{a_{k}b_{k}}$ and $\Phi_{c_{k}d_{k}}$ are the incoming and outgoing $\mathfrak{so}(6)$ flavours of the $k^{th}$ spin chain site in the ``scattering" with the auxiliary particle.  The indexes $a$ and $b$ indicate the incoming and outgoing flavours in the auxiliary space, they take on values $\{1,2,3,4\}$ of the $\mathbf{4}$ representation. The trace of this monodromy, the transfer matrix, is obtained by identifying the indexes $a$ and $b$ and summing over the four fundamental flavours. The set of inhomogeneities $\{\theta_{k}\}$ must be taken to zero to describe the spin chain with Hamiltonian \eqref{eq:so6Hamiltonian}. However, we keep them finite as their presence do not affect our construction of the spectrum.

In section \ref{sec:ABA} we build a Bethe basis that diagonalizes the transfer matrix of \eqref{eq:mono4} and the Hamiltonian \eqref{eq:so6Hamiltonian}. This construction yields a wing-vertex model  that renders a representation of the  Bethe states as we present in the following section.
\subsection{The $\mathfrak{so}(6)$ vertex model}\label{sec:vertexmodel}
In this section we introduce a vertex model obtained  from the ABA in section \ref{sec:ABA}. The Bethe states can be obtained as partition functions of this vertex model when imposing appropriate  boundary conditions.
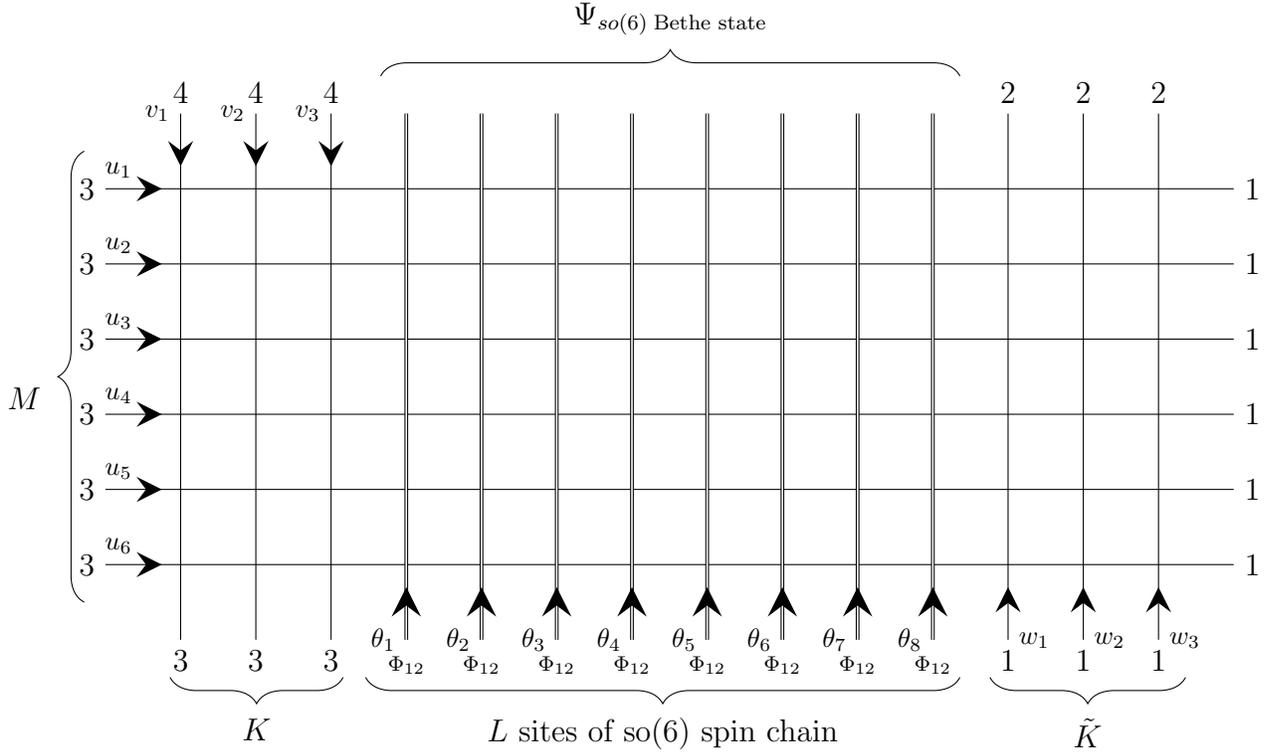
\begin{figure}[H]
\begin{center}
\begin{tikzpicture}[scale=1]
\draw [decoration={markings,mark=at position .05 with
    {\arrow[scale=3,>=stealth]{>}}},postaction={decorate}] (0,1) node[left]{$3$} node[above]{\footnotesize $\quad u_{6}$} to (15,1) node[right]{$1$};
\draw [decoration={markings,mark=at position .05 with
    {\arrow[scale=3,>=stealth]{>}}},postaction={decorate}] (0,2) node[left]{$3$} node[above]{\footnotesize $\quad u_{5}$} to (15,2)  node[right]{$1$};
\draw [decoration={markings,mark=at position .05 with
    {\arrow[scale=3,>=stealth]{>}}},postaction={decorate}]
(0,3) node[left]{$3$} node[above]{\footnotesize $\quad u_{4}$} to (15,3)  node[right]{$1$};
\draw [decoration={markings,mark=at position .05 with
    {\arrow[scale=3,>=stealth]{>}}},postaction={decorate}] (0,4) node[left]{$3$} node[above]{\footnotesize $\quad u_{3}$} to (15,4)  node[right]{$1$};
\draw [decoration={markings,mark=at position .05 with
    {\arrow[scale=3,>=stealth]{>}}},postaction={decorate}] (0,5) node[left]{$3$} node[above]{\footnotesize $\quad u_{2}$} to (15,5)  node[right]{$1$};
\draw [decoration={markings,mark=at position .05 with
    {\arrow[scale=3,>=stealth]{>}}},postaction={decorate}] (0,6) node[left]{$3$} node[above]{\footnotesize $\quad u_{1}$} to (15,6)  node[right]{$1$};
\draw [decoration={markings,mark=at position .1 with
    {\arrow[scale=3,>=stealth]{>}}},postaction={decorate}] (1,7) node[above]{$4$} node[left]{\footnotesize $\quad v_{1}$} -- (1,0)node[below]{$3$} ;
\draw [decoration={markings,mark=at position .1 with
    {\arrow[scale=3,>=stealth]{>}}},postaction={decorate}] (2,7) node[above]{$4$} node[left]{\footnotesize $\quad v_{2}$} -- (2,0)node[below]{$3$} ;
\draw [decoration={markings,mark=at position .1 with
    {\arrow[scale=3,>=stealth]{>}}},postaction={decorate}] (3,7) node[above]{$4$} node[left]{\footnotesize $\quad v_{3}$} -- (3,0)node[below]{$3$} ;;
\draw [decoration={markings,mark=at position .1 with
    {\arrow[scale=2,>=stealth]{>}}},postaction={decorate},double] (4,0) node[below]{$\underset{\Phi_{12}}{\;}$} node[left]{\footnotesize $\quad \theta_{1}$} to (4,7);
\draw [decoration={markings,mark=at position .1 with
    {\arrow[scale=2,>=stealth]{>}}},postaction={decorate},double] (5,0) node[below]{$\underset{\Phi_{12}}{\;}$} node[left]{\footnotesize $\quad \theta_{2}$} to (5,7);
    \draw [decoration={markings,mark=at position .1 with
    {\arrow[scale=2,>=stealth]{>}}},postaction={decorate},double] (6,0) node[below]{$\underset{\Phi_{12}}{\;}$} node[left]{\footnotesize $\quad \theta_{3}$} to (6,7);
    \draw [decoration={markings,mark=at position .1 with
    {\arrow[scale=2,>=stealth]{>}}},postaction={decorate},double] (7,0) node[below]{$\underset{\Phi_{12}}{\;}$} node[left]{\footnotesize $\quad \theta_{4}$} to (7,7);
    \draw [decoration={markings,mark=at position .1 with
    {\arrow[scale=2,>=stealth]{>}}},postaction={decorate},double] (8,0) node[below]{$\underset{\Phi_{12}}{\;}$} node[left]{\footnotesize $\quad \theta_{5}$} to (8,7);
    \draw [decoration={markings,mark=at position .1 with
    {\arrow[scale=2,>=stealth]{>}}},postaction={decorate},double] (9,0) node[below]{$\underset{\Phi_{12}}{\;}$} node[left]{\footnotesize $\quad \theta_{6}$} to (9,7);
    \draw [decoration={markings,mark=at position .1 with
    {\arrow[scale=2,>=stealth]{>}}},postaction={decorate},double] (10,0) node[below]{$\underset{\Phi_{12}}{\;}$} node[left]{\footnotesize $\quad \theta_{7}$} to (10,7);
    \draw [decoration={markings,mark=at position .1 with
    {\arrow[scale=2,>=stealth]{>}}},postaction={decorate},double] (11,0) node[below]{$\underset{\Phi_{12}}{\;}$} node[left]{\footnotesize $\quad \theta_{8}$} to (11,7);
\draw [decoration={markings,mark=at position .1 with
    {\arrow[scale=3,>=stealth]{>}}},postaction={decorate}](14,0) node[below]{$1$} node[right]{\footnotesize $w_{3}$} to (14,7) node[above]{$2$};
\draw [decoration={markings,mark=at position .1 with
    {\arrow[scale=3,>=stealth]{>}}},postaction={decorate}](13,0) node[below]{$1$}  node[right]{\footnotesize $w_{2}$} to (13,7) node[above]{$2$};
\draw [decoration={markings,mark=at position .1 with
    {\arrow[scale=3,>=stealth]{>}}},postaction={decorate}] (12,0) node[below]{$1$}  node[right]{\footnotesize $w_{1}$} to (12,7) node[above]{$2$};
\draw [decorate,decoration={brace,amplitude=10pt},xshift=-4pt,yshift=0pt]
(3.8,7.5) -- (11.5,7.5) node [black,midway,yshift=0.4cm,above] 
{$\Psi_{so(6)\;\text{Bethe state}}$};
\draw [decorate,decoration={brace,amplitude=10pt,mirror},xshift=-4pt,yshift=0pt]
(1,-.5) -- (3.3,-.5) node [black,midway,yshift=-0.4cm,below] 
{$K$};
\draw [decorate,decoration={brace,amplitude=10pt,mirror},xshift=-4pt,yshift=0pt]
(3.6,-.5) -- (11.5,-.5) node [black,midway,yshift=-0.4cm,below] 
{$L$ sites of so(6) spin chain};
\draw [decorate,decoration={brace,amplitude=10pt,mirror},xshift=-4pt,yshift=0pt]
(11.9,-.5) -- (14.5,-.5) node [black,midway,yshift=-0.4cm,below] 
{$\tilde{K}$};
\draw [decorate,decoration={brace,amplitude=10pt},xshift=-8pt,yshift=0pt]
(0,.5) -- (0,6.5) node [black,midway,xshift=-0.8cm,below] 
{$M$};
\end{tikzpicture}
\end{center}
\caption{so(6) vertex model with $L=8\,$,$\,K=3\,$,$\,M=6\,$ and $\tilde{K}=3$. The arrows indicate the direction of flavour injection.}
\label{fig:vertexmodel}
\end{figure}
\subsubsection{$R$-matrices in the lattice}
 The vertex model is given by the winged lattice in figure \ref{fig:vertexmodel}. This lattice is composed by simple lines in the $\mathbf{4}$ fundamental representation spanned by flavours $\{1,2,3,4\}$ and double lines in the $\mathbf{6}$-antisymmetric representation of $\mathfrak{su}(4)$ spanned by \eqref{eq:Zletters}. The simple lines represent three types of auxiliary spaces with spectral parameters given by three sets of roots or rapidities $\{v\}_{K},\{u\}_{M},\{w\}_{\tilde{K}}$. These auxiliary sets are in one to one correspondence with the nodes of the $\mathfrak{so}(6)$ or $\mathfrak{su}(4)$ Dynkin diagram:
\begin{equation}
\text{$
\begin{tikzpicture}
\node[circle,draw,minimum size=.75cm] (K1) at  (0,0) {};
\node[circle,draw,minimum size=.75cm] (K2) at  (2,0) {};
\node[circle,draw,minimum size=.75cm] (K3) at  (4,0) {};
\draw (K1) node[above=0.5]{$\{v\}_{K}$} -- (K2) node[above=0.5]{$\{u\}_{M}$}  -- (K3) node[above=0.5]{$\{w\}_{\tilde{K}}$} ;
\end{tikzpicture}
$}
\end{equation}
To each of the simple lines we associate a probe particle with rapidities: $v$ for the vertical lines on the left wing  , $u$ for the horizontal lines and $w$ for the vertical lines on the right wing of figure \ref{fig:vertexmodel}. These probe particles ``scatter" among each other forming  wing auxiliary lattices. This scattering or crossing of lines is controlled by the $R$-matrix:
\begin{equation}\label{eq:R44matrix}
R_{\mathbf{4}\mathbf{4}}(u,v)\, = \frac{u-v}{u-v-i}\; I_{\mathbf{4}\mathbf{4}} - \frac{i}{u-v-i}\; P_{\mathbf{4}\mathbf{4}}
\end{equation}
which also provides the Boltzmann weights of the vertex model in the wing lattices as:
\begin{equation}\label{eq:latticeR44matrix}
\text{$
\begin{tikzpicture}[baseline=8mm] 
\draw[->-=.4] (0,1) node [left]{$a$} node [above]{$u$}  to (2,1)  ;
\draw[->-=.4] (1,0) node [below]{$b$} node [left]{$v$} to (1,2) ;
\end{tikzpicture}\, = \, \frac{u-v}{u-v-i}\left(
\begin{tikzpicture}[baseline=8mm] 
\draw[->-=.4] (0,1) node [left]{$a$}  to (2,1)  node [right]{$a$} ;
\draw[->-=.4,blue] (1,0) node [below]{$b$}  to (1,2) node [above]{$b$};
\end{tikzpicture}
\right)
 - 
\frac{i}{u-v-i}\,
\left(
\begin{tikzpicture}[baseline=8mm] 
\draw[->-=.5] (0,1) node [left]{$a$} .. controls(1,1) ..  (1,2) node [above]{$a$};
\draw[->-=.5,blue] (1,0) node [below]{$b$} .. controls(1,1) ..  (2,1) node [right]{$b$};
\end{tikzpicture}
\right)
$}
\end{equation}
This $R$-matrix is composed of two terms: the identity $I$ that keeps the flavours in their original direction and the permutation $P$ that swaps the directions the flavours follow. On the right hand side of equation \eqref{eq:latticeR44matrix} the arrows indicate the flow of the flavour, while the auxiliary root $u$($v$) is always attached to the horizontal(vertical) direction.

The novel part of our vertex model is given by the core of the lattice, where simple ($\mathbf{4}$) and double ($\mathbf{6}$) lines crossed. The Boltzmann weights associated to these crossed lines are given by the $R_{\mathbf{4}\mathbf{6}}$-matrix:
\begin{align}\label{eq:R46matrix}
R_{\mathbf{4}\mathbf{6}}(u,\theta)\, &=\frac{u-\theta+i/2}{u-\theta-i/2}\;I_{\mathbf{4}\mathbf{6}}-\frac{i}{u-\theta - i/2}\;P_{\mathbf{4}\mathbf{6}}\nonumber \\
\begin{tikzpicture}[scale=0.8,baseline=6mm]
\draw[->-=.4] (0,1) node [left]{$a$} node[above]{$u$}  to (2,1);
\draw[->-=.4,double] (1,0) node [below]{$\underset{\Phi_{bc}}{\;}$} node[left]{$\theta$}  to (1,2);
\end{tikzpicture}  & =  \, \frac{u-\theta+i/2}{u-\theta-i/2}\left(
\begin{tikzpicture}[scale=.8,baseline=6mm]
\draw[->-=.4] (0,1) node [left]{$a$}  to (2,1)  node [right]{$a$} ;
\draw[->-=.4,blue] (0.95,0) node [below]{$\underset{\color{black}\Phi_{\color{blue}b\color{red}c}}{\;}$}  to (0.95,2);
\draw[->-=.4,red] (1.05,0) to (1.05,2) node [above]{$\underset{\color{black}\Phi_{\color{blue}b\color{red}c}}{\;}$};
\end{tikzpicture}
\right)
 - 
\frac{i}{u-\theta-i/2}\,
\left(
\begin{tikzpicture}[scale=.8,baseline=6mm]
\draw[->-=.5] (0,.95) node [left]{$a$} .. controls(.95,1.03) ..  (.95,2) node [above]{$\underset{\color{black}\Phi_{a\color{blue}b}}{\;}$};
\draw[->-=.5,blue] (1,0)  node [below]{$\underset{\color{black}\Phi_{\color{blue}b\color{red}c}}{\;}$}  to (1,2);
\draw[->-=.5,red] (1.05,0) .. controls(1.05,.97) ..  (2,1) node [right]{$c$};
\node(minus) at (3,1){$-$};
\draw[->-=.5] (4,.95)  node [left]{$a$} .. controls(4.95,1.03) ..  (4.95,2) node [above]{$\underset{\color{black}\Phi_{a\color{red}c}}{\;}$} ;
\draw[->-=.5,red] (5,0) node [below]{$\underset{\color{black}\Phi_{\color{red}c\color{blue}b}}{\;}$} to (5,2);
\draw[->-=.5,blue] (5.05,0) .. controls(5.05,.97) ..  (6,1) node [right]{$b$} ;
\end{tikzpicture}
\right)
\end{align}
Now the permutation term $P_{\mathbf{4}\mathbf{6}}$ has two pieces corresponding to the two possibilities of swapping the flavour index $a$ with the indexes of $\Phi_{bc}$.
\subsubsection{The boundary conditions and Bethe state}
   In order to obtain the $\mathfrak{so}(6)$ Bethe state from this vertex model we fix the boundaries of the wings as shown in figure \ref{fig:vertexmodel}. With these restrictions the auxiliary spaces on the wings become effectively two-dimensional and the wing lattices render two $6$-vertex models: $wing_{3,4}$ and $wing_{1,2}$, associated to $\mathfrak{su}(2)$ representations spanned by flavours $\{3,4\}$ and $\{1,2\}$ respectively. 

To complete the boundary conditions we restrict the bottom of the double lines to have incoming flavour $Z\equiv\Phi_{12}$. This choice makes the vertex models $wing_{3,4}$ and $wing_{1,2}$ play the role of reservoirs. In this way the first wing injects $M-K$ units of flavour $3$ and $K$ units of flavour $4$ to the double-lined lattice. The second wing absorbs $M-\tilde{K}$ units of flavour $1$ and $\tilde{K}$ of flavour $2$.

Considering these boundary conditions we follow the flavour rules and Boltzmann weights in \eqref{eq:latticeR44matrix} and \eqref{eq:R46matrix} to construct the $\mathfrak{so}(6)$ Bethe state, which can finally be read off from the top of the lattice in figure  \ref{fig:vertexmodel}. The Bethe state obtained from this vertex model has $L$ sites and global charges:
\begin{equation}\label{eq:chargeo6}
\mathfrak{so}(6) \text{ charges }:\qquad [M-2 K,\, L+K-2 M+\tilde{K},\,M-2 \tilde{K}]
\end{equation}
The Bethe state can be expressed as a linear combination of states in the $\mathfrak{so}(6)$ coordinate basis with charges \eqref{eq:chargeo6} and length $L$. The states of this basis are composed of all allowed combinations of letters \eqref{eq:Zletters}, considering their individual charges are: 
\begin{equation}\label{eq:Zcharges}
Z :\, [0,1,0]\,,\; X :\,  [1,-1,1]\,,\; Y :\,  [-1,0,1]\,,\; \bar{Y} :\,  [1,0,-1]\,,\; \bar{X} :\,  [-1,1,-1]\,,\; \bar{Z} :\,  [0,-1,0]\,.
\end{equation}
As an example, for $L=2$ , $K=1$, $M=2$ , $\tilde{K}=1$ the total charge is $[0,0,0]$ and  the correspondent coordinate basis is given by: $\{|Z\bar{Z}\rangle\,,|\bar{Z}Z\rangle\,,|X\bar{X}\rangle\,,|\bar{X}X\rangle\,,|Y\bar{Y}\rangle\,,|\bar{Y}Y\rangle\}$. The coefficient of one of these coordinate states, in the linear combination that renders the Bethe state, is determined by imposing the corresponding letters as boundary conditions at the top of the double lines of figure \eqref{fig:vertexmodel}. Then we should consider all the possible paths the flavour can follow, consistent with the boundary conditions. Finally the coefficient is given by the sum of the Boltzmann weights associated to each possible path. 

Following the rules of this vertex model it is possible to determine the general form of the Bethe state as a linear combination of the coordinate basis. We present this in section \ref{sec:CBA} as a Coordinate Bethe Ansatz(CBA). In the following section we present the origin of this vertex model from the  ABA.

\subsection{ The Algebraic Bethe Ansatz (ABA)}\label{sec:ABA}
\subsubsection{The monodromy $T_{\mathbf{4}}$ and its elements}

The ``scattering" of a probe particle in the $\mathbf{4}$ representation with the spin chain sites in the $\mathbf{6}$ representation is given by a product of $R_{\mathbf{4}\mathbf{6}}$ matrices and renders the $T_{\mathbf{4}}$ monodromy matrix:
\begin{equation}\label{eq:T46}
\begin{tikzpicture}[scale=1.2]
\node(O1) at (-3,.9){$T_{\mathbf{4}}(\lambda)= R_{\mathbf{4}\mathbf{6}}(\lambda,\theta_{L})\cdots R_{\mathbf{4}\mathbf{6}}(\lambda,\theta_{1})\;=\;$};
    \draw [thick,->-=.05,->-=.20,->-=.35,->-=.5,->-=.65,->-=.80,->-=0.95,black] (0,1.0) coordinate (a_1)node [below]{$\overset{\lambda}{\textbf{4}}$}   to (7.0,1.0) coordinate (d_1) ;

     \draw [double,->-=.166,->-=.833,black](1.0,0) coordinate (i_1) node [below,black]{$\overset{\theta_{1}}{\textbf{6}}$}  to (1.0,2.0) coordinate (l_1) ;
      
     \draw [double,->-=.166,->-=.833,black] (2.0,0) coordinate (i_2) node [below,black]{$\overset{\theta_{2}}{\textbf{6}}$}  to (2.0,2.0) coordinate (l_2) ;
     
     \draw [double,->-=.166,->-=.833,black] (3.0,0) coordinate (i_3) node [below,black]{$\overset{\theta_{3}}{\textbf{6}}$} to (3.0,2.0) coordinate (l_3);
     
     \draw [double,->-=.166,->-=.833,black] (4.0,0) coordinate (i_4) node [below,black]{$\overset{\theta_{4}}{\textbf{6}}$} to (4.0,2.0) coordinate (l_4);
     
     \draw [double,->-=.166,->-=.833,black] (5.0,0) coordinate (i_5) node [below,black]{$\overset{\theta_{5}}{\textbf{6}}$} to (5.0,2.0) coordinate (l_5);
     
     \draw [double,->-=.166,->-=.833,black] (6.0,0) coordinate (i_6) node [below,black]{$\overset{\theta_{6}}{\textbf{6}}$} to (6.0,2.0) coordinate (l_6);
     
       
    \coordinate (j_1) at (intersection of a_1--d_1 and i_1--l_1);
    \coordinate (j_2) at (intersection of a_1--d_1 and i_2--l_2);
    \coordinate (j_3) at (intersection of a_1--d_1 and i_3--l_3);
    
    \coordinate (k_1) at (intersection of a_1--d_1 and i_4--l_4);
    \coordinate (k_2) at (intersection of a_1--d_1 and i_5--l_5);
    \coordinate (k_3) at (intersection of a_1--d_1 and i_6--l_6);
    \fill[black] (j_1) circle (2pt);
    \fill[black] (j_2) circle (2pt);
    \fill[black] (j_3) circle (2pt);
    \fill[black] (k_1) circle (2pt);
    \fill[black] (k_2) circle (2pt);
    \fill[black] (k_3) circle (2pt);
\end{tikzpicture}
\end{equation}

From the point of view of the auxiliary space the monodromy is a $4\times 4$ matrix, whose elements are operators that act exclusively over the spin chain:
\begin{equation}\label{eq:Telements}
T_{\mathbf{4}}(\lambda) = \left(\begin{matrix}
A_{11} & A_{12} & \color{blue}B_{13} & \color{blue}B_{14}\\
A_{21} & A_{22} & \color{blue}B_{23} & \color{blue}B_{24}\\
\color{red}C_{31} & \color{red}C_{32} & D_{33} & D_{34}\\
\color{red}C_{41} & \color{red}C_{42} & D_{43} & D_{44}\\
\end{matrix}\right)
\end{equation}
In order to obtain the elements of this matrix in the graphical representation in \eqref{eq:T46}, we simply fix the boundaries of the horizontal line to take specific flavour values $\{1,2,3,4\}$. 
The correspondent transfer matrix, given by the trace of the monodromy \eqref{eq:Telements}, is:
\begin{equation}\label{eq:transfer}
\mathcal{T} = A_{11}+A_{22}+D_{33}+D_{44}
\end{equation}
Now our aim is to construct the ABA to find the eigenvalues and eigenstates of this transfer matrix.  For this we start by identifying one of its trivial eigenstates, the pseudo-vacuum:
\begin{equation}\label{eq:vacuum}
|\Omega_{L}\rangle \equiv |Z^{L}\rangle \equiv |\Phi_{12}^{L}\rangle
\end{equation}
We consider this pseudo-vacuum as a reference state to start the construction of the ABA. With this choice the action of the monodromy matrix organizes into $2\times2$ blocks. The pseudo-vacuum diagonalizes the $A$ and $D$ blocks:
\begin{align}\label{eq:ADvacuum}
\mathbb{A}(\lambda)|\Omega_{L}\rangle &=  \begin{pmatrix}
A_{11}&A_{12}\\
A_{21}&A_{22}  
\end{pmatrix} |\Omega_{L}\rangle = \begin{pmatrix}
a(\lambda)&0\\
0 & a(\lambda)  
\end{pmatrix} |\Omega_{L}\rangle\nonumber\\
 \mathbb{D}(\lambda)|\Omega_{L}\rangle & = \begin{pmatrix}
D_{33} & D_{34}\\
D_{43} & D_{44}  
\end{pmatrix} |\Omega_{L}\rangle = \begin{pmatrix}
d(\lambda) & 0\\
0 & d(\lambda)  
\end{pmatrix} |\Omega_{L}\rangle
\end{align}
with (considering $\theta_{k}=0$):
\begin{equation}
a(\lambda) = 1 \qquad \text{and} \qquad d(\lambda) = \left(\frac{\lambda+i/2}{\lambda-i/2}\right)^{L}
\end{equation}
and it is annihilated by the $\mathbb{C}$-block elements:
\begin{equation}
\mathbb{C}(\lambda)|\Omega\rangle = \begin{pmatrix}
C_{31} & C_{32}\\
C_{41} & C_{42}  
\end{pmatrix}  |\Omega\rangle = 0 
\end{equation}
The action of the $B$-operators over \eqref{eq:vacuum} is much less trivial. They create magnon-states or plane waves when acting over the pseudo-vacuum. The operator $B_{jk}$ injects flavour $k\in\{3,4\}$ and absorbs flavour $j\in\{1,2\}$ from the state it acts over. When acting over the pseudovacuum \eqref{eq:vacuum} it creates a magnon of type $\Phi_{1k}$ when $j=2$ or type $\Phi_{2k}$ when $j=1$. For instance the operator $B_{13}$ creates a $X\equiv \Phi_{23}$ magnon as:
\begin{equation}\label{B-op}
|\Psi_{X}\rangle = \sum_{n=1}^{L}\psi_{n}(u)\,|Z\cdots \overset{\overset{n}{\downarrow}}{X} \cdots\,Z \rangle \,=\sum_{n=1}^{L}\,
 \begin{tikzpicture}[baseline=8mm]
    \draw [blue,->-=.15,->-=.45,->-=.75] (0,1) coordinate (a_1)node [above]{${\color{black} u}$} node[left]{${\color{black} 3}$}  to (3.5,1) coordinate (b_1) ;   
    \draw [blue,->-=.2] (b_1)    .. controls(4,1) .. (4,2)node[above]{ ${\color{black} \overset{\overset{n}{\downarrow}}{\text{\small $\Phi_{23}$}}}$} coordinate (c_1)  ;     
     \foreach \x in {1,2,3,5,6} 
      {
        \draw[double,->-=.166,->-=.833] (\x,0) node [below,black]{\small $\Phi_{12}$}  to (\x,2)  ; 
      } 
   
     \draw [->-=.5] (4.05,0)  node [below,black]{\small $\Phi_{12}$}  to (4.05,2)      ;
     
     \draw [red] (4.1,0) coordinate (d_1) .. controls(4.1,1) .. (4.5,1) coordinate (e_1) ;   
    \draw [red,->-=.0,->-=.4,->-=.8] (e_1)   to (7,1)node[right]{${\color{black} 1}$}  ; 
\end{tikzpicture}
\end{equation}
with wave-function $\psi_{n}(u)=\left(\prod_{k=1}^{n-1}\frac{u-\theta_{k}+i/2}{u-\theta_{k}-i/2}\right)\left(\frac{-i}{u-\theta_{n}-i/2}\right)$ as can be read off from the lattice with Boltzmann weights \eqref{eq:R46matrix}. 

Similarly we can create other magnon-states with different flavours or $\mathfrak{so}(6)$ charges by using operators $B_{23}\,,B_{24}\,$ and $B_{14}$. The relationship between these creation operators and the $\mathfrak{so}(6)$ charges is summarized in the following figure:
\begin{equation}\label{eq:Bblock}
\mathbb{B} = \begin{pmatrix}
B_{13} & B_{14} \\
B_{23} & B_{24}
\end{pmatrix}\,,\qquad 
\begin{tikzpicture}[scale=0.8,baseline=6mm]
\node(Z) at (1,1){$Z$}; 
\node(X) at (4,1){$X$};
\node(Y) at (7,3){$Y$};
\node(Yb) at (7,-1){$\bar{Y}$};
\node(Xb) at (10,1){$\bar{X}$};
\node(Zb) at (13,1){$\bar{Z}$};
\draw [thick,->-=.6,bend left=20] (Z) to  node[rectangle,midway,below] {$\color{blue}B_{13}$} (X);
\draw [thick,->-=.6,bend left=27] (Z) to node[rectangle,midway,below] {$\color{blue}B_{24}
$} (Xb);
\draw [thick,->-=.6,bend right=27] (X) to node[rectangle,midway,above] {$\color{blue}B_{24}
$} (Zb);
\draw [thick,->-=.6,bend left=20] (Z) to node[rectangle,midway,above] {$\color{blue}B_{23}$} (Y);
\draw [thick,->-=.6,bend right=20] (Z) to node[rectangle,midway,below] {$\color{blue}B_{14}$} (Yb);
\draw[thick,->-=.6,bend right=20] (Xb) to node[rectangle,midway,above] {$\color{blue}B_{13}$} (Zb);
\draw [thick,->-=.6,bend right=20] (Yb) to node[rectangle,midway,below] {$\color{blue}B_{23}$} (Zb);
\draw[thick,->-=.6,bend left=20] (Y) to node[rectangle,midway,above] {$\color{blue}B_{14}$} (Zb);
\end{tikzpicture}
\end{equation}
\subsubsection{The Bethe Ansatz}
The $B$ creation operators play a key role in the construction of the spectrum of the transfer matrix. The states created by repeated action of $B$-operators over the reference state \eqref{eq:vacuum} serve as a basis to propose a general Ansatz for the eigenstates of the transfer matrix as:
\begin{equation}\label{eq:1ansatz}
|\Psi\rangle =  \psi_{a_{1}\cdots a_{M}}\, \psi^{\tilde{a}_{1}\cdots \tilde{a}_{M}}\, B_{\tilde{a}_{1}a_{1}}(u_{1})\cdots B_{\tilde{a}_{M}a_{M}}(u_{M})\, |\Omega_{L}\rangle 
\end{equation}
where the indexes $a_{k}$ and $\tilde{a}_{k}$ take on flavour values $\{3,4\}$ and $\{1,2\}$ respectively, and the unconstrained tensors  $\psi_{a_{1}\cdots a_{M}}$ and $\tilde{\psi}^{\tilde{a}_{1}\cdots\dot{a}_{M}}$ weight the contributions of states constructed by different choices of $B_{\tilde{a}_{m}a_{m}}$-operators.

We can rewrite the ansatz \eqref{eq:1ansatz} by using $2\times 2$ $\mathbb{B}$-blocks instead of individual $B_{\tilde{a}_{k}a_{k}}$-operators. With this purpose we introduce the wing-auxiliary states $|\psi\rangle$ and $\langle\tilde{\psi} |$ as:
\begin{equation}
\psi_{a_{1}a_{2}\cdots a_{M}} = \langle a_{1}a_{2}\cdots a_{M}|\psi\rangle 
\end{equation}
\begin{equation}
\tilde{\psi}^{\tilde{a}_{1}\tilde{a}_{2}\cdots \tilde{a}_{M}} =  \langle \tilde{\psi}\,|\,\tilde{a}_{1}\,\tilde{a}_{2}\cdots \tilde{a}_{M} \rangle 
\end{equation}
with states $\langle a_{1}a_{2}\cdots a_{M}|$ and $|\tilde{a}_{1}\tilde{a}_{2}\cdots \tilde{a}_{M}\rangle$ forming coordinate basis in the tensor product of two-dimensional subspaces: $\mathbf{2}_{1}\otimes\mathbf{2}_{2}\cdots\otimes \mathbf{2}_{M}$ and $\tilde{\mathbf{2}}_{1}\otimes\tilde{\mathbf{2}}_{2}\cdots\otimes \tilde{\mathbf{2}}_{M}$ respectively. We define the $\mathbb{B}_{m}$-blocks as: 
\begin{equation}
\mathbb{B}_{m}(u)  = |\tilde{a}_{m}\rangle\, B_{\tilde{a}_{m}a_{m}}(u)\,\langle a_{m} |
\end{equation}
In this way the operator $\mathbb{B}_{m}$-block  acts over the spin chain space and intertwines  between the auxiliary spaces $\mathbf{2}$ (flavours $\{3,4\}$) and $\tilde{\mathbf{2}}$ (flavours $\{1,2\}$) as:
\begin{equation}
\mathbb{B}_{m}\,:\,\quad \mathbf{2}_{m}\otimes \mathbf{6}_{1}\otimes \cdots \otimes \mathbf{6}_{L} \,\longrightarrow \,\tilde{\mathbf{2}}_{m}\otimes \mathbf{6}_{1}\otimes \cdots \otimes \mathbf{6}_{L}
\end{equation}
while its action over other auxiliary spaces $\mathbf{2}_{k}$ and $\tilde{\mathbf{2}}_{k}$ is trivial for $k\neq m$.

Using $\mathbb{B}$-blocks and wing-auxiliary states we reformulate the Bethe Ansatz \eqref{eq:1ansatz} as:
\begin{equation}\label{eq:so6Bethe}
 |\Psi\rangle ={\color{red}\langle \tilde{\psi}|}\,\mathbb{B}_{1}(u_{1})\cdots \mathbb{B}_{M}(u_{M})\, {\color{blue}|\psi\rangle}\otimes |\Omega_{L}\rangle 
\end{equation}

Using the Ansatz \eqref{eq:so6Bethe} we now need to solve the eigenvalue problem:
\begin{equation}\label{eq:eigenvalueproblem}
\mathcal{T}(\lambda)|\Psi\rangle = \Lambda(\lambda) |\Psi\rangle 
\end{equation}
This means we need to find the restrictions over the auxiliary roots $\{u\}$ and the wings states such equation \eqref{eq:eigenvalueproblem} holds. In what follows we sketch the steps to achieve this diagonalization. These will heavily rely on the Yang-Baxter algebra presented in section \ref{sec:YB}.

We first reexpress the transfer matrix \eqref{eq:transfer} by defining block operators $\mathbb{A}$ and $\mathbb{D}$ as:
\begin{equation}
\mathbb{A}_{m}(u)  = |\tilde{a}_{m}\rangle\,A_{\tilde{a}_{m}\tilde{b}_{m}}(u)\,\langle \tilde{b}_{m} |\quad \text{and} \quad  \mathbb{D}_{m}(u) = |a_{m}\rangle\,D_{a_{m}b_{m}}\langle b_{m} |
\end{equation}
with non-trivial action over the spaces:
\begin{equation}
\mathbb{A}_{a}\,:\,\quad \tilde{\mathbf{2}}_{a}\otimes \mathbf{6}_{1}\otimes \cdots \otimes \mathbf{6}_{L} \,\longrightarrow \,\tilde{\mathbf{2}}_{a}\otimes \mathbf{6}_{1}\otimes \cdots \otimes \mathbf{6}_{L}
\end{equation}
\begin{equation}
\mathbb{D}_{a}\,:\,\quad \mathbf{2}_{a}\otimes \mathbf{6}_{1}\otimes \cdots \otimes \mathbf{6}_{L} \,\longrightarrow \,\mathbf{2}_{a}\otimes \mathbf{6}_{1}\otimes \cdots \otimes \mathbf{6}_{L}
\end{equation}
In this language the transfer matrix is now a sum of traces of $\mathbb{A}$ and $\mathbb{D}$ blocks and the eigenvalue problem has two pieces associated to these blocks:
\begin{equation}\label{eq:ADplus}
\mathcal{T}(\lambda) |\Psi\rangle =\,Tr_{a}\,{\color{red}\mathbb{A}_{a}(\lambda)}|\Psi\rangle + Tr_{a}\,{\color{blue}\mathbb{D}_{a}(\lambda)} |\Psi\rangle
\end{equation}
where ``$a$" labels an auxiliary space $\tilde{\mathbf{2}}$ for the $\mathbb{A}$-block and $\mathbf{2}$ for the $\mathbb{D}$-block.

Now starting with equation \eqref{eq:ADplus} the strategy is to commute the $\mathbb{A}$ and $\mathbb{D}$ blocks through the product of $\mathbb{B}$-blocks until we reach the pseudo-vacuum that satisfies \eqref{eq:ADvacuum}. This is possible using the commutation relations provided by the Yang-Baxter algebra (see section \ref{sec:YB}). Once we follow this procedure the result has two type of terms: wanted and unwanted. From the wanted terms we can reproduce the eigenvalue equation \eqref{eq:eigenvalueproblem} and read off the correspondent transfer matrix eigenvalue $\Lambda$. On the other hand the unwanted terms spoil the eigenvalue equation and their vanishing is a necessary condition to satisfy \eqref{eq:eigenvalueproblem}. Here we only show the wanted terms:
\begin{align}\label{eq:ADonlyWanted}
\mathcal{T}(\lambda) |\Psi\rangle & = \, \Phi_{0}(\lambda)\, Tr_{a}\,\,{\color{red}\langle\tilde{\psi}|\,T_{a}(u_{1\cdots M}|\lambda)}\,\mathbb{B}_{1}(u_{1})\cdots \mathbb{B}_{M}(u_{M})\, {\color{red}\mathbb{A}_{a}(\lambda)}\,{\color{blue}|\psi\rangle}\,\otimes|\Omega_{L}\rangle \nonumber\\
&\qquad + \Theta_{0}(\lambda)\, Tr_{a}\,{\color{red}\langle\tilde{\psi}|}\,\mathbb{B}_{1}(u_{1})\cdots \mathbb{B}_{M}(u_{M})\,{\color{blue} \mathbb{D}_{a}(\lambda)\,T_{a}(\lambda|u_{M\cdots 1})|\psi\rangle} \otimes |\Omega_{L}\rangle  \nonumber\\
&\qquad + \text{unwanted terms from commuting $\mathbb{A}$ and $\mathbb{B}$-blocks}\nonumber\\
&\qquad + \text{unwanted terms from commuting $\mathbb{D}$ and $\mathbb{B}$-blocks}
\end{align}
with:
\begin{equation}
\Phi_{0}(\lambda|\{u\}) = \prod_{j=1}^{M}\frac{\lambda - u_{j} + i}{\lambda - u_{j}} \qquad \text{and} \qquad  \Theta_{0}(\lambda|\{u\}) =\prod_{j=1}^{M}\frac{\lambda - u_{j}-i}{\lambda - u_{j}} 
\end{equation}
As a by-product of the commutations of $\mathbb{A}$-$\mathbb{B}$ and $\mathbb{D}$-$\mathbb{B}$ blocks in \eqref{eq:ADonlyWanted} we obtain two auxiliary nested $\mathfrak{su}(2)$  monodromies\footnote{This  is analogous to the appearance of a nested $\mathfrak{su}(2)$  monodromy in the $\mathfrak{su}(3)$ Bethe Ansatz. But now we have two copies of nested monodromies, one for each wing $\langle\tilde{\psi} |$ and $|\psi \rangle$} acting in the spaces of $\langle \tilde{\psi}|$ and $|\psi\rangle$ respectively:
\begin{equation}\label{eq:T12}
T_{a}(u_{1\cdots M}|\lambda) \equiv \, R_{1a}(u_{1},\lambda)\,\cdots \, R_{Ma}(u_{M},\lambda)\qquad \text{with} \quad a\equiv \tilde{\mathbf{2}}_{a}
\end{equation}
\begin{equation}\label{eq:T34}
T_{a}(\lambda|u_{M\cdots 1}) \equiv \, R_{aM}(\lambda,u_{M})\,\cdots \, R_{a1}(\lambda,u_{1})\qquad \text{with} \quad a\equiv \mathbf{2}_{a}
\end{equation}
with $R$-matrices given by \eqref{eq:R44matrix} but now restricted to act over $\mathfrak{su}(2)$ subspaces. Namely they read
\begin{equation}\label{eq:22R-matrix}
R_{\mathbf{2}\mathbf{2}}(u,v)\, = \frac{u-v}{u-v-i}\; I_{\mathbf{2}\mathbf{2}} - \frac{i}{u-v-i}\; P_{\mathbf{2}\mathbf{2}}\,.
\end{equation}

Furthermore we can directly act with $\mathbb{A}$ over the pseudovacuum in the first line of \eqref{eq:ADonlyWanted},  since it does not act non-trivially over the wing states. Similarly in the second line we can commute $\mathbb{D}_{a}$ and  the nested monodromy $T_{a}$ in the presence of the trace  and act over the pseudovacuum. Using the diagonalization properties \eqref{eq:ADvacuum} of pseudovacuum we can simplify 
 \eqref{eq:ADonlyWanted} and obtain wing transfer matrices as traces of \eqref{eq:T12} and \eqref{eq:T34} :
\begin{align}\label{eq:transferwing}
\mathcal{T}(\lambda) |\Psi\rangle  & = \, \Phi_{0}(\lambda)\,a(\lambda)\, \boxed{\color{red}\langle\tilde{\psi}|\,Tr_{a}\,T_{a}(u_{1\cdots M}|\lambda)}\color{black}\,\,\mathbb{B}_{1}(u_{1})\cdots \mathbb{B}_{M}(u_{M})\color{blue}|\psi\rangle\,\color{black}\otimes |\Omega_{L}\rangle + \cdots\nonumber\\
&\qquad + \Theta_{0}(\lambda)\,d(\lambda)\, \color{red}\langle\tilde{\psi}|\color{black} \,\mathbb{B}_{1}(u_{1})\cdots \mathbb{B}_{M}(u_{M})\,\color{blue} \,\boxed{ Tr_{a}\,T_{a}(\lambda|u_{M\cdots 1})|\psi\rangle }\color{black}\otimes |\Omega_{L}\rangle + \cdots 
\end{align}
We now see that to reproduce the eigenvalue equation \eqref{eq:eigenvalueproblem} from the ``wanted" terms, the wing states must be eigenstates of the corresponding nested $\mathfrak{su}(2)$ transfer matrices of the monodromies \eqref{eq:T12} and \eqref{eq:T34}. This auxiliary problem is simply solved by the standard $\mathfrak{su}(2)$ ABA \cite{su2Faddeev}: 
\begin{equation}\label{eq:wingBethe}
 \langle\tilde{\psi} | = \langle\Omega_{\tilde{\mathbf{2}}}|\,\mathcal{C}(w_{1})\cdots \mathcal{C}(w_{\tilde{K}}) \qquad \text{and}\qquad  |\psi\rangle = \mathcal{B}(v_{1})\cdots \mathcal{B}(v_{K})|\Omega_{\mathbf{2}}\rangle  
\end{equation}
where $\mathcal{B}$ and $\mathcal{C}$ are creation and annihilation operators extracted from the monodromies \eqref{eq:T34} and \eqref{eq:T12} respectively. They act over $\mathfrak{su}(2)$ vacuum states of the wings given by:
\begin{equation}
\langle \Omega_{\tilde{\mathbf{2}}}| \equiv \langle 1^{M}| \qquad \text{and} \qquad |\Omega_{\mathbf{2}}\rangle \equiv |3^{M}\rangle
\end{equation}
In addition the sets of auxiliary roots $\{v\}$ and $\{w\}$ must be on-shell, that is they must fulfil $\mathfrak{su}(2)$ Bethe equations with the set $\{u\}$ as inhomogeneities. Assuming these conditions the $\mathfrak{su}(2)$ Bethe states \eqref{eq:wingBethe} diagonalize the wing transfer matrices as:
\begin{equation}
\langle\tilde{\psi}|\,Tr_{a}\,T_{a}(u_{1\cdots M}|\lambda)=\tilde{\Lambda}^{\mathfrak{su}(2)}(\lambda)\, \langle\tilde{\psi}|\qquad \text{and}\qquad Tr_{a}\,T_{a}(\lambda|u_{M\cdots 1})|\psi\rangle = \Lambda^{\mathfrak{su}(2)}(\lambda)\, |\psi\rangle 
\end{equation}
With the wing-states on-shell , \eqref{eq:transferwing} becomes the eigenvalue equation \eqref{eq:eigenvalueproblem}  up to unwanted terms:
\begin{equation}
\mathcal{T}(\lambda)|\Psi\rangle = \left(\Phi_{0}\,a\,\tilde{\Lambda}^{\mathfrak{su}(2)} + \Theta_{0}\,d\,\Lambda^{\mathfrak{su}(2)}\right)\, |\Psi\rangle + \text{unwanted terms}
\end{equation}
More explicitly the transfer matrix eigenvalue is given in terms of the spectral parameter $\lambda$ and the sets of auxiliary roots $\{u\}_{M}$ , $\{v\}_{K}$ and $\{w\}_{\tilde{K}}$ :
\begin{align}\label{eq:Lambda}
\Lambda(\lambda) &=   \left( \prod_{j=1}^{M}\frac{\lambda - u_{j} + i}{\lambda - u_{j}}\right)\left(\prod_{k=1}^{\tilde{K}}\frac{\lambda-w_{k}-i}{\lambda-w_{k}}  +  \prod_{j=1}^{M}\frac{\lambda - u_{j}}{\lambda - u_{j}+ i}\; \prod_{k=1}^{\tilde{K}}\frac{\lambda - w_{k}+i}{\lambda - w_{k}}\right)\nonumber\\
&\qquad +\,\left(\frac{\lambda +i/2}{\lambda -i/2}\right)^{L} \;\left(\prod_{j=1}^{M}\frac{\lambda - u_{j}-i}{\lambda - u_{j}}\right)\left(\prod_{k=1}^{K}\frac{\lambda-v_{k}+i}{\lambda - v_{k}} + \prod_{j=1}^{M} \, \frac{\lambda - u_{j}}{\lambda-u_{j}-i}\;\prod_{k=1}^{K}\,\frac{\lambda - v_{k}-i}{\lambda - v_{k}} \right)
\end{align}

\begin{figure}[H]
\begin{center}
\begin{tikzpicture}[scale=1]
\node[](betheso6) at (-2,2) {$|\Psi\rangle\;=\;$};
\node[draw, rectangle, minimum width=1cm, minimum height=3cm,top color=white,bottom color=blue,rounded corners](bluewing) at (0,2) {$\psi$};
\node[draw, rectangle, minimum width=4cm, minimum height=3cm,rounded corners](core) at (3,2) {$\prod_{m=1}^{M}\mathbb{B}_{m}(u_{m})$};
\node[draw, rectangle, minimum width=1cm, minimum height=3cm,top color=white,bottom color=red,rounded corners](redwing) at (6,2) {$\tilde{\psi}$};
\node[draw=red!80, rectangle, inner sep=2pt, minimum width=4cm, minimum height=.7cm,rounded corners](vacuum) at (3,-.2) {$\Omega\equiv Z^{L}$};
\foreach \x in {.15,.30,.45,.60,.75,.90} {
   \draw[->-=.5,color=blue] ($(bluewing.north east)!\x!(bluewing.south east)$) to ($(core.north west)!\x!(core.south west)$);
   \draw[->-=.5,color=red] ($(core.north east)!\x!(core.south east)$) -- ($(redwing.north west)!\x!(redwing.south west)$);
  }
  \foreach \x in {.1,.2,.3,.4,.5,.6,.7,.8} {
   \draw[double,->-=.6,color=red] ($(vacuum.north west)!\x!(vacuum.north east)$) to ($(core.south west)!\x!(core.south east)$);}
\end{tikzpicture}
\end{center}
\caption{The so(6) Bethe state}
\label{fig:BetheState}
\end{figure}
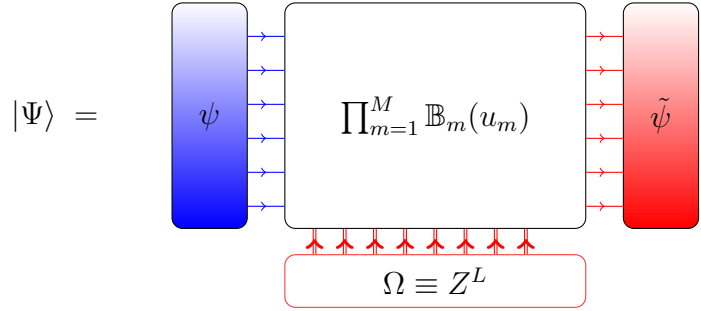

The vanishing of the unwanted terms puts constrains over the set of roots $\{u\}$. These constrains constitute the Bethe equations of the $\mathfrak{so}(6)$ middle node. Alternatively we can arrive to the same conditions by imposing the vanishing of the spurious poles at $\lambda = u_{1\cdots M}$ of the transfer matrix eigenvalue \eqref{eq:Lambda}. This latter method to obtain Bethe equations is the so called analytic Bethe Ansatz:
\begin{equation}
\underset{\lambda=u_{m}}{\mathbf{Res}}\Lambda(\lambda) = 0 \quad \longrightarrow \quad \left(\frac{u_{m}+i/2}{u_{m}-i/2}\right)^{L}=\prod_{j\neq m}^{M}\frac{u_{m}-u_{j}+i}{u_{m}-u_{j}-i}\,\prod_{k=1}^{\tilde{K}}\frac{u_{m}-w_{k}-i}{u_{m}-w_{k}}\,\prod_{k=1}^{K}\frac{u_{m}-v_{k}}{u_{m}-v_{k}+i}
\end{equation}
To obtain the standard form of $\mathfrak{so}(6)$ Bethe equations we must perform the shifts\footnote{ This is equivalent to define the $\mathfrak{su}(2)$ monodromies \eqref{eq:T12} and \eqref{eq:T34} with the Lax  pair instead of the $R$-matrix. These two objects differ by a shift of $i/2$ in the spectral parameter }:
\begin{equation}\label{eq:shift}
 w\rightarrow w-i/2 \qquad \text{and}\qquad v\rightarrow v+i/2
\end{equation}
In summary, the $\mathfrak{so}(6)$ Bethe state is given by the Ansatz in \eqref{eq:so6Bethe} with wing states given by the nested $\mathfrak{su}(2)$ Bethe states \eqref{eq:wingBethe} and with the sets of auxiliary roots $\{u\}\,$,$\{v\}$ and $\{w\}$ on-shell. The structure of the Bethe Ansatz presented in figure \ref{fig:BetheState} is equivalent to the vertex model in figure \ref{fig:vertexmodel}.

\subsection{The Yang-Baxter algebra}\label{sec:YB}
In this section we present the Yang-Baxter algebra of our $\mathfrak{so}(6)$ model. This algebra provides the technical steps to find the wanted and unwanted terms of the Bethe Ansatz in section \ref{sec:ABA}, as well as for our final result \eqref{eq:fullcB}  for the scalar product \eqref{eq:c123brute}.

The $R$ matrices in \eqref{eq:R44matrix} and \eqref{eq:R46matrix} fulfil the Yang-Baxter equation:
\begin{equation}
R_{\mathbf{4}_{a}\mathbf{4}_{b}}(u,v)\,R_{\mathbf{4}_{a}\mathbf{6}}(u,\theta)\,R_{\mathbf{4}_{b}\mathbf{6}}(v,\theta)\, = \, R_{\mathbf{4}_{b}\mathbf{6}}(v,\theta) \,R_{\mathbf{4}_{a}\mathbf{6}}(u,\theta)\,R_{\mathbf{4}_{a}\mathbf{4}_{b}}(u,v)
\end{equation}
where $a$ and $b$ label two different spaces in the $\mathbf{4}$ fundamental representation.

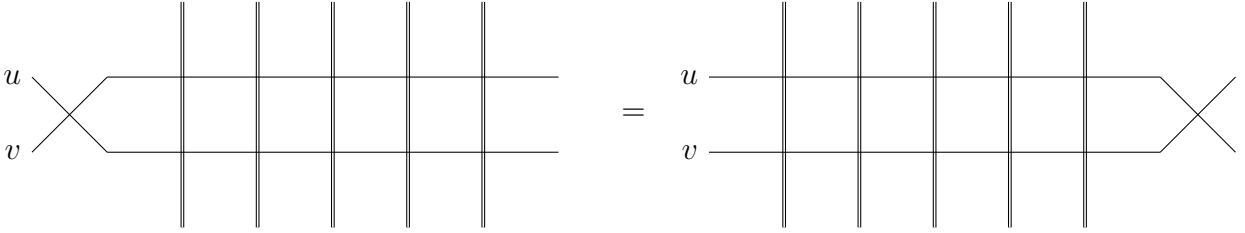
\begin{figure}[H]
\begin{center}
\begin{tikzpicture}
\draw [-,double]  (1,1) to (1,4);
\draw [-,double]  (2,1) to (2,4);
\draw [-,double]  (3,1) to (3,4);
\draw [-,double]  (4,1) to (4,4);
\draw [-,double]  (5,1) to (5,4);
\draw [-]  (0,2) coordinate (u) to (6,2) node[right]{};
\draw [-]  (0,3) coordinate (v) to (6,3) node[right]{};
\draw [-]  (-1,3) node[left]{$u$} to (u);
\draw [-]  (-1,2) node[left]{$v$} to (v);
\node(O1) at (7,2.5){$=$};
\draw [-,double]  (9,1) to (9,4);
\draw [-,double]  (10,1) to (10,4);
\draw [-,double]  (11,1) to (11,4);
\draw [-,double]  (12,1) to (12,4);
\draw [-,double]  (13,1) to (13,4);
\draw [-]  (8,2) node[left]{$v$}  to (14,2) coordinate (v3);
\draw [-]  (8,3) node[left]{$u$} to (14,3) coordinate (v4);
\draw [-]  (v3) to (15,3) node[right]{};
\draw [-]  (v4) to (15,2) node[right]{};
\end{tikzpicture}
\end{center}
\caption{$R\,T\,T\,=\,T\,T\,R$ relation}
\label{fig:RTT}
\end{figure}

This can be straightforwardly generalized to a Yang-Baxter relation for the monodromy in \eqref{eq:T46}, the so called $RTT$ relation\footnote{ In our graphical representations the order of action of operators should be read from left to right. While in our equations we respect the usual order of operator action, that is operators on the right act first. In this way the left-hand side in figure \ref{fig:RTT} represents the right-hand side on equation \eqref{eq:RTT}. }:
\begin{equation}\label{eq:RTT}
R_{ab}(u,v)\,T_{a}(u)\,T_{b}(v) =\, T_{b}(v)\,T_{a}(u)\,R_{ab}(u,v)
\end{equation}

Furthermore, taking the trace over the $4$-dimensional auxiliary spaces in \eqref{eq:RTT} we obtain the commutation relation for the transfer matrices:
\begin{equation}
[\mathcal{T}(u),\mathcal{T}(v)] = 0 \qquad \text{with}\quad \mathcal{T}(u) = Tr_{a}(T_{a}(u))
\end{equation}
This latter relation gives a family of conserved local charges in convolution when expanding the transfer matrix around $u=-i/2$ (without inhomogeneities).

The $RTT$ relation also provides the algebra of the monodromy elements in \eqref{eq:Telements}, known as the Yang-Baxter algebra. This is a set of commutation relations that can be obtained by specifying the boundary conditions in the four-dimensional auxiliary spaces: $a:(i)\rightarrow(k)$ and $b:(j)\rightarrow(l)$ as follows:
\begin{equation}\label{eq:RTTindex}
\left(R_{ab}(u,v)\right)^{^{(k)(l)}}\left(T_{a}(u)\right)_{(i)}\left(T_{b}(v)\right)_{(j)} = \left(T_{b}(v)\right)^{(l)} \left(T_{a}(u)\right)^{(k)} \left(R_{ab}(u,v)\right)_{(i)(j)} 
\end{equation}
where the lower indexes in parenthesis indicate the initial flavours and the upper indexes correspond to the final flavours. We leave implicit the intermediate flavours indexes over which we must sum over.

Expanding the $R$-matrices in \eqref{eq:RTTindex}  into identity and permutation as in \eqref{eq:R44matrix} we obtain the following algebra of operators:
\begin{equation}\label{eq:YangBaxterAlgebra}
[T_{k\,i}(u),T_{l\,j}(v)] = \left(\frac{-i}{u-v}\right)\, \left(\,T_{l\,i}(v)\,T_{k\,j}(u)\, -\, T_{l\,i}(u)\,T_{k\,j}(v) \,\right)
\end{equation}
where $T_{13}\equiv B_{13}$ and likewise for other operators in \eqref{eq:Telements}.

The Yang-Baxter algebra \eqref{eq:YangBaxterAlgebra} plays a key role in the construction of the ABA in section \ref{sec:ABA} and also in the computation of the scalar product that gives the tree level structure constant in section \ref{sec:scalarproduct}. In what follows we provide some of the details involved in these calculations.
 
\subsubsection{The wanted and unwanted terms of the Bethe Ansatz}
In the ABA construction we use the Yang-Baxter algebra organized into blocks. For this we restrict the four-dimensional auxiliary spaces in \eqref{eq:RTT} to the subspaces $\mathbf{2}$ (flavours $\{3,4\}$) or $\tilde{\mathbf{2}}$ (flavours $\{1,2\}$), instead of strictly fixing the boundary conditions. For instance to derive the $\mathbb{D}$-block and $\mathbb{B}$-block commutation relation we restrict the auxiliary spaces as: $a:\,\mathbf{2}_{(3,4)}\rightarrow \mathbf{2}_{(3,4)}$ and $b:\,\mathbf{2}_{(3,4)}\rightarrow \tilde{\mathbf{2}}_{(1,2)}$:
\begin{equation}
\left(R_{ab}(u,v)\right)^{(3,4)(1,2)}\left(T_{a}(u)\right)_{(3,4)}\left(T_{b}(v)\right)_{(3,4)} = \left(T_{b}(v)\right)^{(1,2)} \left(T_{a}(u)\right)^{(3,4)} \left(R_{ab}(u,v)\right)_{(3,4)(3,4)} 
\end{equation}
expanding the $R$-matrix of the left hand side we obtain:
\begin{equation}\label{eq:DB}
\mathbb{D}_{a}(u)\mathbb{B}_{b}(v) = \left(\frac{u-v-i}{u-v}\right)\, \mathbb{B}_{b}(v)\mathbb{D}_{a}(u)\, R_{ab}(u,v) + \left(\frac{i}{u-v}\right)\, \mathbb{B}_{b}(u)\mathbb{D}_{a}(v)\,P_{ab}
\end{equation}
Under these restrictions now $a$ and $b$ label two-dimensional spaces. 

Similarly, by making another appropriate choice of boundary conditions:  $a:\,\tilde{\mathbf{2}}_{(1,2)}\rightarrow \tilde{\mathbf{2}}_{(1,2)}$ and $b:\,\mathbf{2}_{(3,4)}\rightarrow \tilde{\mathbf{2}}_{(1,2)}$, we obtain the $\mathbb{A}$-$\mathbb{B}$ commutation relation:
\begin{equation}\label{eq:AB}
 \mathbb{A}_{a}(u) \mathbb{B}_{b}(v) = \left(\frac{u-v+i}{u-v}\right)\, R_{ba}(v,u)\, \mathbb{B}_{b}(v) \mathbb{A}_{a}(u)  + \left(-\frac{i}{u-v}\right)\,P_{ab}\, \mathbb{B}_{b}(u)\mathbb{A}_{a}(v)  
\end{equation}
Using these commutation relations, \eqref{eq:DB}  and \eqref{eq:AB}, we can compute the wanted and unwanted terms as a result of commuting $\mathbb{A}$ and $\mathbb{D}$ through the product of $\mathbb{B}$-blocks in the ansatz \eqref{eq:so6Bethe}. These results are given by:
\begin{align}
\color{red}{\mathbb{A}_{a}(\lambda)}\color{black}\,\mathbb{B}(u_{1\cdots M}) &= \Phi_{0}(\lambda|\{u\}) \, \color{red}{T_{a}(u_{1\cdots M}|\lambda)}\color{black}\,\mathbb{B}(u_{1\cdots M})\,{\color{red}\mathbb{A}_{a}(\lambda)}\nonumber\\
&\quad + \,\sum_{k=1}^{M} \Phi_{k}(\lambda|\{u\})\, \color{red} T_{a}(u_{1\cdots M}|u_{k})\,\color{black} T_{k}(u_{k}|u_{k-1\cdots 1})\,\, \mathbb{B}_{k}(\lambda)\, \mathbb{B}(u_{1\cdots \hat{k}\cdots M})\, T_{k}(u_{1\cdots k-1}|u_{k})\,\color{red} \mathbb{A}_{a}(u_{k})
\end{align}
where we use the short-hand notation:
\begin{equation}
\mathbb{B}(u_{1\cdots M})\equiv\prod_{m=1}^{M}\mathbb{B}_{m}(u_{m})\qquad\text{and}\qquad  \mathbb{B}(u_{1\cdots \tilde{k}\cdots M})\equiv\prod_{\underset{m=1}{m\neq k}}^{M}\mathbb{B}_{m}(u_{m})
\end{equation} 
as well as:
\begin{align}
T_{k}(u_{k}|u_{k-1\cdots 1}) &= R_{k\,k-1}(u_{k},u_{k-1})\,\cdots\, R_{k\,2}(u_{k},u_{2})\, R_{k\,1}(u_{k},u_{1})\nonumber\\
T_{k}(u_{1\cdots k-1}|u_{k}) &= R_{1\,k}(u_{1},u_{k})\,R_{k\,1}(u_{k},u_{1})\,\cdots\, R_{k-1\,k}(u_{k-1},u_{k})
\end{align}
The $\Phi_{k}$ coefficients are:
\begin{equation}
\Phi_{0}(\lambda|\{u\}) = \prod_{j=1}^{M}\frac{u_{j}-\lambda-i}{u_{j}-\lambda}  \qquad\text{and}\qquad \Phi_{k}(\lambda|\{u\}) = \frac{i}{u_{k}-\lambda}\;\prod_{\underset{j\neq k}{j=1}}^{M} \frac{u_{j}-u_{k}-i}{u_{j}-u_{k}}
\end{equation}
Similarly we commute a $\mathbb{D}$-block through a product of $\mathbb{B}$-blocks as:
\begin{align}
{\color{blue} \mathbb{D}_{a}(\lambda)}\, \mathbb{B}(u_{1\cdots M}) &= \Theta_{0}(\lambda|\{u\})\, \mathbb{B}(u_{1\cdots M})\,{\color{blue} \mathbb{D}_{a}(\lambda)\, T_{a}(\lambda|u_{M\cdots 1})}\nonumber\\
&\quad +\,\sum_{k=1}^{M}\,\Theta_{k}(\lambda|\{u\})\, T_{k}(u_{k}|u_{k-1\cdots 1})\,\mathbb{B}_{k}(\lambda)\,\mathbb{B}(u_{1\cdots \hat{k}\cdots M})\,T_{k}(u_{1\cdots k-1}|u_{k})\,\color{blue} \mathbb{D}_{a}(u_{k}) \, T_{a}(u_{k}|u_{M\cdots 1})
\end{align}
with coefficients:
\begin{equation}
\Theta_{0}(\lambda|\{u\}) =\prod_{j=1}^{M}\frac{\lambda - u_{j}-i}{\lambda - u_{j}} \qquad\text{and}\qquad \Theta_{k}(\lambda|\{u\}) =\frac{i}{\lambda - u_{k}} \; \prod_{\underset{j\neq k}{j=1}}^{M} \frac{u_{k}-u_{j}-i}{u_{k}-u_{j}}
\end{equation}

\subsubsection{The $C$-commutation relations for the scalar product}\label{sec:Ccom}
When computing the scalar product \eqref{eq:c123brute}, using the ABA, we will need of the commutation relations of the annihilation $\mathbb{C}$-block and the other elements of the monodromy. These can be obtained from the Yang-Baxter algebra \eqref{eq:YangBaxterAlgebra} as:
\begin{align}
  [\mathbb{C}_{a}(u),\mathbb{B}_{b}(v)] &= \left(\frac{-i}{u-v}\right)\,\left(\mathbb{A}_{b}(v)\,\mathbb{D}_{a}(u)\,P_{ab}-\,P_{ab}\, \mathbb{A}_{a}(u)\mathbb{D}_{b}(v)\right)\label{eq:CB}\\
  [\mathbb{C}_{a}(u),\mathbb{A}_{b}(v)] &= \left(\frac{-i}{u-v}\right)\,\left(\mathbb{A}_{b}(v)\,\mathbb{C}_{a}(u)\,P_{ab}-P_{ab}\,\mathbb{A}_{a}(u)\,\mathbb{C}_{b}(v)\right)\label{eq:CA} \\
  [\mathbb{C}_{a}(u),\mathbb{D}_{b}(v)] &= \left(\frac{-i}{v-u}\right)\,\left(P_{ab}\,\mathbb{D}_{b}(v)\,\mathbb{C}_{a}(u) - \mathbb{D}_{a}(u)\,\mathbb{C}_{b}(v) \,P_{ab}  \right)\label{eq:CD}
\end{align}
\subsection{The coordinate Bethe Ansatz (CBA)}\label{sec:CBA}
As explained in section \ref{sec:vertexmodel} we can expand  the Bethe states in terms of a coordinate basis as:
\begin{equation}\label{statecoordinatebasis}
|\Psi_{\mathfrak{so}(6)}\rangle = \sum_{\text{coordinate basis}}\, \Psi_{ZX\cdots}(\{{\red v}\},\{u\},\{{\blue w}\})\,|Z\,X\,\cdots \rangle
\end{equation}
where $|Z\,X\cdots\rangle$ stands for an element of the coordinate basis and the coefficient $\Psi_{ZX\cdots}$ is its correspondent wave-function.  This wavefunction is obtained as a partition function in the lattice in figure \ref{fig:vertexmodel}, with top boundary conditions imposed by the correspondent element of the coordinate basis.
\subsubsection{The coordinate basis as strings of auxiliary roots}
To introduce the wave-function of a given element in the coordinate basis we first define a one to one map between the elementary fields \eqref{eq:Zletters} and a set of rapidities: 
\begin{equation}\label{eq:tostring}
Z\equiv \theta  \qquad X\equiv \overset{u}{\theta}\qquad Y \equiv \overset{\overset{{\blue w}}{u}}{\theta}\qquad \bar{Y} \equiv \overset{\overset{{\red v}}{u}}{\theta} \qquad \bar{X}\equiv \overset{\overset{{\red v}\;{\blue w}}{u}}{\theta}\qquad \bar{Z}\equiv \overset{\overset{{\red v}\,{\blue w}}{u_{1}\,u_{2}}}{\theta}\,,
\end{equation}
where $\theta$'s are the inhomogeneities defined for each spin chain site. To make manifest the structure of the nested Bethe ansatz, here we represented the fields by stacking the roots: $u$ is the root at the middle node whereas ${\blue w}$ and ${\red v}$ are the nested roots associated with the left and the right nodes respectively.

Using this representation, we can re-express the coordinate basis as a collection of (sets of) rapidities, for instance: 
\begin{equation}
|Z_{(\theta_{1})}\,X_{(\theta_{2})}\,\bar{Z}_{(\theta_{3})}\,\bar{Y}_{(\theta_{4})}\rangle  \equiv |\theta_{1}\;\,\overset{u_{1}}{\theta_{2}}\;\,\overset{\overset{{\red v_{1}}\,{\blue w_{1}}}{u_{2}\,u_{3}}}{\theta_{3}}\;\,\overset{\overset{{\red v_{2}}}{u_{4}}}{\theta_{4}}\rangle
\end{equation}
Here we assigned a numeration to the auxiliary roots in the order of appearance\footnote{Since we later sum over all permutations of auxiliary roots this enumeration becomes irrelevant.}. In what follows, we call such a collection of rapidities a {\it string}. The full wave function \eqref{statecoordinatebasis} can then be written as
\begin{equation}\label{statestringbasis}
|\Psi_{\mathfrak{so}(6)}\rangle = \sum_{{\sf s}\in \,\substack{\text{all possible}\\ \text{strings}}}\, \Psi_{{\sf s}}(\{{\red v}\},\{u\},\{{\blue w}\})\,|{\sf s} \rangle
\end{equation}

\subsubsection{The wave-function}
The wavefunction $\Psi_{{\sf s}}$ for the string ${\sf s}$ is given by a sum over weighted permutations over all the auxiliary roots:
\begin{align}\label{eq:firstso6}
\Psi_{{\sf s}}= \,\sum_{\pi \in Per(K_{1})}\!\!\!S(\{{\red v}\}_{\pi}) \sum_{\sigma \in Per(K_{2})}\!\!\!S(\{u\}_{\sigma})\,\sum_{\tilde{\pi} \in Per(K_{3})}\!\!\!S(\{{\blue w}  \}_{\tilde{\pi}})\,\times \, \Psi^{{\rm bare}}_{{\sf s}}(\{{\red v}\}_{\pi},\{u\}_{\sigma},\{{\blue w}\}_{\tilde{\pi}})\,,
\end{align}
where the notation $\{\ast \}_{\sigma}$ denotes that the set $\ast$ is permuted according to the permutation $\sigma$. The multiparticle $S$-matrix $S(\{u\}_{\sigma})$ brings the ordered momenta $\{u\}$ to the ordering $\{u\}_{\sigma}$ and is given by a factorized product of two-body S-matrices as in the examples:
\begin{align}
S(\{u_{3},u_{2},u_{1}\}) &= S(u_{1},u_{2})S(u_{1},u_{3})S(u_{2},u_{3})\quad \text{with}\quad S(u_{a},u_{b})=\frac{u_{a}-u_{b}-i}{u_{a}-u_{b}+i}\nonumber\\
S(\{u_{3},u_{1},u_{2}\}) & = S(u_{1},u_{3})\,S(u_{2},u_{3})
\end{align}
In a spin chain with $L$ sites, the correspondent ``bare" wavefunction $\Psi^{{\rm bare}}_{{\sf s}}$ is given by
\begin{equation}\label{barewavedefinition}
\Psi^{{\rm bare}}_{{\sf s}}(\{{\red v}\},\{u\},\{{\blue w}\}) = \prod_{n=1}^{L}\, \Phi({\sf s}_{\,n})\,,
\end{equation}
where the individual wave function $\Phi({\sf s}_{n})$, defined for the $n$-th element of the string ${\sf s}$, is given by
\begin{align}\label{individialwavefunction}
\begin{aligned}
\Phi(\,\theta\,) &= 1\\
 \Phi(\,\overset{u}{\theta_{n}}\,) &= \varphi_{n}(u|\{\theta\})\\
\Phi(\,\overset{\overset{{\blue w}}{u_{m}}}{\theta_{n}}\,) &= \varphi_{n}(u_{m}|\{\theta\})\times \tilde{\varphi}_{m}({\blue w}|\{u\})\\
\Phi(\,\overset{\overset{{\red v}}{u_{m}}}{\theta_{n}}\,) &= \varphi_{n}(u_{m}|\{\theta\})\times \tilde{\varphi}_{m}({\red v}|\{u\})\\
\Phi(\,\overset{\overset{{\red v}\,{\blue w}}{u_{m}}}{\theta_{n}}\,) &= (-1)\times\varphi_{n}(u_{m}|\{\theta\})\times \tilde{\varphi}_{m}({\red v}|\{u\})\times \tilde{\varphi}_{m}({\blue w}|\{u\})\\
\Phi(\,\overset{\overset{{\red v}\,{\blue w}}{u_{m}\,u_{m+1}}}{\theta_{n}}\,) &= \frac{1}{2}\times \varphi_{n}(u_{m}|\{\theta\})\times\varphi_{n}(u_{m+1}|\{\theta\})\\
& \qquad\times \Big(\tilde{\varphi}_{m}({\blue w}|\{u\})-\tilde{\varphi}_{m+1}({\blue w}|\{u\})\Big)\times \Big(\tilde{\varphi}_{m}({\red v}|\{u\})-\tilde{\varphi}_{m+1}({\red v}|\{u\})\Big)\;\;
\end{aligned}
\end{align}
where we include the label of the rapidities ($u_{m}$) only when this is necessary to express the correspondent wavefunction. The factors $\varphi$ and $\tilde{\varphi}$ are one-particle wave functions and are given by:
\begin{align}
\varphi_{n}(u|\{\theta\}) &= \left(\prod_{l=1}^{n-1}\frac{u-\theta_{l}+i/2}{u-\theta_{l}-i/2}\right)\times \underbrace{\frac{1}{u-\theta_{n}-i/2}}_{\text{occupation factor}}\nonumber\\
\tilde{\varphi}_{m}(w|\{u\}) &= \left(\prod_{l=1}^{m-1}\frac{w-u_{l}+i/2}{w-u_{l}-i/2}\right)\times \frac{1}{w-u_{m}-i/2}
\end{align}
Needless to say, when the rapidities are permuted in \eqref{barewavedefinition}, we should also permute the rapidities in the definitions of the wave functions, which are given by the right hand sides of \eqref{individialwavefunction}, accordingly.

\subsection{ The scalar product: tree level structure constant}\label{sec:scalarproduct}
As we saw in the introduction of this Appendix, the tree level planar three-point function in figure \ref{fig:wickcontraction} is given by the scalar product between a rotated BMN vacuum and a Bethe state:
\begin{align}\label{eq:mathcalA}
\mathcal{C}_{123} &= \langle \tilde{Z}^{l}| \Psi(\{v\},\{u\}_{M},\{w\})\rangle_{l-Bethe}\nonumber\\
&= \langle \tilde{Z}^{l}|\otimes {\color{red}\langle \tilde{\psi}|}\,\mathbb{B}_{1}(u_{1})\cdots \mathbb{B}_{M}(u_{M})\, {\color{blue}|\psi\rangle}\otimes |\Omega_{l}\rangle 
\end{align}
where Bethe state is given by the Ansatz in figure \ref{fig:BetheState} but with the number of sites or elementary fields equal to the bridge ``$l$".  
\subsubsection{A global rotation}
In order to compute the scalar product \eqref{eq:mathcalA} using the machinery of the ABA, we first need to express the rotated vacuum in this language. This is achieved by means of a global rotation of the original vacuum:
\begin{equation}
|\tilde{Z}^{L}\rangle = e^{\mathsf{b}}\,|Z^{L}\rangle
\end{equation}
The generator ``$\mathsf{b}$" of this global rotation can be simply obtained from the $B$-block by taking the trace and sending to $\infty$ the spectral parameter ``$u$";
\begin{equation}
\lim_{u\rightarrow \infty}\, Tr\,\mathbb{B}(u) = \frac{i}{u}\, \mathsf{b}
\end{equation}
this lowering generator is composed by the elements:
\begin{equation}
\mathsf{b} = \mathsf{b}_{13}+\mathsf{b}_{24}
\end{equation}
When $\mathsf{b}_{13}$ acts over the vacuum generates a $X$ excitation, when $\mathsf{b}_{24}$ acts generates  -$\bar{X}$ and when both act over the same site a $\bar{Z}$ excitation is generated. In this way we generate the rotated vacuum:
\begin{equation}
\tilde{Z} \equiv Z+\bar{Z}+X-\bar{X}
\end{equation}
In the scalar product \eqref{eq:mathcalA} we must use instead the bra state for which we use the $C$-block as:
\begin{equation}\label{eq:climit}
\langle \tilde{Z}^{l}| = \langle \Omega |\,e^{\mathsf{c}}\qquad\text{with}\qquad  \lim_{u\rightarrow\infty}\,Tr\,\mathbb{C}(u) = \frac{i}{u}\, \mathsf{c} 
\end{equation}
\subsubsection{The scalar product}
Now we outline the steps we take to compute the scalar product \eqref{eq:mathcalA}. The first step is to notice that the Bethe state has a defined $\mathfrak{so}(6)$ charge determined by the number of (finite) auxiliary roots $\{u\}_{M}$,$\{v\}$ and $\{w\}$. While in the expansion of global rotation $e^{\mathsf{c}}=1+\mathsf{c}+\cdots$ only the term $\mathsf{c}^{M}$ matches this $\mathfrak{so}(6)$ charge . So the scalar product \eqref{eq:mathcalA} can be simplify to:
\begin{equation}\label{eq:scalarproduct}
  \mathcal{C}_{123} =\frac{1}{M!}\times  \langle\Omega |\otimes \color{red}\langle\tilde{\psi}| \, \color{black}\mathsf{c}^{M}\, \color{black}\,\mathbb{B}_{1}(u_{1})\cdots \mathbb{B}_{M}(u_{M})\color{blue}|\psi\rangle\color{black}\otimes |\Omega\rangle   
\end{equation}
The next step is to commute all the $\mathsf{c}$ operators through the $B$-blocks, such as we can use their annihilation properties:
\begin{equation}
\mathsf{c}\,|\Omega\rangle=0\qquad\text{and}\qquad \langle\Omega|\,\mathbb{B}(u)\,=0
\end{equation}
The commutator of $\mathsf{c}$ and $B$ can be found from the Yang-Baxter algebra. Taking the limit \eqref{eq:climit} of \eqref{eq:CB}  we obtain:
\begin{equation}
[\mathsf{c},\mathbb{B}_{a}(u)] = \mathbb{D}_{a}(u)-\mathbb{A}_{a}(u)\,
\end{equation}
Since $\mathbb{A}$ and $\mathbb{D}$ blocks are generated we also need of their commutators with $\mathsf{c}$, which can be extracted in a similar way from \eqref{eq:CA} and \eqref{eq:CD} as:
\begin{equation}
 [\mathsf{c},\mathbb{D}_{a}(u)] =-\mathbb{C}_{a}(u)\,,\quad [\mathsf{c},\mathbb{A}_{a}(u)] =\mathbb{C}_{a}(u)\quad\text{and}\quad \quad [\mathsf{c},\mathbb{C}_{a}(u)] =0
\end{equation}
The final set of commutators we need are between $\mathbb{C}$ and $\mathbb{A}$, $\mathbb{B}$, $\mathbb{D}$ blocks. These are given in section \ref{sec:Ccom}.

All in all the result of commuting $\mathsf{c}^{M}$ through a product of $M$ $B$-block operators is given by a sum  over bi-partitions:
\begin{align}\label{eq:fullcB}
\frac{1}{M!}\times \mathsf{c}^{M}\;\mathbb{B}_{1}(u_{1})\cdots \mathbb{B}_{M}(u_{M}) &= \sum_{\alpha\cup \bar{\alpha}=\{u\}}\,(-1)^{|\alpha|}\,h_{\alpha,\bar{\alpha}}\, R_{\bar{\alpha},\alpha}\,\mathbb{A}_{\alpha}\,\mathbb{D}_{\bar{\alpha}}\, R^{\alpha,\bar{\alpha}}\nonumber\\
&\qquad\quad + \mathbb{C}\text{-terms}+\mathsf{c}\text{-terms}
\end{align}
where we use the short-hand notation:
\begin{equation}
\mathbb{A}_{\alpha} = \prod_{u\in\alpha} \, \mathbb{A}(u)\,,\qquad \mathbb{D}_{\bar{\alpha}} = \prod_{u\in\bar{\alpha}} \, \mathbb{D}(u)\quad\text{and}\quad h_{\alpha,\bar{\alpha}}  = \prod_{u\in\alpha,v\in\bar{\alpha}} \frac{u-v-i}{u-v}
\end{equation}
and the $\mathbb{C}$ and $\mathsf{c}$-terms are products of operators that annihilate the $\mathfrak{so}(6)$ pseudo-vacuum.

The matrix operator $R^{\alpha,\bar{\alpha}}$ is a product of $\mathfrak{su}(2)$ $R$-matrices \eqref{eq:22R-matrix} that changes the order of the roots $\{u_{1}\cdots u_{M}\}$ to the order $\{\alpha,\bar{\alpha}\}$, while the operator $R_{\bar{\alpha},\alpha}$ takes the roots from the ordering $\{\bar{\alpha},\alpha\}$ to the ordering $\{u_{1}\cdots u_{M}\}$. 

\begin{figure}[H]
\begin{center}
\subfigure[Scattering from $\{u\}$ to $\{\alpha,\bar{\alpha}\}$]
{\label{fig:inital}
\begin{tikzpicture}
\node[draw, circle,minimum size=1cm](Rmatrix) at (0,0) {$R^{\alpha,\bar{\alpha}}$};
\node[draw,rectangle,minimum width=5cm,rounded corners] (ulist) at (0,-2) {$u_{1}\;\; u_{2}\;\; u_{3}\quad\cdots\quad u_{M-1}\quad u_{M}$};
\node[draw,rectangle,minimum width=2.6cm,minimum height=.5cm,rounded corners](alpha) at (-1.5,2) {$\alpha$};
\node[draw,rectangle,minimum width=2.6cm,rounded corners](balpha) at (1.5,2) {$\bar{\alpha}$};
\foreach \a in {1,2,...,9}{
\draw[->-=.6] ($(ulist.north west)!\a*.1!(ulist.north east)$) to (\a*360/20+180:.61cm);
}
\foreach \a in {1,2,3,4,5}{
\draw[->-=.6] (\a*360/30+100:.61cm) to ($(alpha.south east)!\a*.16!(alpha.south west)$);
}
\foreach \a in {1,2,3,4}{
\draw[->-=.6] (\a*360/30:.61cm) to ($(balpha.south east)!\a*.20!(balpha.south west)$);
}
\end{tikzpicture}
}
\qquad\qquad
\subfigure[Scattering from $\{\bar{\alpha},\alpha\}$ to  $\{u\}$]
{\label{fig:final}
\begin{tikzpicture}
\node[draw, circle,minimum size=1cm](Rmatrix) at (0,0) {$R_{\bar{\alpha},\alpha}$};
\node[draw,rectangle,minimum width=5cm,rounded corners] (top) at (0,2) {$u_{1}\;\; u_{2}\;\; u_{3}\quad\cdots\quad u_{M-1}\quad u_{M}$};
\node[draw,rectangle,minimum width=2.6cm,minimum height=.5cm,rounded corners](balpha) at (-1.5,-2) {$\bar{\alpha}$};
\node[draw,rectangle,minimum width=2.6cm,rounded corners](alpha) at (1.5,-2) {$\alpha$};
\foreach \a in {1,2,...,9}{
\draw[->-=.6] (\a*360/20:.61cm) node {} to ($(top.south east)!\a*.1!(top.south west)$);
}
\foreach \a in {1,2,3,4}{
\draw[->-=.6] ($(balpha.north west)!\a*.20!(balpha.north east)$) to (\a*360/30+200:.61cm);
}
\foreach \a in {1,2,3,4,5}{
\draw[->-=.6] ($(alpha.north west)!\a*.16!(alpha.north east)$) to (\a*360/30+280:.61cm);
}
\end{tikzpicture}
}
\end{center}
\caption{multi-scattering $R$-matrices in \eqref{eq:fullcB} and \eqref{eq:c123prefinal}}
\label{fig:NiceImage}%
\end{figure}
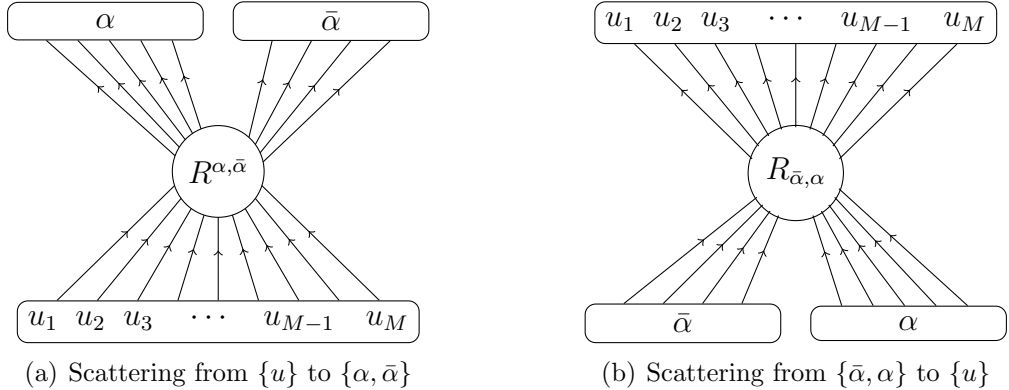

For instance, when $\alpha = \{u_{2},u_{4}\}\,$ and $\,\bar{\alpha}=\{u_{1},u_{3},u_{5}\}$, the multiparticle scattering operators are:
\begin{align}
\{u_{1},u_{2},u_{3},u_{4},u_{5}\}\overset{R^{\alpha,\bar{\alpha}}}{\longrightarrow} \{\overbrace{u_{2},u_{4}}^{\alpha},\overbrace{u_{1},u_{3},u_{5}}^{\bar{\alpha}}\}:&\qquad R^{\alpha,\bar{\alpha}} = R_{14}\,R_{12}\,R_{34}\nonumber\\
\{\underbrace{u_{1},u_{3},u_{5}}_{\bar{\alpha}},\underbrace{u_{2},u_{4}}_{\alpha}\}\overset{R_{\bar{\alpha},\alpha}}{\longrightarrow} \{u_{1},u_{2},u_{3},u_{4},u_{5}\} :&\qquad R_{\bar{\alpha},\alpha} =  R_{32}\,R_{54}\, R_{52}
\end{align}
where we use the short-hand notation $R_{ab}=R_{ab}(u_{a},u_{b})$

Finally we can compute the scalar product \eqref{eq:scalarproduct}  using \eqref{eq:fullcB} and the eigenstate relations of the pseudovacuum  \eqref{eq:ADvacuum}:
\begin{equation}\label{eq:c123prefinal}
\mathcal{C}_{123} = \sum_{\alpha\cup \bar{\alpha}=\{u\}}\,(-1)^{|\alpha|}\,a_{\alpha}\,d_{\bar{\alpha}}\,h_{\alpha,\bar{\alpha}} \,\times\,\underbrace{\color{red}\langle\tilde{\psi}|\color{black}\, R_{\bar{\alpha},\alpha}\, R^{\alpha,\bar{\alpha}}\,\color{blue}|\psi\rangle\color{black} }_{\mathcal{M}atrix_{\alpha,\bar{\alpha}}}
\end{equation}
with $a_{\alpha}=\prod_{u\in\alpha}a(u)=1$ and  $d_{\bar{\alpha}}=\prod_{u\in\bar{\alpha}}\,d(u)=\prod_{u\in\bar{\alpha}}\,\left(\frac{u+i/2}{u-i/2}\right)^{L}$. Needless to say, this expression is already strongly resembling the all loop expressions in the main text including the involved matrix part as proposed in \cite{C123Paper}. 
\subsubsection{Wings on-shell: simplifying the matrix part $\mathcal{M}atrix_{\alpha,\bar{\alpha}}$ }
The matrix part can be further simplify considering that the wing-states are (on-shell) $\mathfrak{su}(2)$ Bethe states. In this case the action of the $R$-matrices has the simple effect of reshuffling the inhomogeneities $\{u\}$ of the wing-states so we obtain:
\begin{equation}\label{eq:prestates}
{\color{red}\langle\tilde{\psi}|}\, R_{\bar{\alpha},\alpha}\,=  {\color{red}\langle\tilde{\psi}_{\bar{\alpha},\alpha}|} \qquad \text{and}\qquad R^{\alpha,\bar{\alpha}}\,{\color{blue}|\psi \rangle } =\,{\color{blue}|\psi_{\alpha,\bar{\alpha}}\rangle}  
\end{equation}
where $|\psi_{\alpha,\bar{\alpha}}\rangle$ is the wing state $|\psi\rangle$  with the inhomogeneities order as in the top of figure \ref{fig:inital} and $\langle\psi_{\bar{\alpha},\alpha}|$ is the wing $\langle\tilde{\psi} |$ with the ordering as in the bottom of figure \ref{fig:final}. In this way the matrix part is simply given by the scalar product of the states \eqref{eq:prestates}. To simplify this scalar product it is necessary to place the inhomogeneities of the two states in the same ordering. This can be achieved by using again the multi-scattering $R$ matrix:
\begin{equation}
|\psi_{\alpha,\bar{\alpha}}\rangle = R_{\bar{\alpha},\alpha}^{\alpha,\bar{\alpha}}\,|\psi_{\bar{\alpha},\alpha}\rangle
\end{equation}
The special feature of this reordering scattering matrix is that it can be expressed as a product of nested $\mathfrak{su}(2)$ transfer matrices with spectral parameters $u\in\bar{\alpha}$ and inhomogeneities $\{u\}$ :
\begin{equation}
 R_{\bar{\alpha},\alpha}^{\alpha,\bar{\alpha}} = \prod_{u\in\bar{\alpha}}\,\mathcal{T}^{su(2)}(u)
\end{equation}
Since our wing states are on-shell we can replace the transfer matrices by the correspondent eigenvalue and obtain the periodicity relation:
\begin{equation}
|\psi_{\alpha,\bar{\alpha}}\rangle = \left(\prod_{u\in\bar{\alpha}}\mathcal{T}^{su(2)}(u)\right)\,|\psi_{\bar{\alpha},\alpha}\rangle = \left(\prod_{u\in\bar{\alpha}}\Lambda^{su(2)}(u)\right)\,|\psi_{\bar{\alpha},\alpha}\rangle 
\end{equation}
So for on-shell Bethe states the cost of reordering is just a phase. Then the matrix part is given by:
\begin{equation}
\text{$\mathcal{M}$atrix}_{\alpha,\bar{\alpha}}\, =  \left(\prod_{u\in\bar{\alpha}}\Lambda^{su(2)}(u)\right)\,\langle \tilde{\psi}_{\bar{\alpha},\alpha}|\psi_{\bar{\alpha},\alpha}\rangle
\end{equation}
We now have the scalar product of two on-shell Bethe states with inhomogeneities in the same ordering. Considering their orthogonality property we know this scalar product vanishes unless the set of wing roots are identical $\{v\}=\{w\}$\footnote{Here we refer to the shifted wing roots $v\rightarrow v+i/2$ and $w\rightarrow w-i/2$ which  appear in the standard form of the so(6) Bethe equations. }. Under this condition the scalar product is given by the Gaudin-determinant. This determinant is invariant under permutations of the inhomogeneities so it can be taken out of the sum over partitions. The final expression for the unnormalized tree level structure constant is:
\begin{equation}
\boxed{\mathcal{C}^{\mathfrak{so}(6)}_{123} = Gaudin_{\mathfrak{su}(2)\text{-}wing}\times \sum_{\alpha\cup \bar{\alpha}=\{u\}}\,(-1)^{|\alpha|}\,h_{\alpha,\bar{\alpha}}\,\times \prod_{u\in\bar{\alpha}}\,\left(\frac{u+i/2}{u-i/2}\right)^{L}\,\Lambda^{\mathfrak{su}(2)}(u)  }
\end{equation} 
where $\Lambda^{\mathfrak{su}(2)}(u) = \prod_{k=1}^{K}\frac{u - v_{k}+i/2}{u-v_{k}-i/2}$ , after performing the shift \eqref{eq:shift}.

\section{Nested Hexagons in the $\mathfrak{su}(1,1|2)$ Sector}\label{su112}

In this appendix, we study the hexagon amplitudes in the $\mathfrak{su}(1, 1|2)$ sector and provide a detailed derivation of the formulas given in section \ref{sec3}. This is a large sector of operators, the largest closed subsector containing both the $\mathfrak{sl}(2)$ and $\mathfrak{su}(2)$ diagonal sectors, see e.g.~\cite{BeisertThesis}. Still, the analysis in this sector turns out to be remarkably simple. In the following, we review the Bethe ansatz wave functions for these operators \cite{Beisert:2005tm} and employ them to the computation of the hexagon amplitudes.
\subsubsection*{Wave functions}

The states of interest are usual BMN operators above the BPS vacuum $|0\mathcal{i} = \textrm{tr}\, Z^{L}$,
\beq\label{generic}
|\chi_{A_{1}\dot{A}_{1}}\ldots \chi_{A_{K}\dot{A}_{K}}\mathcal{i}\,,
\eeq
with each excitation $\chi$ lying in a $(\textbf{1}|\textbf{1}) \otimes (\dot{\textbf{1}}|\dot{\textbf{1}})$ irrep of the residual symmetry subalgebra $\mathfrak{su}(1|1) \oplus \dot{\mathfrak{su}}(1|1)$. Equivalently, a magnon $\chi$ can take 4 possible values, out of the 16 ones,
\beq
Y = \phi \otimes \dot{\phi}\, , \qquad \Psi = \psi \otimes \dot{\phi}\, , \qquad \dot{\Psi} = \phi\otimes \dot{\psi}\, , \qquad DZ = \psi \otimes \dot{\psi}\, ,
\eeq
with $Y$ a complex scalar, $\Psi$ and $\dot{\Psi}$ two gauginos, and $D$ a lightcone covariant derivative. The dynamics factorizes along the two wings and a generic scattering eigenstate can be written as a tensor product of a left and a right wave function, built out of $\phi|\psi$ and $\dot{\phi}|\dot{\psi}$ respectively. We recall below the structure of these wave functions, focusing on the left part of the state. The right part defines an isomorphic problem and all the formulae hereafter apply to it after ``dotting" whatever can be dotted. (One must use $\dot{\textbf{u}} = \textbf{u}$, for the main roots, since these are inhomogeneities common to the two factorized auxiliary spin chains.)  

Off-shell Bethe states for the inhomogeneous $\mathfrak{su}(1|1)$ spin chain are constructed in the usual manner, using the restriction of the full $\mathfrak{su}(2|2)$ S-matrix \cite{Beisert:2006qh} to the $\phi|\psi$ subspace as a fundamental $R$-matrix. The analysis is quite similar to the standard algebraic Bethe ansatz for the $\mathfrak{su}(2)$ spin chain with the difference that the $B$ operator here is \textit{fermionic}. As a starting point, one must choose a pseudovacuum state. We shall work in the $\mathfrak{su}(2)$ grading for which the reference state,  the so-called level II vacuum, is chosen to be made of scalars only
\beq
|0\mathcal{i}^{II}_{\textbf{u}} = |\phi_{1}\phi_{2} \ldots \phi_{K}\mathcal{i} \, .
\eeq
The subscript $\textbf{u}$ reminds us that this vacuum state depends on the ordering of the lattice. Acting with a $B$ operator produces a fermionic ``spin wave" along the chain%
\footnote{The $B(y)$ operator is obtained by scattering a probe particle with rapidity $y = x^{-}_{0}$ through the chain, with the boundary condition that it starts as a fermion and ends as a boson. When defining $B$ in this way, using the S-matrix of \cite{Beisert:2006qh}, we also strip out an inessential overall factor. The latter factor is invariant under permutation of the spin chain inhomogeneities and thus does not affect the fundamental property of the $B$ operator.}
\beq
B_{\textbf{u}}(y)|0\mathcal{i}^{II}_{\textbf{u}} = \sum_{n=1}^{K} \Psi_{n}(y) |\phi_{1} \ldots \psi_{n} \ldots \phi_{K}\mathcal{i}\, ,
\eeq
with the wave function \cite{Beisert:2005tm}
\beq\label{single-w}
\Psi_{n}(y) = \frac{a_{n}}{y-x^{+}_{n}} \prod_{j=1}^{n-1}S^{II, I}(y, x_{j})\, .
\eeq
Here $a_{n} = \sqrt{i(x_{n}^{-}-x_{n}^{+})}$ is a free parameter of the representation (for the relative normalization between boson and fermion), which we have fixed to its unitary value, for convenience. The phase
\beq\label{Sii-i}
S^{II, I}(y, x_{j}) = \frac{y - x^{-}_{j}}{y-x^{+}_{j}}\, ,
\eeq
can be seen as the S-matrix for bringing the fermion through the lattice site $x_{j}$. The scattering among the fermionic waves happens to be trivial $S^{II, II}(y_{1}, y_{2}) = 1$, up to the statistics. Hence the multiparticle wave function is the totally antisymmetric product of the individual spin waves,
\beq\label{multi}
B_{\textbf{u}}(y_{1})\ldots B_{\textbf{u}}(y_{F})|0\mathcal{i}^{II}_{\textbf{u}} = \sum_{n_{1}<\ldots < n_{F}} \textrm{det}\, \Psi_{n_{i}}(y_{j})\, |\phi_{1} \ldots \psi_{n_{1}} \ldots \psi_{n_{F}} \ldots \phi_{K}\mathcal{i}\, .
\eeq

Note that all these wave functions fulfill, as a consequence of the Yang-Baxter equation, the so-called compatibility condition \cite{Beisert:2005tm}
\beq\label{cc}
\mathcal{S}_{\pi}|\Psi\mathcal{i} = A_{\pi}|\Psi^{\pi}\mathcal{i}\, ,
\eeq
where $\pi$ is an arbitrary permutation of the labels $\textbf{u}$ and $|\Psi^{\pi}\mathcal{i} = B_{\pi(\textbf{u})}(y_{1})\ldots B_{\pi(\textbf{u})}(y_{F})|0\mathcal{i}^{II}_{\pi(\textbf{u})}$ is the state obtained through this relabelling. The overall factor in (\ref{cc}) is the eigenvalue of the scattering matrix $\mathcal{S}_{\pi}$ on the spin chain vacuum,
\beq
A_{\pi} = \prod_{ij \in \pi}A_{ij}\, ,
\eeq
with the product running over the pairwise scattering events in the permutation $\pi$ and with $A_{12}$ the scattering phase for two identical bosons, see \cite{Beisert:2006qh}.%
\footnote{Put differently, $\mathcal{S}_{\pi}$ intertwines between $B_{\textbf{u}}$ and $B_{\pi(\textbf{u})}$, i.e., $\mathcal{S}_{\pi}B_{\textbf{u}}(y) = B_{\pi(\textbf{u})}(y)\mathcal{S}_{\pi}$, and $\mathcal{S}_{\pi}|0\mathcal{i}^{II}_{\textbf{u}} = A_{\pi} |0\mathcal{i}^{II}_{\pi(\textbf{u})}$.}

Lastly, imposing periodic boundary conditions for the spin waves generates the Bethe ansatz equations of the level II rapidities $\textbf{y} = \{y_{j}\}$. Looking back at the wave function (\ref{single-w}) and recalling that fermions do not interact in this model, one easily infers that this choice corresponds to%
\footnote{Just impose $\Psi_{K+1}(y) = \Psi_{1}(y)$ for $x_{K+1} = x_{1}$.}
\beq\label{BAEy}
1 = \prod_{k=1}^{K} S^{II, I}(y_{j}, x_{k}) = \prod_{k=1}^{K} \frac{y_{j}-x^{-}_{k}}{y_{j}-x^{+}_{k}}\, .
\eeq
Below, we shall refer to the level II (or wing) wave function $\Psi$ as being on shell when all the $\textbf{y}$ rapidities are subject to the BAEs (\ref{BAEy}). Note that this implies a condition on the $\textbf{y}$'s, while the $\textbf{u}$'s shall remain arbitrary.

\subsubsection*{Hexagon amplitudes}

Equipped with the wave functions, we can attack the problem of evaluating the hexagon amplitudes in the $\mathfrak{su}(1,1|2)$ subsector. We shall focus on configurations with an equal number of left and right fermions on each hexagon, since these are the sole configurations for which the form factors are nonzero. Symmetrywise a left fermion adds $1/2$ to the left Lorentz spin of the operator and similarly for a right fermion. Hence the overall condition that $F=\dot{F}$ is just saying that operators in the OPE of two scalars are in symmetric (traceless) Lorentz representations.

We begin with the simplest set-up where all the magnons are sitting on the same hexagon, i.e., $\alpha = \textbf{u}$. In this case there is only one hexagon form factor to compute, the one for the full state,
\beq\label{calA}
\mathcal{A}_{\textbf{u}\varnothing} = (-1)^{\mathfrak{f}}\, \mathcal{h} \mathfrak{h}\mathcal{k}\Psi\mathcal{i}|\dot{\Psi}\mathcal{i}\, ,
\eeq
where $(-1)^{\mathfrak{f}}$ is a grading factor for bringing the original state (\ref{generic}) to the above factorized form, with all the left magnons standing on the left of the right magnons. When $F=\dot{F}$, we can just write $\mathfrak{f} = F(F-1)/2$.

The rule for evaluating the hexagon form factor $\mathcal{h} \mathfrak{h}\mathcal{k}\Psi\mathcal{i}|\dot{\Psi}\mathcal{i}$ is to first scatter the left part from its ingoing to its outgoing configuration and then contract the outcome with the right part using the left-right inner product
\beq\label{LRp}
(\chi_{K} \ldots \chi_{1}, \dot{\chi}_{1}\ldots \dot{\chi}_{K}) = \prod_{j=1}^{K} \delta_{\chi_{j}\dot{\chi}_{j}}\, ,
\eeq
where $\delta_{\chi\dot{\chi}} = 1$ if the two excitations are the same, and $\delta_{\chi\dot{\chi}} = 0$ otherwise. The latter orthogonality condition between bosons and fermions, together with the fact that the number of fermions is preserved by the S-matrix (in the $\mathfrak{su}(1,1|2)$ subsector), imply that the amplitude (\ref{calA}) is zero if the numbers of left and right fermions are different. As alluded before, this is a consequence of the diagonal symmetry preserved by the hexagon, and the argument applies to each partition separately in a generic split configurations with $\alpha, \bar{\alpha} \neq \varnothing$.

To avoid possible confusion, we shall use round brackets $(., .)$, as in (\ref{LRp}), to denote the left-right overlap. This one is a real bilinear form on the tensor product of the left and right Hilbert spaces $\mathbb{H}\otimes \dot{\mathbb{H}}$. It is different from the usual Hermitian inner product $\mathcal{h}.| .\mathcal{i}$, defined separately on each Hilbert space, which is of course sesquilinear. Since the left and right Hilbert spaces are isomorphic we can also overlap left and right states using the Hermitian inner product. The relation to the round product is then given by
\beq
\mathcal{h}\Psi| \dot{\Psi}\mathcal{i} = (\Psi^{\dagger}, \dot{\Psi})\, ,
\eeq
where $\Psi^{\dagger}$ is the Hermitian conjugate of the state $\Psi$, described by the complex conjugate wave function on the transposed lattice.

Introducing the permutation $\pi$ that reverses the ordering of the spin chain, we can write the hexagon amplitude as
\beq
\mathcal{h} \mathfrak{h}\mathcal{k}\Psi\mathcal{i}|\dot{\Psi}\mathcal{i} = (\mathcal{S}_{\pi}\Psi, \dot{\Psi})\, .
\eeq
Its evaluation is straightforward. Indeed, $\mathcal{S}_{\pi}$ simply maps $\Psi$ to an outgoing state $\overleftarrow{\Psi}$, that is, to a state to be read from the right to the left,
\beq
|\overleftarrow{\Psi}\mathcal{i} = \sum_{n_{1}< \ldots <n_{F}}\overleftarrow{\Psi}_{\textbf{n}}(\textbf{y}) |\phi_{K} \ldots \psi_{n_{F}} \ldots \psi_{n_{1}} \ldots \phi_{1}\mathcal{i}\, ,
\eeq
with a multiparticle wave function given as before, though in terms of the outgoing wave
\beq\label{out-g}
\overleftarrow{\Psi}_{n}(y) = \frac{a_{n}}{y-x^{-}_{n}} \prod_{j=1}^{n-1}S^{I, II}(x_{j}, y)\, ,
\eeq 
where $S^{I, II} = 1/S^{II, I}$. More accurately, $\mathcal{S}_{\pi}$ does act like that, up to an overall scalar factor,
\beq\label{main-loc}
\mathcal{S}_{\pi} |\Psi\mathcal{i} =  (-1)^{\mathfrak{f}} \, S_{\Psi} |\overleftarrow{\Psi}\mathcal{i}\, ,
\eeq
with $S_{\Psi}$ defined by
\beq\label{overall}
S_{\Psi} = A_{\pi} \prod_{j, k}S^{II, I}(y_{j}, x_{k})\, .
\eeq
It follows from it that we can write the hexagon form factor as the overlap
\beq
\mathcal{h} \mathfrak{h}\mathcal{k}\Psi\mathcal{i}|\dot{\Psi}\mathcal{i} = (-1)^{\mathfrak{f}} \, S_{\Psi} \times (\overleftarrow{\Psi}|\dot{\Psi}) = (-1)^{\mathfrak{f}} \, S_{\Psi} \times \mathcal{h}\Psi|\dot{\Psi}\mathcal{i}\, ,
\eeq
with the last equality holding for a real state, that is such that $\overleftarrow{\Psi}_{\textbf{n}}(\textbf{y}) = \Psi_{\textbf{n}}(\textbf{y})^*$. Notice that the prefactor $S_{\Psi}$ has a simple interpretation. If we think of our state as being made of the union of the $\textbf{y}$- and $\textbf{u}$- ``diagonalized excitations'', then $S_{\Psi}$ is the S-matrix obtained upon reversing the ordered set $\textbf{y}\cup \textbf{u}$. It is not really significant for the computation of the structure constant however. The reason is that the hexagon amplitude only determines the structure constant up to an overall factor that implements the change of normalization between infinite- and finite-volume Bethe states.%
\footnote{See \cite{Pozsgay:2007kn,Pozsgay:2007gx} for a discussion about this prescription in the context of diagonal S-matrix theory.} This one contains, in particular, the factor $1/\sqrt{S_{\Psi}S_{\dot{\Psi}}}$ that removes the dependence on the ordering of the rapidities and cancels (\ref{overall}) when $\Psi = \dot{\Psi}$. The grading factor $(-1)^{\mathfrak{f}}$ also cancels out in (\ref{calA}). Hence, in the end, we are left with the product $(\overleftarrow{\Psi}|\dot{\Psi})$ for the partition $(\alpha, \bar{\alpha}) = (\textbf{u}, \varnothing)$, in line with our main formula.

The proof of (\ref{main-loc}) is immediate. Permuting the $\textbf{u}$ labels in the wave function (\ref{single-w}), with $\pi(i) = K+1-i$, yields
\beq\label{act-wave}
\Psi^{\pi}_{n}(y) = \frac{a_{\pi(n)}}{y-x^{+}_{\pi(n)}} \prod_{j=1}^{n-1}S^{II, I}(y, x_{\pi(j)}) = \prod_{j=1}^{K}S^{II, I}(y, x_{j}) \times \overleftarrow{\Psi}_{\pi(n)}(y)\, ,
\eeq
after using $S^{II, I}S^{I, II} = 1$. It translates into (\ref{main-loc}) for the multiparticle wave function (\ref{multi}), with the factor $(-1)^{\mathfrak{f}}$ coming from the transformation of the determinant upon reversing of the ordering of the $\textbf{y}$ variables. Note that there is no need to impose the BAEs on the $\textbf{y}$'s to derive (\ref{act-wave}) and (\ref{main-loc}).

In the general case, we have to partition the chain into two non-empty subsets, $\alpha\cup \bar{\alpha} = \textbf{u}$. For simplicity, but with no loss of generality, we consider a partition that preserves the ordering of the full set $\textbf{u}$, that is, $\alpha = \{x_{1}, \ldots , x_{l}\}$ and $\bar{\alpha} = \{x_{l+1}, \ldots , x_{K}\}$, with $l=|\alpha|$. The action of the S-matrix $\mathcal{S}_{\pi} = \mathcal{S}_{\pi_\alpha}\mathcal{S}_{\pi_{\bar{\alpha}}}$ now depends on whether the fermion is on the $\alpha$ or on the $\bar{\alpha}$ partition, since $\pi_{\alpha}$ reverses the order of $\alpha$ only, and similarly for $\pi_{\bar{\alpha}}$. If $1 \leqslant n \leqslant l$ we find the same result as in (\ref{act-wave}), 
\beq
\Psi^{\pi}_{n}(y) = \prod_{j=1}^{l}S^{II, I}(y, x_{j}) \times \overleftarrow{\Psi}_{\pi(n)}(y)\, ,
\eeq
with the product being restricted to the $x$'s in the subset $\alpha$, but if $n > l$ we get an extra phase in front,
\beq
\Psi^{\pi}_{n}(y) = \prod_{j=1}^{K}S^{II, I}(y, x_{j})\times \prod_{j=1}^{l}S^{II, I}(y, x_{j}) \times \overleftarrow{\Psi}_{\pi(n)}(y)\, .
\eeq
Remarkably, when the roots $\textbf{y}$ are on shell, this difference becomes immaterial and the flipped wave function $\Psi^{\pi}$ is found to be proportional to the outgoing one $\overleftarrow{\Psi}$ again. After taking care of the way the determinant changes under the reordering of the $\textbf{y}$'s, we get
\beq
(-1)^{\mathfrak{f}_{\alpha}+\mathfrak{f}_{\bar{\alpha}}} A_{\pi_{\alpha}}A_{\pi_{\bar{\alpha}}}\times \prod_{j=1}^{l}\prod_{k=1}^{F}S^{II, I}(y_{k}, x_{j}) \times (\overleftarrow{\Psi}|\dot{\Psi})\, ,
\eeq
for a multi-fermion state on the split configuration. Here, $A_{\pi_\alpha} = \prod_{i<j \in \alpha}A_{ij}$ and $\mathfrak{f}_{\alpha} = F_{\alpha}(F_{\alpha}-1)/2$, with $F_{\alpha}$ the number of fermions on the subchain $\alpha$, and similarly for $\bar{\alpha}$. The grading factors disappear in the full amplitude, giving
\beq\label{su2}
\mathcal{A}_{\alpha\bar{\alpha}} = A_{\pi_{\alpha}}A_{\pi_{\bar{\alpha}}}\times S^{I, II}(\bar{\alpha}, \textbf{y}) \times (\overleftarrow{\Psi}|\dot{\Psi})\, ,
\eeq
after using the BAEs (\ref{BAEy}) and the unitarity of the S-matrix, $S^{I, II}(y, x) = 1/S^{II, I}(x, y)$, to rewrite the $\textbf{y}$-dependent factor. Formula (\ref{su2}) is the $\mathfrak{su}(2)$ counterpart of the one given in Section~\ref{NH} in the $\mathfrak{sl}(2)$ grading. They both convey the same message, namely, that the amplitude for the split configuration is proportional to the inner product $\mathcal{h}\Psi|\dot{\Psi}\mathcal{i}$ up to the phase $S^{I, II}(\bar{\alpha}, \textbf{y})$ for the scattering of the roots in $\bar{\alpha}$ with the higher level rapidities $\textbf{y}$. The overall $A$ factors in (\ref{su2}) accounts for the difference of gradings between (\ref{su2}) and (\ref{higherrankA}). They can be absorbed in the dynamical parts of the hexagon form factors using the relation between pure $\mathfrak{sl}(2)$ and pure $\mathfrak{su}(2)$ hexagon amplitudes \cite{C123Paper}
\beq
h_{ij}|_{\mathfrak{su}(2)} = A_{ij} h_{ij}|_{\mathfrak{sl}(2)}\, .
\eeq

\subsubsection*{Gaudin norm}

To conclude, let us mention that we can easily obtain the Gaudin determinant for the norm of the state in this simple model.

It stems from the fact that the multiparticle wave functions (\ref{multi}) are Slater determinants. Hence their overlap is itself a determinant
\beq\label{inner-p}
(\overleftarrow{\Psi}|\dot{\Psi}) = \sum_{n_{1}<\ldots < n_{F}}\overleftarrow{\Psi}_{\textbf{n}}(\textbf{y})\Psi_{\textbf{n}}(\dot{\textbf{y}}) = \textrm{det}\, M\, ,
\eeq
where $M$ is the $F\times F$ matrix made out of the individual overlaps,
\beq\label{Mij}
M_{ij} = \sum_{n=1}^{K} \overleftarrow{\Psi}_{n}(y_{j}) \Psi_{n}(\dot{y}_{i})\, .
\eeq
It immediately leads to the orthogonality property of on-shell Bethe states. Namely, when both the left and right rapidities $\textbf{y}$ and $\dot{\textbf{y}}$ are on shell, the product (\ref{inner-p}) vanishes, except if the two sets are the same, $\textbf{y} = \dot{\textbf{y}}$. This is because $M_{ij} = 0$ if $\dot{y}_{i}$ and $y_{j}$ are on shell and distinct. Indeed, plugging the ingoing and outgoing wave functions, (\ref{single-w}) and (\ref{out-g}), in (\ref{Mij}), with the phase (\ref{Sii-i}) and with $a_{n}^2 = i(x_{n}^{-}-x_{n}^{+})$, yields
\beq
\begin{aligned}
M_{ij} &= \sum_{n=1}^{K} \frac{a_{n}^2}{(y_{j}-x^{-}_{n})(\dot{y}_{i}-x^{+}_{n})}\prod_{k=1}^{n-1}\frac{S^{II, I}(\dot{y}_{i}, x_{k})}{S^{II, I}(y_{j}, x_{k})} \\
&= \frac{i}{y_{j}-\dot{y}_{i}}\sum_{n=1}^{K} \left(1-\frac{S^{II, I}(\dot{y}_{i}, x_{n})}{S^{II, I}(y_{j}, x_{n})}\right)\prod_{k=1}^{n-1}\frac{S^{II, I}(\dot{y}_{i}, x_{k})}{S^{II, I}(y_{j}, x_{k})} \\
& = \frac{i}{y_{j}-\dot{y}_{i}}\left(1-\prod_{k=1}^{K}\frac{S^{II, I}(\dot{y}_{i}, x_{k})}{S^{II, I}(y_{j}, x_{k})}\right)\, ,
\end{aligned}
\eeq
where in the last equality we used that the sum telescopes. Hence, when the two rapidities are on-shell the term in bracket vanishes, see (\ref{BAEy}), and so does $M_{ij}$, if $\dot{y}_{i} \neq y_{j}$.

When the two states are the same, $\Psi = \dot{\Psi}$ (or $\textbf{y} = \dot{\textbf{y}}$, with the rapidities ordered in the same way), the matrix $M$ is diagonal and the norm of the state is the product of the individual norms,
\beq\label{sublevel}
(\overleftarrow{\Psi}|\dot{\Psi}) = \delta_{\textbf{y}\dot{\textbf{y}}} \prod_{i=1}^{F}\mathcal{k}\Psi(y_{i})\mathcal{k}^2\, ,
\eeq
with
\beq\label{dy}
\begin{aligned}
&\mathcal{k}\Psi(y)\mathcal{k}^2 = \sum_{n=1}^{K} \overleftarrow{\Psi}_{n}(y)\Psi_{n}(y) = \sum_{n=1}^{K}\frac{i(x_{n}^{-}-x_{n}^{+})}{(y-x_{n}^{+})(y-x_{n}^{-})} = \sum_{n=1}^{K}i\frac{\partial}{\partial y}\log{S^{II, I}(y, x_{n})}\, .
\end{aligned}
\eeq
We can use the freedom we have to rescale the individual wave functions,  (\ref{single-w}) and (\ref{out-g}), by some function of $y$ to convert $\partial_{y}$ into $\partial_{v}$ in (\ref{dy}), with $v = g(y+1/y)$, and match the convention used in the bulk of the paper, see (\ref{Gaudin}). This choice is irrelevant for the normalized structure constant, which is divided by the square root of the Gaudin determinant $G$ for the full wave function (which includes the level I and all the lower ones).%
\footnote{Implicit here is the assumption that the norm of the level II wave function (\ref{sublevel}) matches with a minor of $G$, as obtained by freezing the level I rapidities $\textbf{u}$.}

\end{document}